\documentclass{jkas}

\def\year{2025} 
\def\volume{} 
\def\issue{} 
\def\beginpage{1} 
\def\received{---} 
\def\accepted{---} 
\def\published{---} 
\date{Received \received; Accepted \accepted; Published \published}

\usepackage{multicol}
\usepackage{multirow}
\usepackage{makecell}
\usepackage{graphicx}
\usepackage{caption}
\usepackage{lipsum}

\title{%
Characterization of the commercial spectrograph system for astronomical observations: PIXIS 1300BX Camera and IsoPlane 320A Spectrograph 
}

\author[1]{Jiwon Jang}{0009-0003-1999-0913}
\author[2]{Changsu Choi}{0000-0001-8642-1241}
\author[1,3]{Ho Seong Hwang}{0000-0003-3428-7612}
\author[4]{Haeun Chung}{0000-0002-3043-2555}
\author[1]{Hyeonguk Bahk}{0009-0002-9878-1126}
\author[1,5]{Dongkok Kim}{0000-0003-4127-6110}
\author[6]{Jae-Woo Kim}{0000-0002-1710-4442}

\affil[1]{Astronomy Program, Department of Physics and Astronomy, Seoul National University, Gwanak-gu, Seoul 08826, Republic of Korea}
\affil[2]{SPEX Inc., R\&D Center: 28165, 215, Yeonge 1-gil 24-35, Osong-eup, Heungdeok-gu, Cheongju-si, Chungcheongbuk-do, Republic of Korea}
\affil[3]{SNU Astronomy Research Center, Seoul National University, Gwanak-gu, Seoul 08826, Republic of Korea}
\affil[4]{University of Arizona, Steward Observatory, 933 N. Cherry Ave., Tucson, AZ 85721, USA.}
\affil[5]{Institute for Data Innovation in Science, Seoul National University, Seoul 08826, Korea}
\affil[6]{Korea Astronomy and Space Science Institute, 776 Daedeok-daero, Yuseong-gu, Daejeon 34055, Republic of Korea}

\def\corrauthor{%
Ho Seong Hwang
\email{hhwang@astro.snu.ac.kr}
}

\def\runningauthor{%
Jang et al.
}

\def\runningtitle{%
Characterization of the commercial spectrograph system for astronomical observations
}

\def\keywords{%
spectroscopic instrument -- characterization -- shutter-less CCD   
}

\def\abstracttext{%
We present the result from a comprehensive laboratory and on-sky characterization of the commercial spectrograph system consisting of a PIXIS 1300BX charge-coupled device (CCD) camera and an IsoPlane 320A spectrograph as part of the preparation of the forthcoming all-sky spectroscopic survey of nearby galaxies (A-SPEC). In the laboratory, we have quantified readout noise, dark current, gain, and full-well capacity via bias, dark, and photon transfer curve analysis at all acquisition modes. To do that, we have developed a gradient correction technique to address row-dependent signal gradients in the image, which are caused by the shutter-less condition of our CCD camera test setup. The technique successfully reproduces the values in the manufacturer specifications. We also have measured quantum efficiency exceeding $80\%$ from 400–800 $\rm nm$ and $\gtrsim90\%$ between 450–750 $\rm nm$, with sub-second persistence decay, making it ideal for rapid, multi-object spectroscopy. Using a set of diffraction gratings (150, 300, and 600 $\rm gr\;mm^{-1}$), we have evaluated the spatial separability of multiple spectra and spectral resolution. We have conducted a test observation with this spectrograph system at the Seoul National University Astronomical Observatory (SAO) 1 m telescope and successfully demonstrated its capability of multi-object spectroscopy with moderate resolution of $R\approx 600-2600$. We release all Python codes for the test and recipes to facilitate further instrument evaluations.}

\begin{document}
\jkashead 
    
\section{Introduction} 
    Advances in detector technology have driven breakthroughs in astronomy. Over the past few decades, the numerous data products from large sky surveys have brought unprecedented improvements to astronomical discoveries. In particular, charge-coupled devices (CCDs) have been widely used for photometry and spectroscopy, primarily owing to their high sensitivity, high linearity, long-term stability, and low readout noise \citep[e.g.,][]{Janesick2001, Howell2006}. Meanwhile, complementary metal‐oxide‐semiconductor (CMOS) image sensors have gained considerable traction in astronomical applications, benefiting from faster readout speeds, reduced power consumption, and widespread commercial availability \citep[e.g.,][]{Fossum1993, Janesick2007}. Advances in CMOS technology—particularly in reducing readout noise and improving sensitivity—made them comparable or even superior to CCDs.   
    
    Nonetheless, CCDs remain highly competitive for both photometric and spectroscopic observations of faint targets, where long exposures render readout speed less critical. Furthermore, a multi-object spectrograph (MOS) using robotic fiber positioners requires additional time for fiber alignment between successive sky fields \citep[e.g.,][]{Fabricant2005,DESICollaboration2016}. In such cases, sensitivity and stability outweigh readout speed. Although modern CMOS sensors have achieved notable improvements, they often exhibit a steeper decline in sensitivity beyond roughly $600\;\rm nm$ \citep{Alarcon2023,Khandelwal2024,Layden2025}, whereas CCDs maintain high quantum efficiency across a broad wavelength range. Therefore, many spectroscopic surveys still adopt CCDs for their observations \citep[e.g.,][]{Gunn1998, Bebek2017}.

    The CCD camera (PIXIS 1300BX) and the spectrograph (IsoPlane 320A) in this study were selected as part of the preparatory work for the all-sky spectroscopic survey of nearby galaxies \citep[A-SPEC;][]{Kwon2025}. These instruments are intended for the commissioning observations of A-SPEC. The main survey itself will be conducted using a spectrograph developed at Australian Astronomical Optics--Macquarie, which is a revised version of the TAIPAN spectrograph \citep{Staszak2016}.
    
    To select the best combination of commercially available spectrographs for our purpose, we imposed several key requirements. First, the spectral resolution needed to be comparable to the scientific requirements of A-SPEC ($R\approx 1600-2000$), with the flexibility to adjust the resolution if necessary. This requirement is met by the IsoPlane 320A through its triple-grating turret, which allows three gratings to be installed and selected interchangeably. Second, the CCD was required to have a high quantum efficiency over the wavelength range relevant to the survey, specifically $\rm QE\gtrsim 80\%$ across $370-870\;\rm nm$. Third, the spectrograph needs to support multi-object spectroscopy with a detector large enough to capture multiple fiber-fed spectra simultaneously. The PIXIS 1300BX, with a detector size $26.8\times26.0\rm\;mm^2$, comfortably covers the required focal-plane size $27\times 22\rm\;mm^2$. At the time of instrument selection in 2021, the combination of IsoPlane 320A and PIXIS 1300BX represented the optimal choice that satisfied all of these criteria within the available budget (approximately 150 million KRW). Although these instruments are used only for commissioning purposes, the performance tests and operational experience gained through this study will be directly applicable to the main A-SPEC instruments in future work. The spectrograph system is currently installed at the 1 m telescope at Seoul National University Astronomical Observatory (SAO).
    
     In this paper, we characterize the CCD camera and the spectrograph intended for A-SPEC surveys, and make our Python test code publicly available. 
     This paper is organized as follows. In Section \ref{sec:instruments}, we summarize the basic specifications of the target instruments as in the datasheet of the manufacturer and outline the laboratory setup. In Section \ref{sec:CCD_test}, we describe the methodology and results of the CCD camera characterization. We also propose a novel approach for managing the shutter-less CCD intended for spectrograph applications in a laboratory environment, incorporating its effect into photometric noise analysis. Our results encompass gradient effect correction, bias level time variability, readout noise, dark current, spatial non-uniformity of thermal noise, photon transfer curve (PTC), linearity, full well capacity (FWC), quantum efficiency (QE), and charge persistence measurements. In Section \ref{sec:spectrograph test}, we present spectrograph characterization results, including spatial separability of multi-fiber spectra and spectral resolution. In Section \ref{sec:on-sky}, we validate performance with on-sky observations of the standard stars as a practical demonstration. Finally, we propose a standardized procedure for evaluating both spectrographs and cameras in future spectroscopic survey programs, and we provide all associated analysis routines as publicly available Python packages, \texttt{EvalCCD} and \texttt{EvalSpec}.

\section{Instruments \label{sec:instruments}}
\subsection{CCD \label{sec:CCD}}
    We test the PIXIS 1300BX Camera\footnote{\url{https://www.teledynevisionsolutions.com/products/pixis/}}, manufactured by Teledyne Princeton Instruments (PI). This camera features a back-illuminated design that makes the detector highly sensitive in both visible and near-infrared (NIR) wavelengths, suppressing the etalon phenomenon (i.e., interference between reflected light from the surface and reflected light from the bottom plane of the silicon substrate) in the NIR regime. It has $1340\times1300$ pixels with a pixel size of $20\;\mu\rm m$. The camera includes thermoelectric cooling down to $-65\;\rm ^\circ C$. The camera features low dark currents and low readout noise, typically measuring $0.01\;e^-\rm\;pixel^{-1}\;s^{-1}$ (at $-60\;\rm ^\circ C$) and $12\;e^-\;\rm rms$ (at $2\;\rm MHz$ readout speed), respectively. While the camera features an electronic shutter, it does not include a mechanical shutter. An external mechanical shutter is only available when the camera is mounted on the spectrograph. The characteristics provided by the manufacturer are summarized in Table \ref{tab:technical_info}.
    \begin{table}[h]
        \caption{Technical information of PIXIS 1300BX\label{tab:technical_info}}
        \centering
        \begin{tabular}{ll}
        \toprule
            Features & Value \\
        \midrule
            CCD format & $1340\times1300$ ($26.8\times26.0\;\rm mm^2$)\\
            Pixel size & $20\;\mu \rm m$\\
            Full well &  $250\;\rm{k}e^-$(single pixel) \\
                      & $1000\;\rm{k}e^-$(output node)\\
            Dark Noise & $0.01\;e^-\rm\;pixel^{-1}\;s^{-1}$ (@$-60\;\rm ^\circ C$)\\
            Readout Noise & $12\;e^-\;\rm rms$ (@$2\;\rm MHz$)\\
            A/D sampling depth & 16 bit\\
            A/D sampling rate & $100\;\rm kHz$ / $2\;\rm MHz$\\
            Shutter type & EM shutter\\
                        & (external shutter port available)\\
            Cooling & $< -65\;\rm ^\circ C$\\
        \bottomrule
        \end{tabular}
    \end{table}
    
\subsection{Optical Test Bench}
\begin{figure*}[ht]
        \centering
        \includegraphics[scale=0.98]{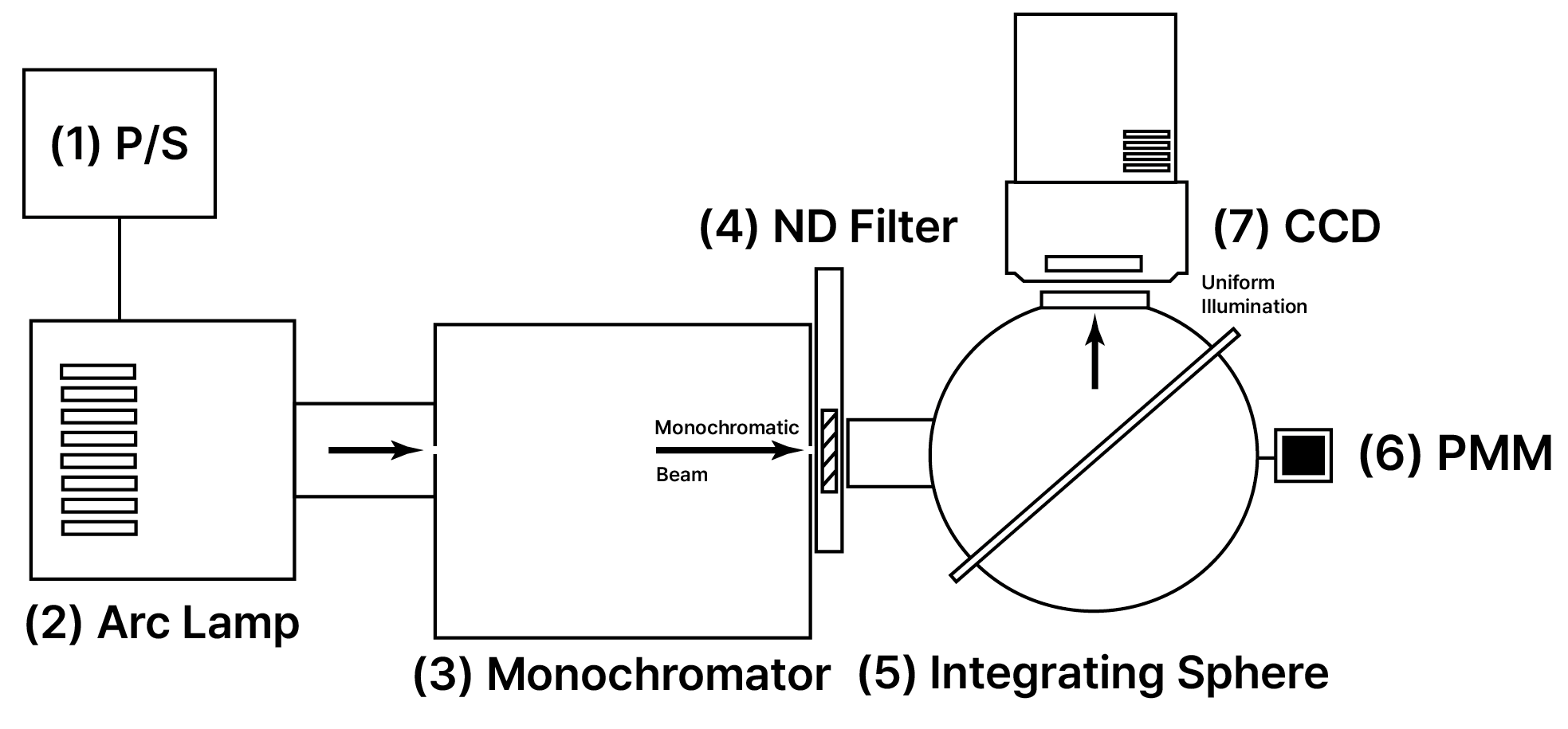}
        \caption{Schematic of optical bench setup for testing CCD camera. (1) power supply, (2) arc lamp, (3) monochromator, (4) ND filter, (5) integrating sphere, (6) power meter, (7) CCD camera. The power meter is drawn along the same axis as the input port for clarity; in our actual setup, it is mounted on the integrating sphere at 90° to both the lamp input and the CCD illumination port.}
        \label{fig:Optical_Bench}
    \end{figure*}

   To conduct the CCD characterization experiments, we set up the optical test bench depicted in Figure \ref{fig:Optical_Bench}. All bias and dark-frame measurements are carried out in a dark room, temperature-controlled room ($16\;\rm ^\circ C$, $40-50\%$ RH). For illuminated tests—such as photon transfer curve, linearity, and charge persistence measurements—we require highly uniform flux on the sensor surface. 
   
   To achieve this, we utilize an integrating sphere, transforming incident light into a uniform output through diffuse reflection. This approach ensures homogeneous illumination without any gradients, enhancing the reliability of our result \citep[e.g.,][]{Labsphere2017,Nayak2001}.
   We couple a continuum Xe lamp\footnote{\url{https://sciencetech-inc.com/shop/category/light-sources-xe-arc-lamp-light-source-9}} into a Newport 6" integrating sphere\footnote{\url{https://www.newport.com/p/819D-SF-6}} via a monochromator\footnote{\url{https://dxg.kr/portfolio/monochromator-spectrograph-monora-200/}} (Figure \ref{fig:Optical_Bench} (2), (3), (5)). The integrating sphere has an exit port with a diameter of $63.5\;\rm mm$ (2.5”), which is large enough to cover the entire CCD sensor; the sensor has a diagonal length of $37.3\;\rm mm$, ensuring uniform illumination across the sensor surface {\bf (see the second paragraph of Section \ref{sec:QE} for more details}).
   We insert a neutral density (ND) filter\footnote{\url{https://www.edmundoptics.com/p/20-od-25mm-dia-uv-nir-nd-filter/30533/}} immediately downstream of the monochromator (Figure \ref{fig:Optical_Bench} (4)) to fine-tune the incident flux moderately for our highly sensitive PIXIS 1300BX Camera. To monitor the intensity of the light source, we mount a Newport 918D-ST wand detector\footnote{\url{https://www.newport.com/p/918D-ST-SL}} with a Newport 1919-R power meter monitor (PMM)\footnote{\url{https://www.newport.com/p/1919-R}} at an auxiliary aperture of the integrating sphere to monitor the absolute light level in real time (Figure \ref{fig:Optical_Bench} (6)). This configuration ensures uniformity and precise flux control, yielding reliable and reproducible camera performance metrics. During quantum-efficiency measurements, we scan the monochromator output across the $300-950\;\rm nm$ range; for all other illuminated tests, we hold the output steady at $2.89\pm0.01\;\rm nW$ and $500\pm1.8\;\rm nm$.  
    
\subsection{Spectrograph and Optical Fiber} 
\begin{figure}[ht]
        \centering
        \includegraphics[width=0.99\linewidth]{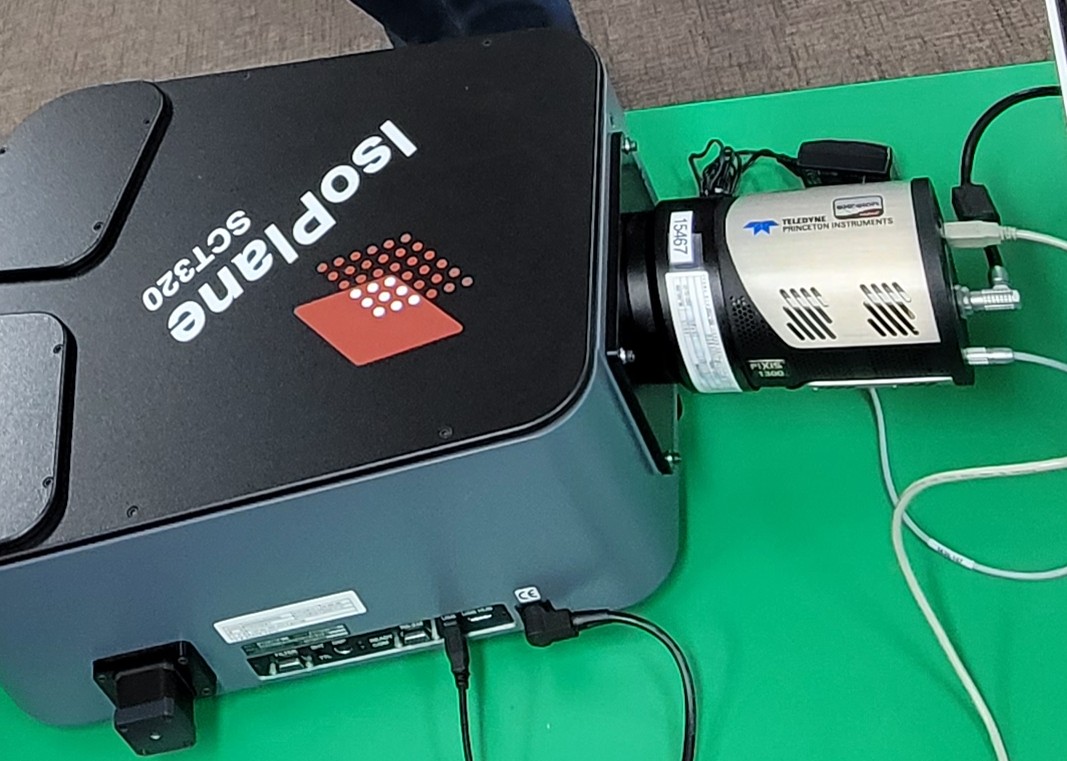}
        \caption{Attachment of PIXIS 1300BX onto IsoPlane 320A.}
        \label{fig:attat}
    \end{figure}

\begin{figure*}
        \centering
        \includegraphics[scale=0.98]{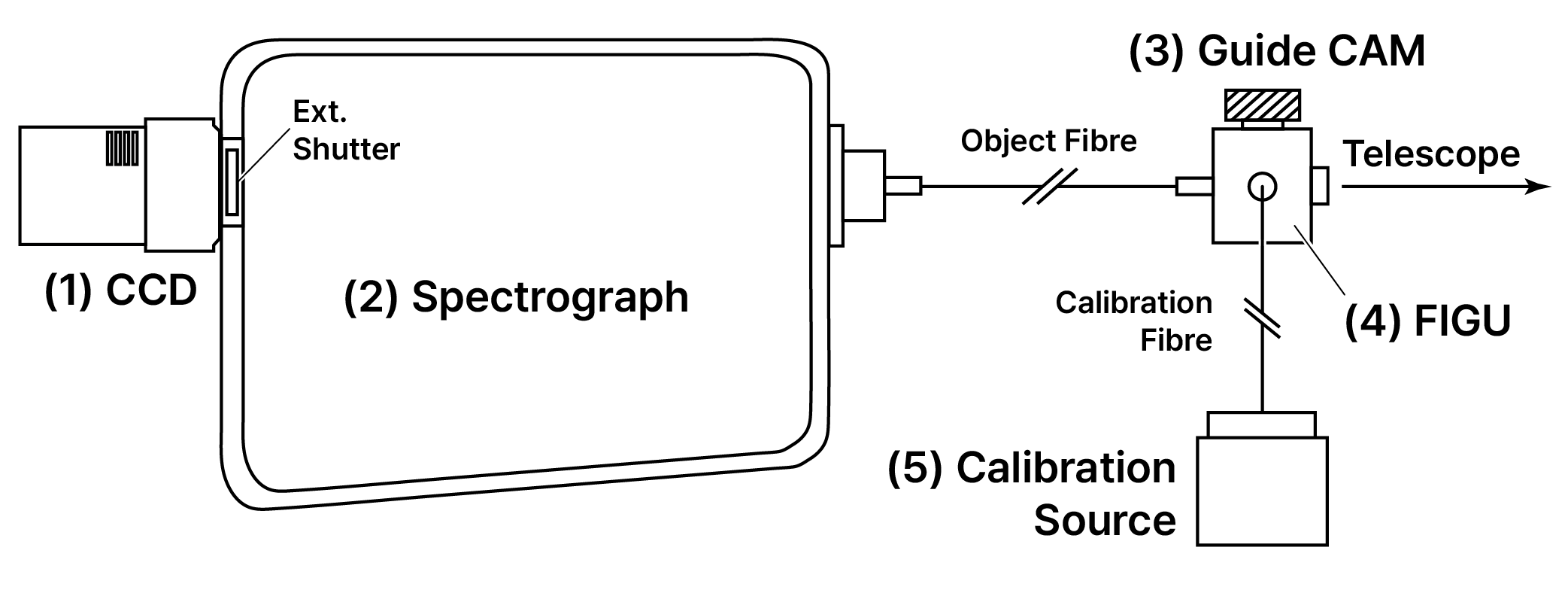}
        \caption{Configuration for testing spectrograph. (1) camera (PIXIS 1300 BX), (2) spectrograph (IsoPlane 320A), (3) guide camera, (4) FIGU, (5) calibration source}
        \label{fig:spectrograph}
    \end{figure*}

\begin{figure}[ht!]
        \centering
        \includegraphics[width=0.99\linewidth]{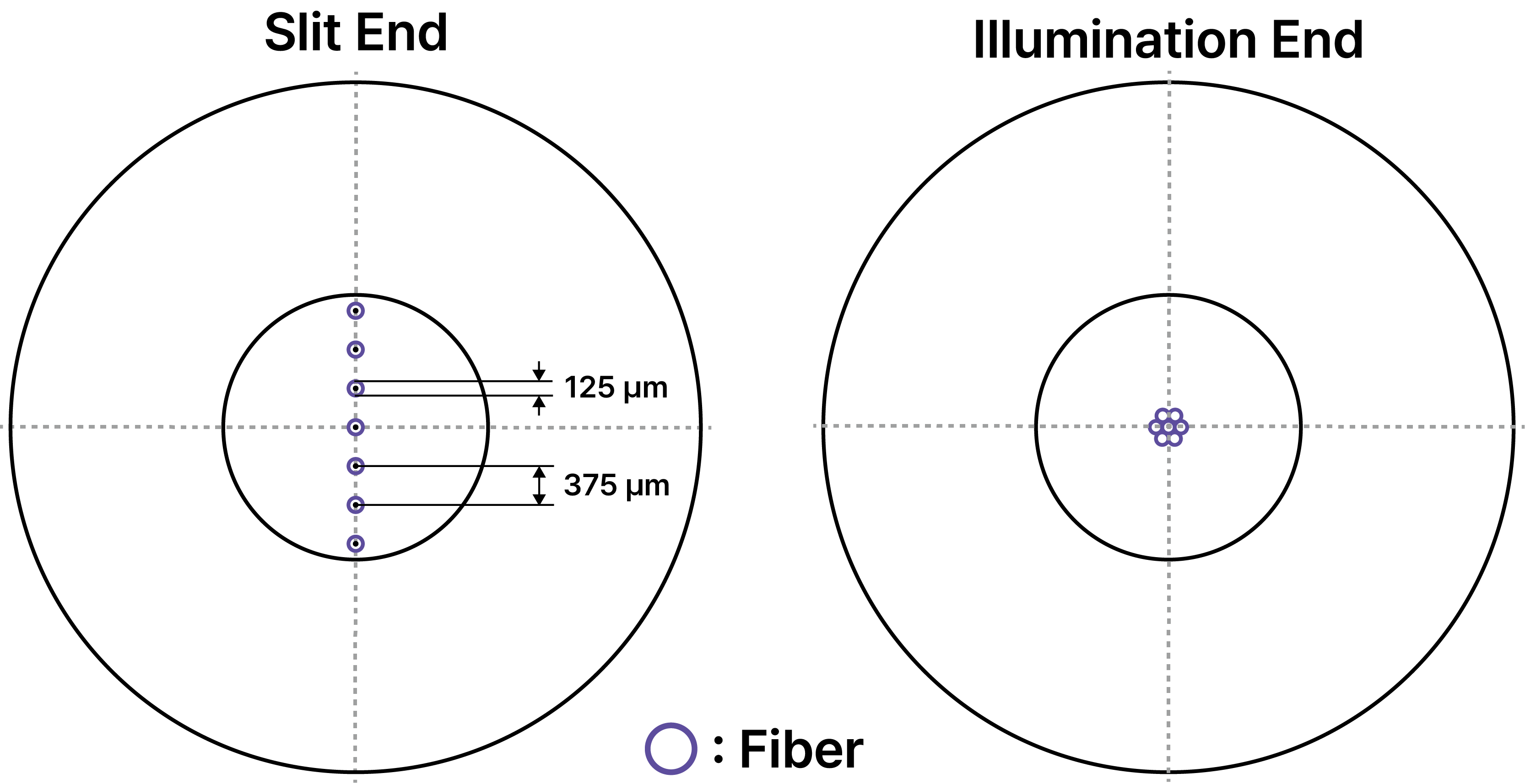}
        \caption{Schematic of the 7-core optical fiber bundle. Left panel depicts the slit end of the fiber, connected to the spectrograph. Right panel depicts the illumination end, connected to FIGU. The violet and black solid circles represent the individual fiber claddings and metal jackets, respectively.}
        \label{fig:fiber_schematic}
    \end{figure}

    We tested the IsoPlane 320A\footnote{\url{https://www.teledynevisionsolutions.com/ko-kr/products/isoplane/}} Spectrograph by coupling our PIXIS 1300BX Camera at its imaging plane (see Figures \ref{fig:attat} and \ref{fig:spectrograph}). The main specifications of IsoPlane 320A are summarized in Table \ref{tab:isoplane320a}. We select three diffraction gratings (groove density of 150, 300, and 600 $\rm gr\;mm^{-1}$) to cover various scientific applications, all calibrated using a Th–Ar lamp\footnote{\url{https://www.shelyak.com/produit/pf0012-etalonnage-eshel/}} (Figure \ref{fig:spectrograph} (5)).
    \begin{table}[ht]
      \centering
      \caption{Technical specifications of the IsoPlane 320A Spectrograph}
      \label{tab:isoplane320a}
      \begin{tabular}{llcc}
        \toprule
        \multicolumn{1}{l}{\textbf{General specifications}} & & &\\
        \midrule
        Focal length            & \rlap{$320\;\rm mm$}  & &      \\
        Aperture ratio          & \rlap{$\rm F/4.6$}    & &       \\
        Astigmatism             & \rlap{Zero} & & \\
        \midrule
        \multicolumn{1}{l}{\textbf{Grating specifications}} 
                                & \multicolumn{3}{c}{Groove density [$\rm gr\,mm^{-1}$]} \\
        \cmidrule(lr){2-4}
        Feature                 & 150    & 300    & 600    \\
        \midrule
        \makecell[l]{Reciprocal Linear\\Dispersion$^{\rm a}$ [$\rm nm\;mm^{-1}$]} 
        & 18.4   & 9.2    & 4.6    \\
        \makecell[l]{Wavelength \\ Coverage$^{\rm b}\;\rm[nm]$}
        & 493    & 247    & 123    \\
        \bottomrule
      \end{tabular}
      \tabnote{$^{\rm a}$ Derived value from datasheet (detailed in Section \ref{sec:spectral resolution})\\ $^{\rm b}$ with a $26.8\;\rm mm$ wide sensor as an imager}
    \end{table}
    For the spatial separability test of the spectrograph, we use a 7-core optical fiber bundle serving as an object fiber. We illustrate the configuration of the independent fibers at both end of our multi-core fiber bundle in Figure \ref{fig:fiber_schematic}. Each fiber has a $100\;\mu \rm m$ diameter core and a $125\;\mu\rm m$ diameter cladding. The fibers are arranged in a hexagonal structure without any spacing at the illumination end, while they are aligned horizontally with a center-to-center spacing of $375\;\mu \rm m$ at the exit face of the bundle. We adjusted the exit face to be oriented parallel to the slit, with the spectrograph input slit width set wider than the individual fiber cores to ensure maximum throughput.
    \begin{figure}[ht!]
        \centering
        \includegraphics[width=0.98\linewidth]{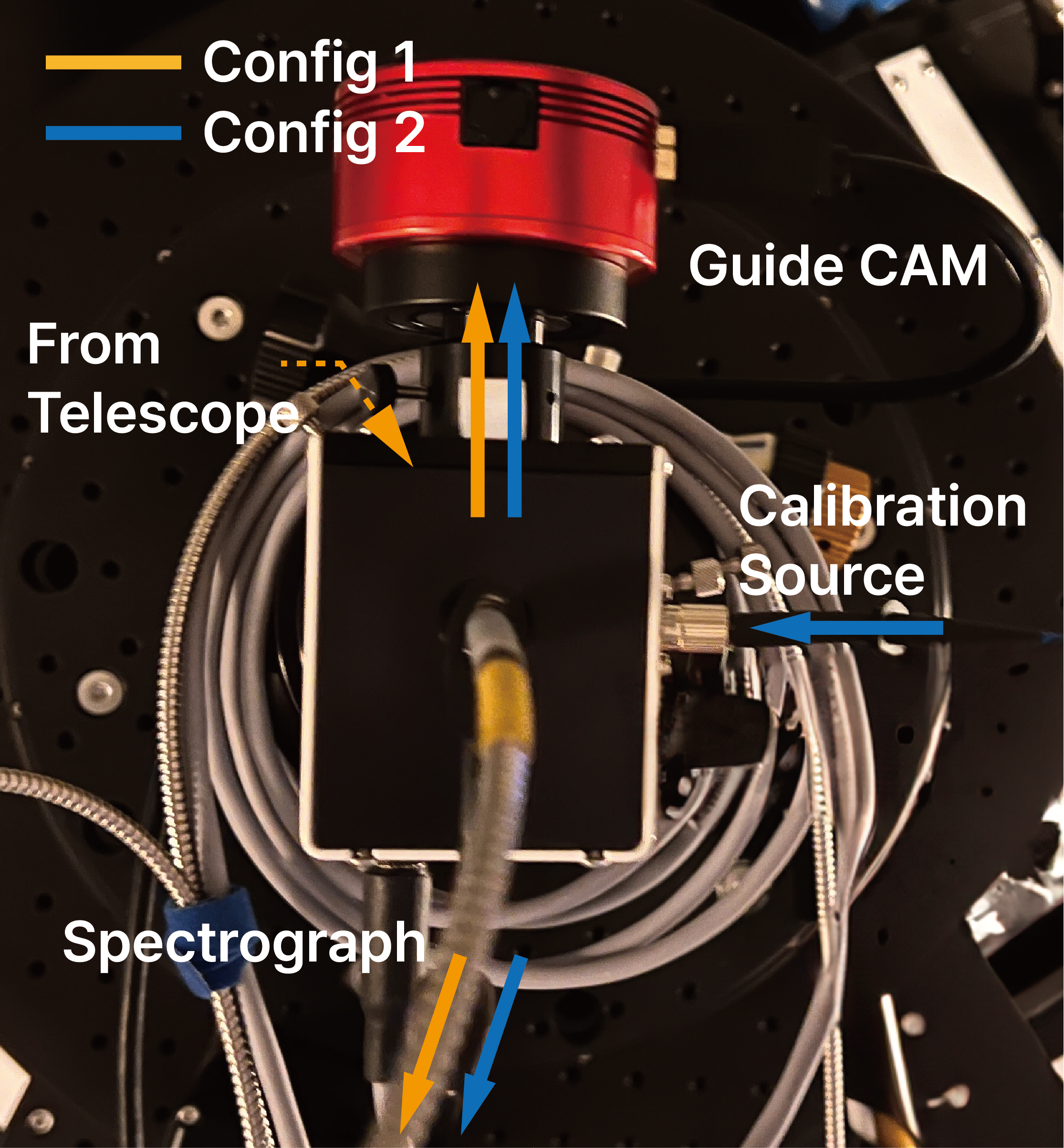}
        \caption{FIGU configuration for on-sky test (Config 1) and wavelength calibration (Config 2). FIGU mechanically switches the ray path by altering the mirror configuration with an electronic magnet.}
        \label{fig:FIGU}
    \end{figure}
    
     To verify spectral resolution and demonstrate scientific calibration during our test observations, it is essential to direct the beam path to achieve optimal illumination of the fiber input face. 
     For this purpose, we employ a fiber guiding unit (FIGU)\footnote{\url{https://www.shelyak.com/produit/pf0008-bonnette-f-6-50\%C2\%B5m/}} (Figure \ref{fig:spectrograph} (3), (4)). FIGU also enables easy modification of the beam path from the object (i.e., light from the telescope) to the calibration source through its configurable mirror (Figure \ref{fig:FIGU}). We use Configuration 2 for wavelength calibration and Configuration 1 for the on-sky test.
     However, our multi-core fiber could not be incorporated with FIGU due to the interface mismatch. We therefore utilize a single-core optical fiber with $50\;\mu \rm m$ core diameter exclusively for the on-sky test.
  
\subsection{Instrument Control Software}
    To control the spectrograph + CCD camera system, we use Teledyne \texttt{LightField}\footnote{\url{https://www.teledynevisionsolutions.com/en-au/products/lightfield/}}, which offers an interactive GUI for integrated control of spectrograph and camera operation. However, \texttt{LightField} cannot automate long sequences of exposures or dynamically vary settings in a pre‑defined order. For standalone CCD tests, we therefore built a custom C++ application on \texttt{PICAM} SDK\footnote{\url{https://www.teledynevisionsolutions.com/en-hk/products/picam-sdk-amp-driver/}}. Our \texttt{PICAM}-based control software fully configures all camera parameters and can schedule sequential exposures at arbitrary time intervals within a specified time range. This automated workflow allows us to probe camera behavior over a wide range of illumination levels and exposure to the presence of a dark signal without manual intervention. 
    
    The general exposure settings include two primary configuration parameters: the gain and the image quality. The gain determines the amplification level of the on-board amplifier, which is available in three modes: low, medium, and high gain. We note that higher gain modes are suitable for fainter signal settings but have lower values in $e^{-}\;\rm ADU^{-1}$. In parallel, the image quality setting determines the dynamic range of each pixel, offering two options: low-noise mode and high-capacity mode. The combination of each gain and image quality mode varies the noise level and gain. Low-noise mode offers a limited dynamic range at a given gain mode, utilizing the full-well capacity range of each pixel, whereas high-capacity mode maximizes the dynamic range in exposure using the full range of potential well capacity. This process automatically adjusts the amplifier gain to ensure a wide dynamic range; we discuss this in detail in Section \ref{sec:linearity}. 
    
    The software also allows the temperature setpoint to be varied within the range of $-65$-$+25\;\rm ^\circ C$. During the cooling process, the software iteratively reports the current sensor temperature at 5 s intervals. Once the temperature reaches the designated setpoint, the feedback process stabilizes the temperature. According to the datasheet, the temperature is maintained with an accuracy of $0.05\;\rm ^\circ C$. However, as the \texttt{PICam} driver only reports integer $^\circ \rm C$ values, which can limit the measurement of temperature variations by more than $1\;\rm^\circ C$. During the test, we set the sensor temperature to $-55\;\rm ^\circ C$, a sufficiently low level to suppress dark current and ensure stability as recommended by the manufacturer. Throughout the experiment, we did not find any significant temperature drifts. 
    
\section{Characterization of the PIXIS 1300BX\label{sec:CCD_test}} 
    In this section, we outline the methodology for assessing a shutter-less CCD camera, specifically, the PIXIS 1300BX, when mounted directly at the spectrograph focus without its mechanical shutter. In most astronomical spectrographs, mechanical shutters are commonly located at the slit entrance rather than directly at the CCD sensor; therefore, we should take this into account when testing the detector solely from its original instruments. We introduce an illumination-matched bias technique. Although the monochromator provides an internal shutter, it cannot be hardware-synchronized with the CCD readout due to incompatibility with the external trigger port of the PIXIS 1300BX. We record bias frames under the identical lamp and filter settings for illumination experiments. Subtracting these illuminated bias frames can remove the row-dependent exposure gradient. 
    
    We then compare our resulting measurements—readout noise, gain, linearity, full-well capacity, and quantum efficiency—directly against the datasheet provided by the manufacturer. This is important because we plan to use the same evaluation setup and procedures for the main instruments of the A-SPEC survey. Therefore, it is essential to test the reliability of our methodologies and pipelines to ensure they can accurately reproduce the well-known manufacturer-provided properties.
     
\subsection{Bias\label{sec:bias}}
     We acquire two types of bias frames in the same camera mode: dark bias and illuminated bias. The dark bias is acquired with zero-second exposures while the light source is turned off, isolating the pure readout effect. In contrast, the illuminated bias is obtained with zero-second exposures under uniform lamp illumination, revealing any extra signal accumulated during readout.
\subsubsection{Gradient Effect Correction \label{sec:gradient}}
    \begin{figure*}[ht!]
        \centering
        \includegraphics[width=0.32\linewidth]{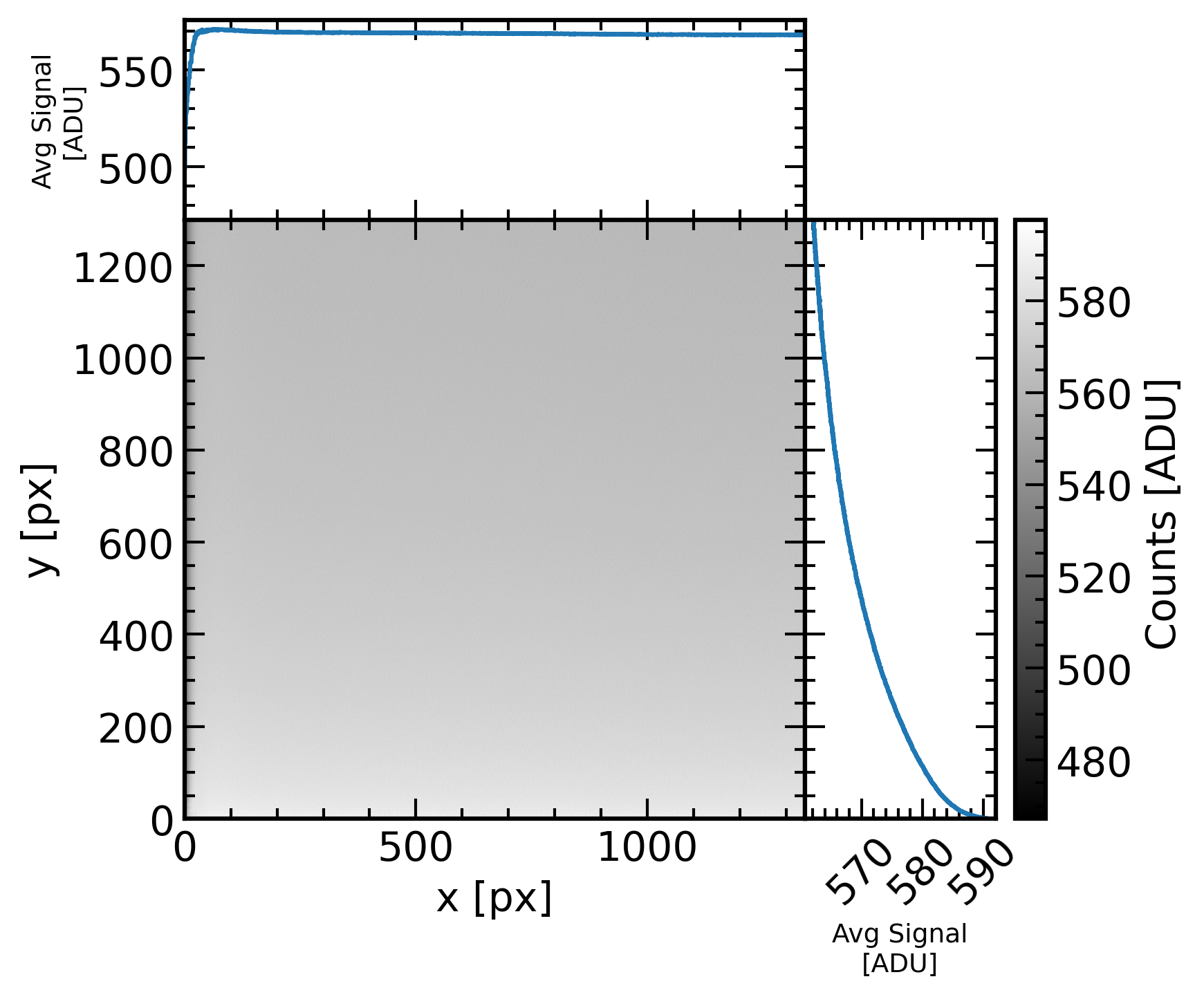}
        \includegraphics[width=0.33\linewidth]{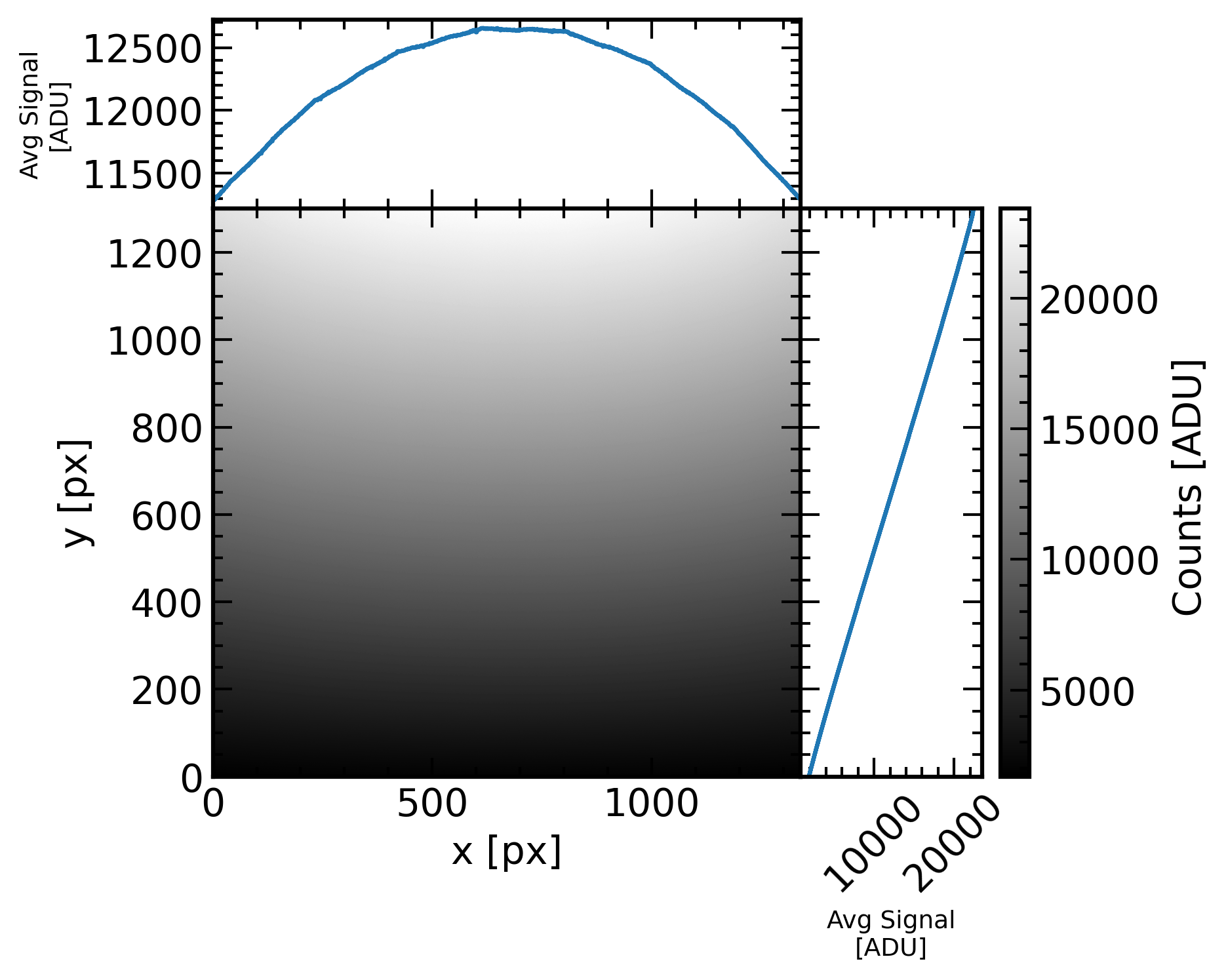}
        \includegraphics[width=0.33\linewidth]{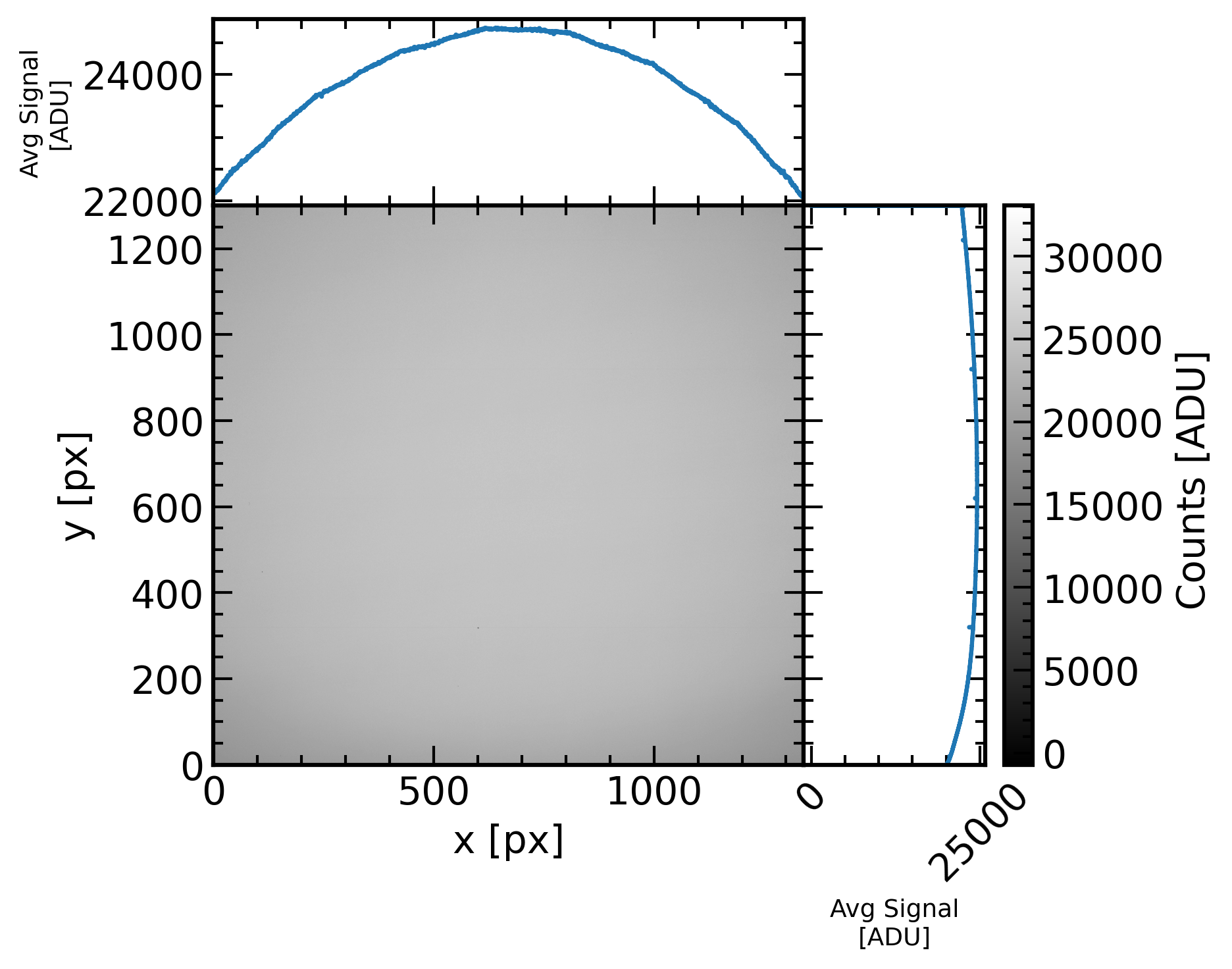}
        \caption{Image of the dark bias (left), the gradient (center), and the difference (right) frame with mean vertical and horizontal profiles. Both frames are obtained at low-noise, high-gain mode.}
        \label{fig:bias_frame}
    \end{figure*}
    Because the PIXIS 1300BX was tested off the spectrograph (i.e., no mechanical shutter at the detector), each pixel in a bare CCD accumulates photons during the sequential readout process, producing an unavoidable exposure-time difference across the vertical axis of the detector. 

    In Figure \ref{fig:bias_frame}, the dark bias frame (left) appears almost homogeneous in the horizontal direction and exhibits a slight vertical slope ($\sim20\;\rm ADU$) even with no illumination. The subtle vertical slope is a well-known phenomenon caused by a gradual change of the DC offset along the row direction \citep{Howell2006}. This pattern, determined by the internal electronics of the CCD, remains consistent over time. We can easily correct this row-dependent offset by subtracting the dark bias frame and applying well-known standard data preprocessing procedures \citep[e.g.,][]{Chromey2010}.

    Subtracting a dark bias from an illumination bias (center) uncovers a strong vertical gradient, reaching $\sim20,000\;\rm ADU$ difference from top to bottom under low-noise, low-gain settings. In contrast, the horizontal profile remains symmetric, consistent with the expected light distribution from the circular aperture of the integrating sphere. The smooth, two-dimensional convex distribution becomes clearer in the difference map between two flat frames taken with different exposure times (right panel).
    
    This substantial gradient emphasizes the importance of accounting for the excessive illumination during the CCD readout process. This effect introduces signals larger than those expected, resulting in the injection of additional shot noise. As a result, this effect may cause systematic biases in subsequent characterizations. Accurate gradient correction is thus essential for precise CCD calibration, even at relatively low illumination levels. In the following sections, we present a method for measuring and correcting this gradient effect. We also demonstrate the effectiveness and necessity of such correction for the various characterizations of the shutter-less CCD: the Photon Transfer Curve (PTC) analysis (in Section \ref{sec:PTC}), Linearity measurements (in Section \ref{sec:linearity}), and Quantum Efficiency (QE) evaluations (in Section \ref{sec:QE}).

    For clarification, this correction is specific to bench-top CCD tests; this phenomenon occurs because we detached the CCD from the spectrograph for separate characterization. By contrast, most scientific observations (i.e., combined operation with a spectrograph) employ a proper mechanical shutter. It does not apply to calibration routines for science exposures of genuine astronomical targets.
    
\begin{figure}[ht]
        \centering
        \includegraphics[width=1.0\linewidth]{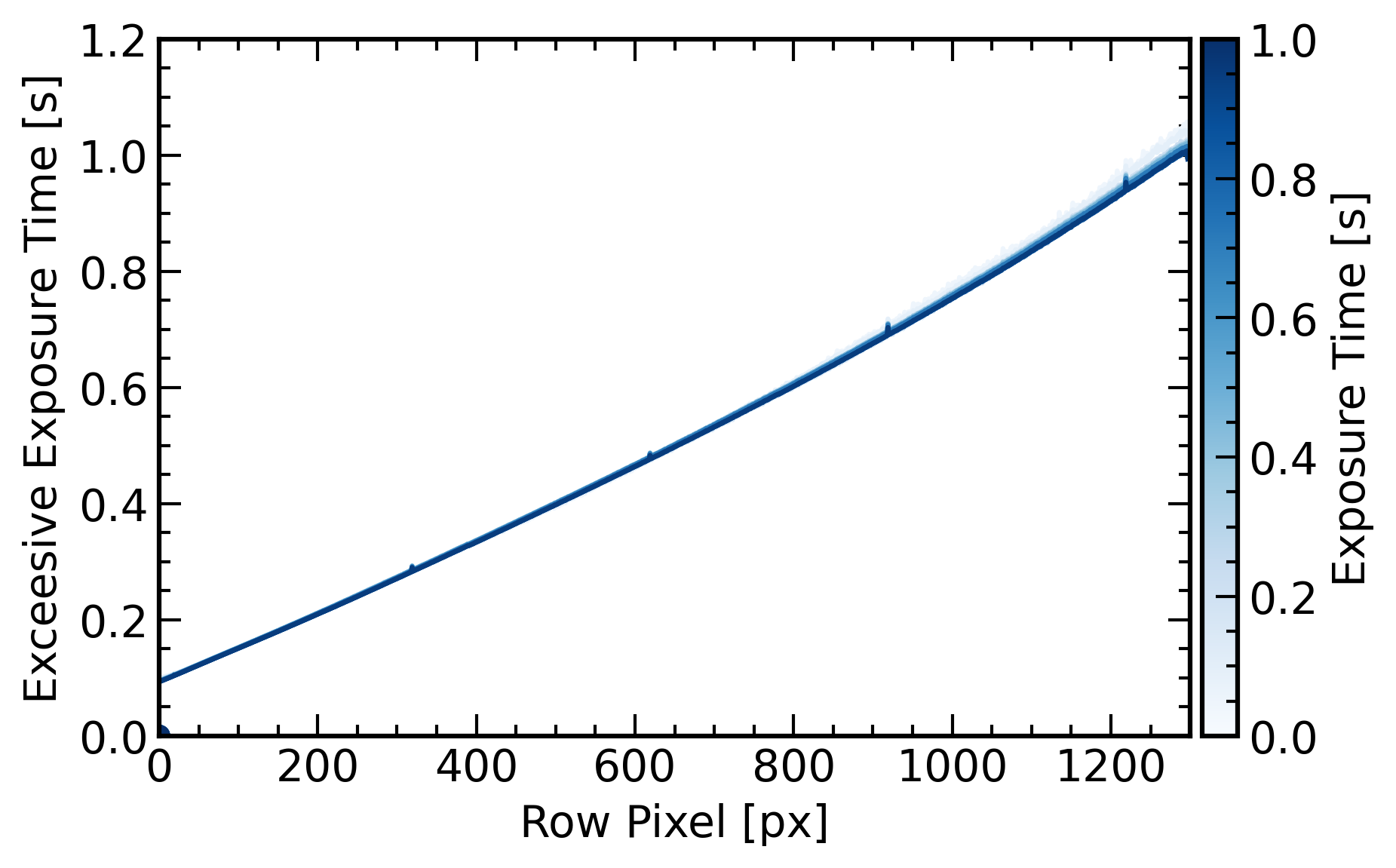}
        \caption{Excessive exposure time in a vertical direction obtained without a mechanical shutter. The color of each curve represents the prearranged exposure time.}
        \label{fig:excessive_exposure}
    \end{figure}

    We characterize the exposure time difference along the readout direction and examine whether these exposure time differences are independent of the nominal (prearranged) exposure time used for the flat frames. We define the Gradient and the Difference at each pixel row by
    \begin{equation}
        \rm Gradient = Bias - Bias^*
        \label{gradient}
    \end{equation}
    \begin{equation}
        \textrm{Diff} = \frac{\rm Flat_1-Flat_2}{\Delta t}
        \label{diff}
    \end{equation}
    The asterisk $(^*)$ denotes a dark bias frame, and $\rm Flat_1$ and $\rm Flat_2$ are independent flat frames in different exposure times with a difference of $\Delta t$. Flat frames satisfy the maximum signal level lies below $50,000\;\rm ADU$, ensuring the linear response of the CCD. In this setup, the difference frame represents the flux distribution of the incident light source. We calculate the excessive exposure time by
    \begin{equation}
        \Delta t_{\rm excess} = {\textrm{Gradient} \over \textrm{Diff}}
    \end{equation}
    In Figure \ref{fig:excessive_exposure}, we find the approximate 0.9 seconds between the highest row and lowest row, consistent with the expected readout time difference at 2 MHz analog-to-digital conversion (ADC) speed over the full pixel array, calculated as follows
    \begin{equation}
        \Delta t_{\rm excess,max} = {N_{\rm row}N_{\rm col} \over f_{\rm ADC}}\approx 0.87\;\rm{s}
    \end{equation}
    where $N_{\rm row}$ and $N_{\rm col}$ are sensor dimensions and $f_{\rm ADC}$ is readout speed of the ADC. The excessive exposure time profile remains consistent across all prearranged exposure times. This confirms that the gradient phenomena are purely caused by the readout procedure, not the intended integration time. Correcting for this gradient requires subtracting the illumination-matched bias frame ($\rm Bias$) from each flat frame.
\subsubsection{Readout Noise and Gradient Noise \label{sec:RDN}}
\begin{figure}[ht]
        \centering
        \includegraphics[width=0.94\linewidth]{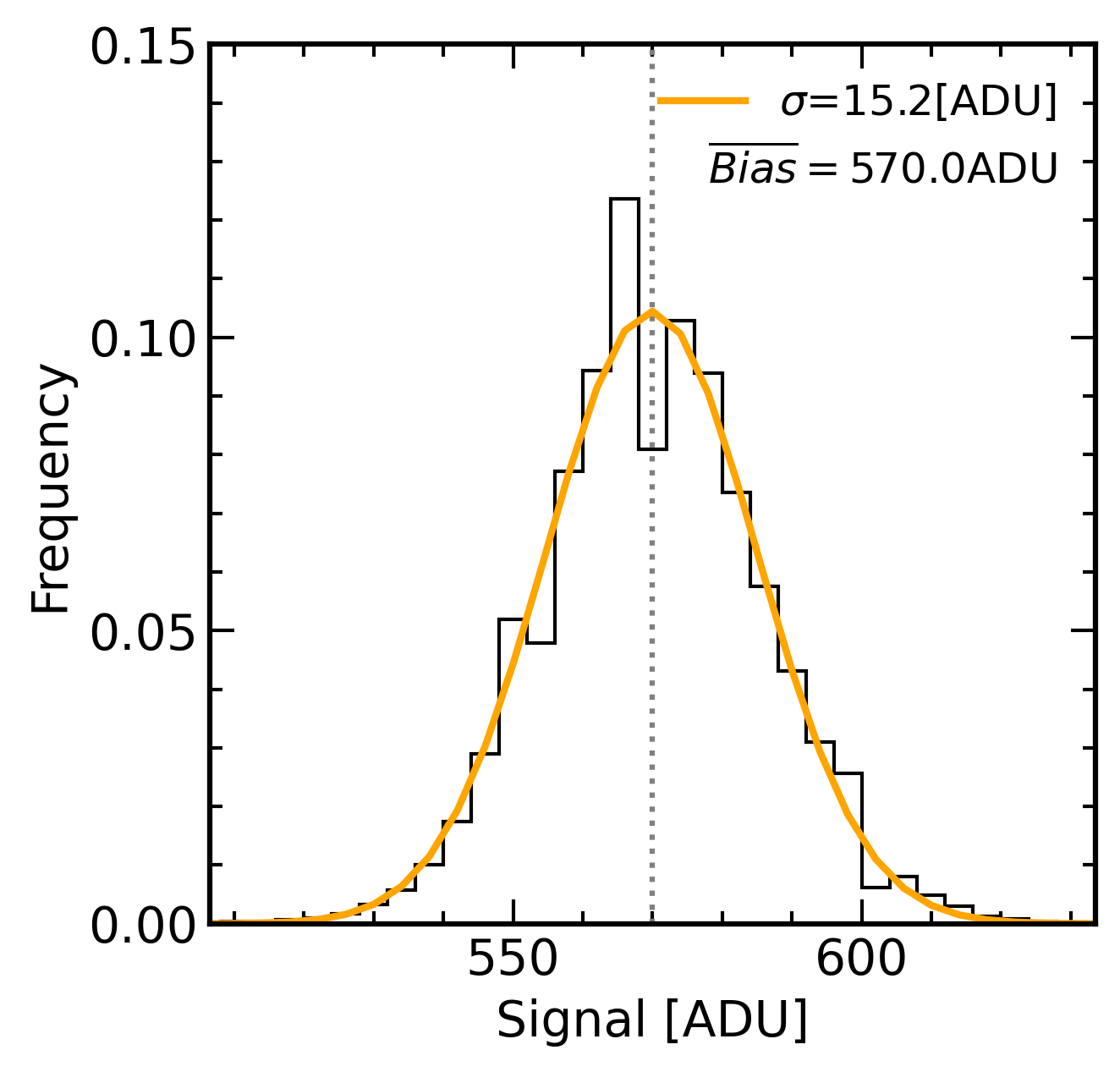}
        \caption{Histogram of pixel variations in a single bias frame with fitted Gaussian profile. This result was obtained at low-noise, high-gain mode.}
        \label{fig:bias_rdn}
    \end{figure}

    In this section, we analyze the statistical properties of the dark bias and the illuminated bias frames. As shown in Figure \ref{fig:bias_rdn}, the pixel-value distribution from a single dark bias frame is well represented by a Gaussian distribution of $\sigma=15.2\;\rm ADU$. A prominent spike at the mean arises from digitization during the ADC process, reflecting quantization of the analog voltage level into a digital output. 

    We quantify the random fluctuations of the camera by comparing 100 independent bias frames under complete darkness and illumination conditions. Following the methodology of \cite{Alarcon2023}, we compute the pixel-by-pixel differences and temporal noises for each pair. By subtracting pairs of frames, we effectively eliminate any fixed spatial patterns (e.g., gradient) and leave only the random fluctuations that characterize temporal noise between the bias frames.
    
\begin{figure*}[t]
        \centering
        \includegraphics[width=0.48\linewidth]{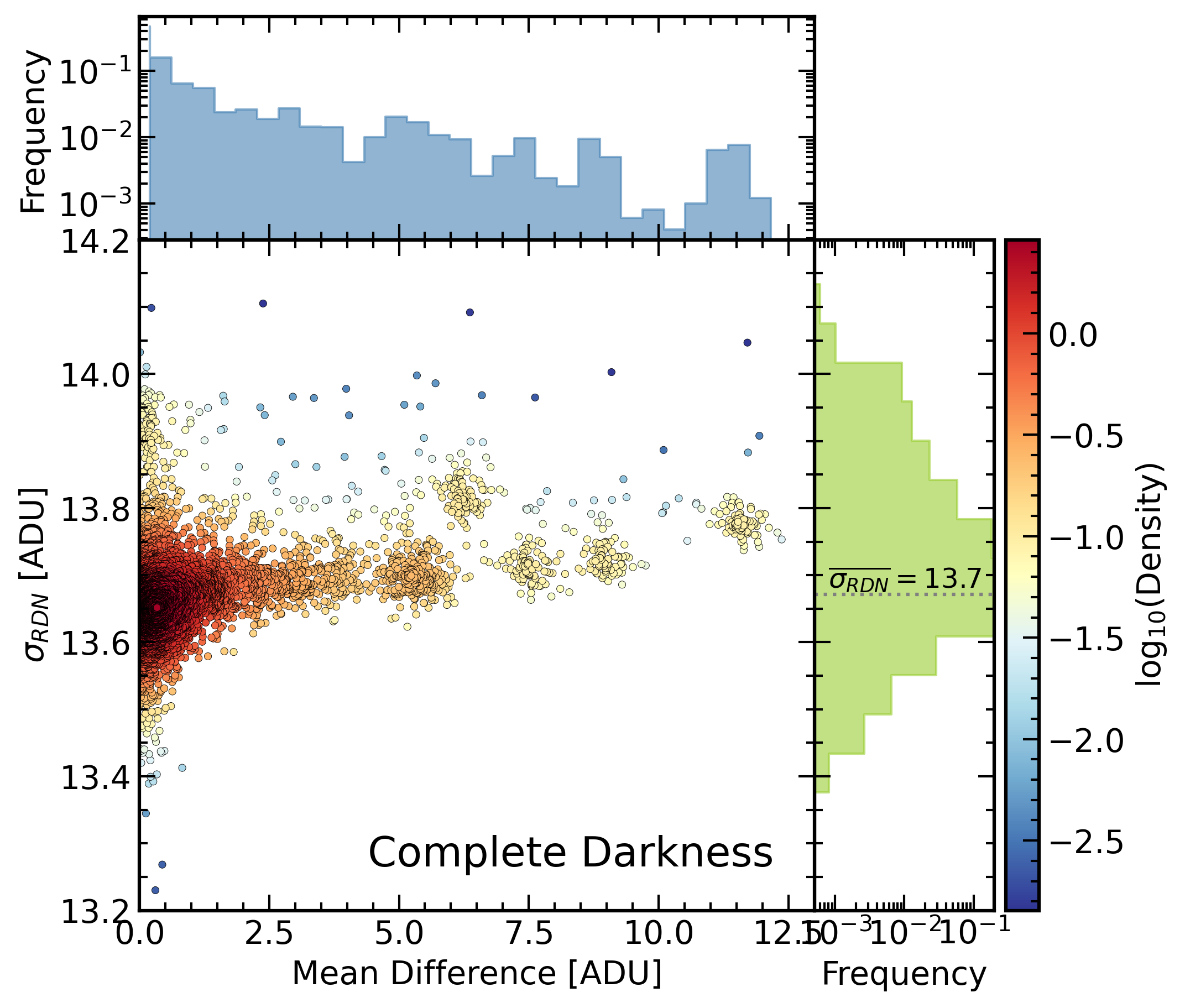}
        \includegraphics[width=0.48\linewidth]{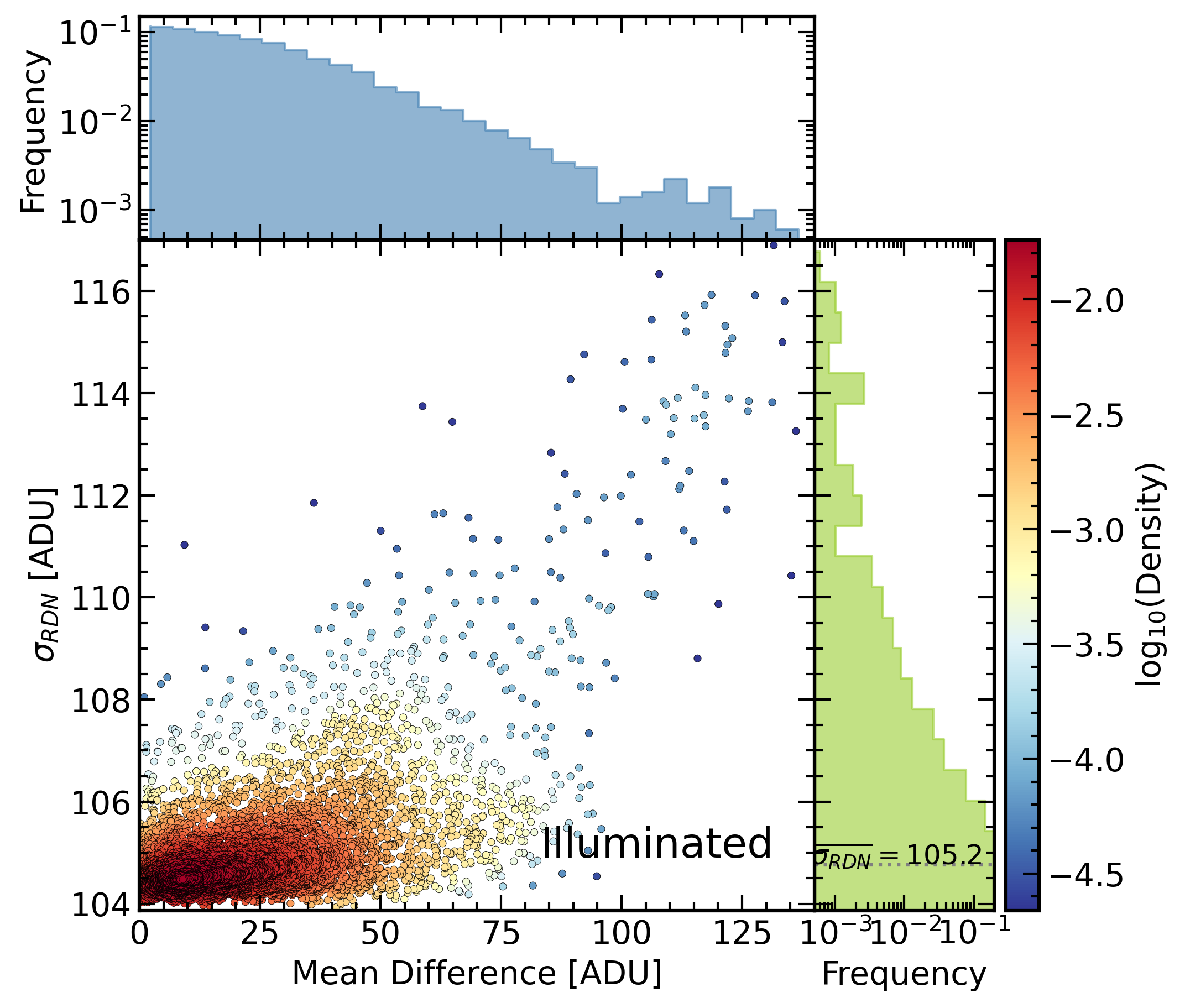}
        \caption{The distribution of mean difference versus readout noise between independent bias frames acquired under \textbf{a)} complete darkness (left panel), \textbf{b)} illumination (right panel) condition. The frames are acquired at low-noise, high-gain mode.}
        \label{fig:rdn}
    \end{figure*}

    Figure \ref{fig:rdn} represents the distribution of mean differences against its temporal noises. The mean differences are mainly concentrated at 0, while notable secondary clusters are displaced from the zero; these offsets suggest frame-to-frame shifts in the mean bias level. We quantify this time-dependent bias drift in Section \ref{sec:bias stability}. However, these shifts exhibit a smaller probability density in 2 orders of magnitude compared to the zero-mean peaks and therefore have a negligible impact on the overall characteristics of the bias frame. 
    
    The behavior of the temporal noise distributions reveals two distinct populations, depending on the presence of illumination: one concentrated at $13.7\;\rm ADU$ and another at $105.2\;\rm ADU$. The scatter under illumination is significantly broader compared to the dark condition. The former case closely aligns with the readout noise $12\;e^{-}\rm\;rms$ once converted via the near-unity gain of low-noise, high-gain mode. The slightly larger readout noise value ($\sigma=15.2\;\rm ADU$) from a single dark bias frame falls to the measured temporal noise from independent pairs $13.7\;\rm ADU$, considering frame-to-frame variability in the bias level. In contrast, the latter case (under illumination) reflects both the RDN and the additional fluctuation introduced by residual illumination, hereafter referred to as gradient noise (GN).

\subsubsection{Bias Level Time Variability \label{sec:bias stability}} 
\begin{figure}[ht]
        \centering
        \includegraphics[width=0.98\linewidth]{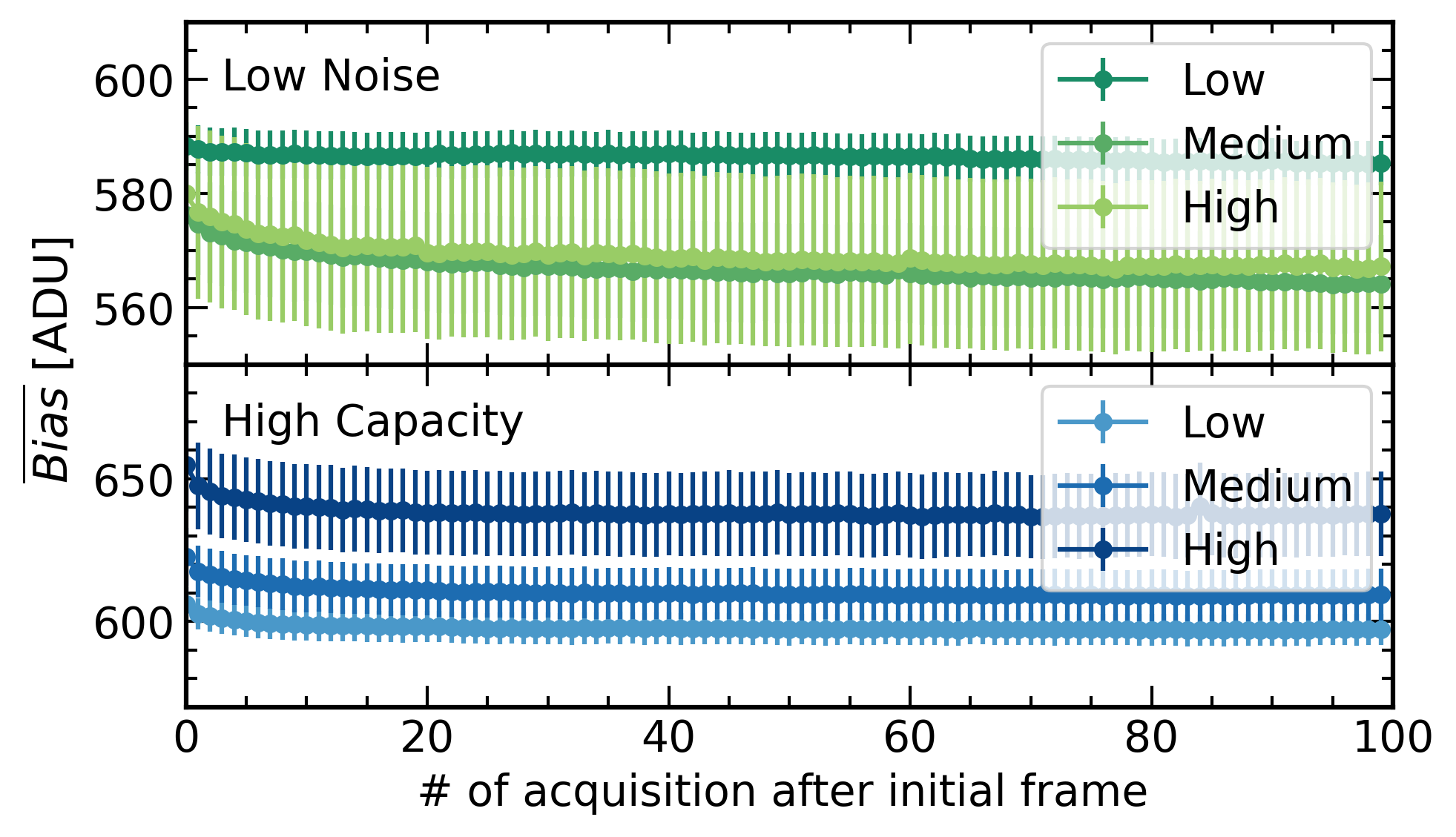}
        \caption{Mean dark bias level variation about the consecutive acquisition at different acquisition settings: high-capacity (upper), low-noise (lower). The error bar represents the standard deviation of pixel value within a single frame.}
        \label{fig:bias_stability}
    \end{figure}

    We inspect the variations in mean dark bias levels over 100 consecutive exposures in all different gain and acquisition modes. As illustrated in Figure \ref{fig:bias_stability}, the mean dark bias level rapidly decreases over the first few frames and then stabilizes. Notably, the low-noise, low-gain modes exhibit slightly less variation than the other modes. The shifts in the mean bias level are much smaller than the readout noises for all cases. We can conclude that there are no significant changes in the mean bias level over time within the sub-readout noise level.

\subsection{Dark \label{sec:dark}}
    We tested the dark frame characteristics in a dark room, varying the exposure time to investigate thermal noise.
\subsubsection{Dark current \label{sec:dark current}}
    \begin{figure}
        \centering
        \includegraphics[width=0.98\linewidth]{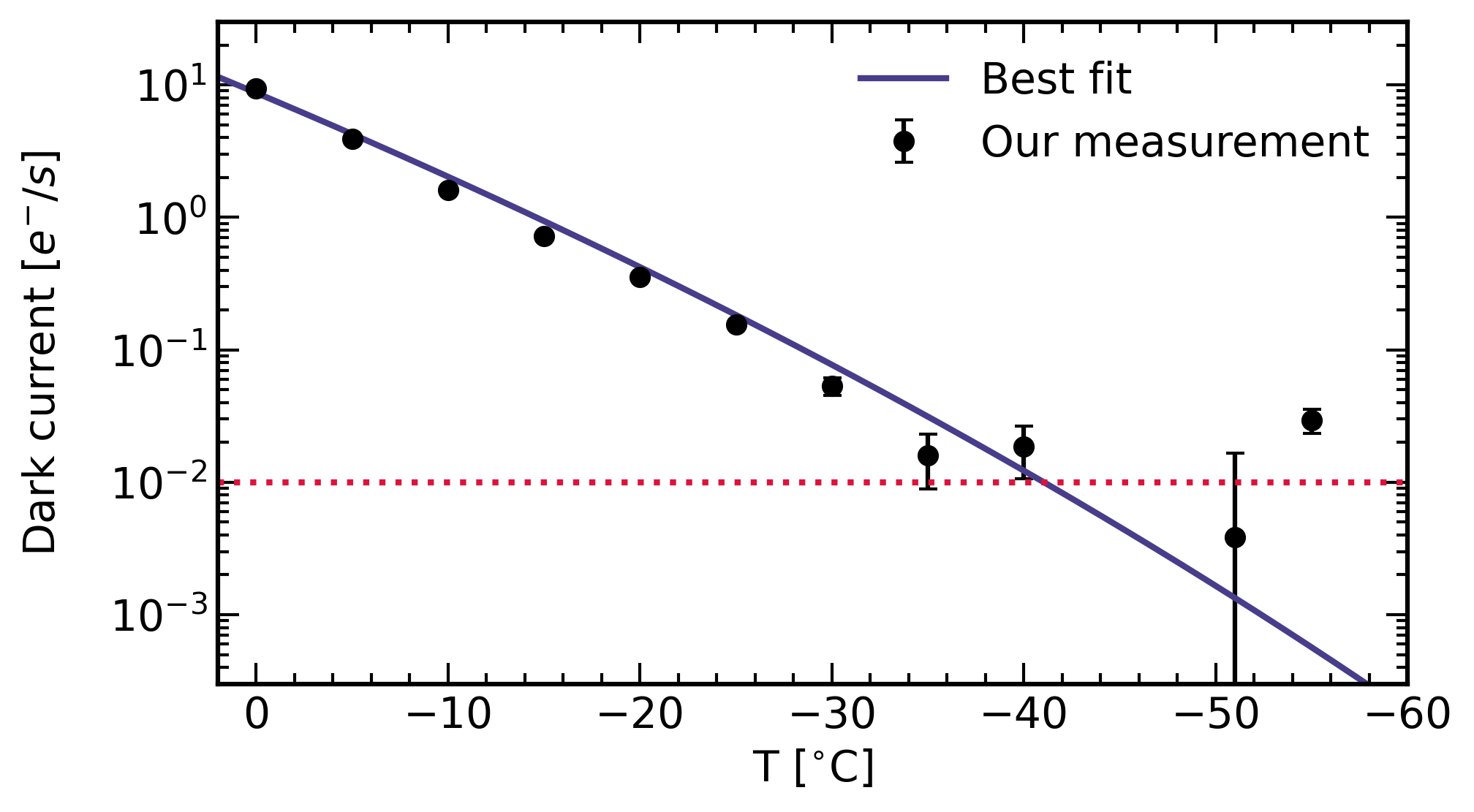}
        \caption{Dark current against temperature at low-noise acquisition mode. Vertical axes are converted into electron units by multiplying gain from the PTC analysis (see Section \ref{sec:PTC}). Each error bar represents $1\sigma$ fitting error of the least-squares fit.}
        \label{fig:dark_current}
    \end{figure}
    To determine the optimal operational temperature for low thermal noise observation, it is crucial to measure dark current, which emerges from the thermal noise of the camera, as a function of temperature. A theoretical model of thermal noise is
    \begin{equation}
        \frac{dN}{dt}=A_0T^{3/2}e^{-E/2k_BT}
        \label{eq:dark_current}
    \end{equation}
    where the amplitude $A_0$ and the band gap $E$ as a free parameter \citep{Chromey2010}. Because the dark current remains constant over time for a given temperature, we can measure it by analyzing the slope of the mean dark signal level deviation over different exposure times. We varied the exposure time of taking dark frames between 0 and 300 s with 60 s intervals; for each exposure time, we obtained 10 frames and took the average value. 
    In Figure \ref{fig:dark_current}, we compare the measured dark current to the fitting line derived from Equation \eqref{eq:dark_current} in low-noise, low-gain mode. This equation well explains our measurement. The value of parameters from fitting resulted as $A_0=(3.19\pm1.16)\times10^{14}\;\rm\;s^{-1}$, $E=(1.77\pm0.11)\;\rm{eV}$. The dark current is less than 0.01 $e^{-}\;\rm pixel^{-1}\;s^{-1}$, indicating that an exposure time of at least 10 minutes is required for the dark current signal to become comparable with readout noise at a lower temperature than $-40\;\rm ^\circ C$. Consequently, the effect of dark current is negligible for that temperature regime.  

\subsubsection{Internal Pattern Noise \label{sec:pattern noise}}
    Thermal noise is not uniform due to the inhomogeneous inner structure of the camera sensor. Especially, the internal wiring patterns and amplifiers can be a major source of heat. We probe internal pattern noise in a high-temperature regime to obtain a stand-out map of the internal pattern with  $5\times5$ average-binned image. By subtracting the median dark current from the measurements, we unveil the residual dark counts, providing a clearer insight into the performance at each temperature.
    \begin{figure*}[ht!]
        \centering
        \includegraphics[width=0.96\linewidth]{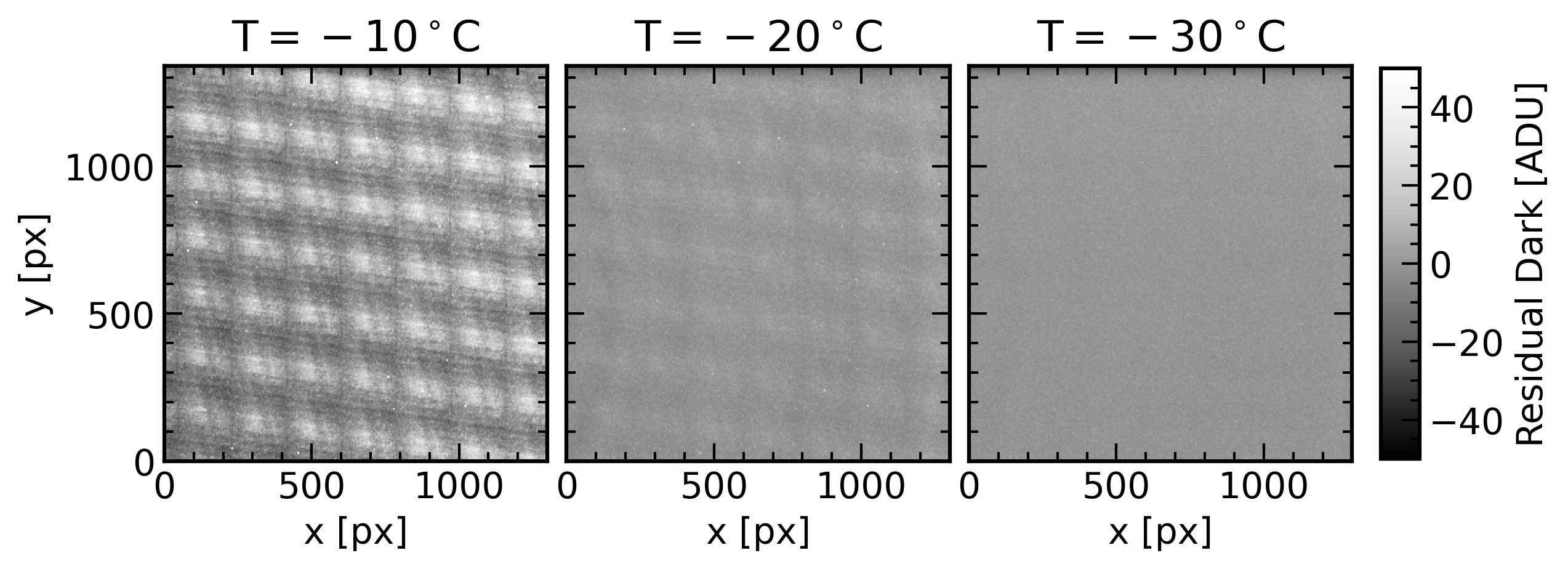}
        \caption{Interline pattern noise map about different temperature conditions. $-10\;\rm ^\circ C$ (left), $-20\;\rm ^\circ C$ (center), $-30\;\rm ^\circ C$ (right). Each frame is obtained at low-noise, low-gain mode, and 300s exposures each.} 
        \label{fig:fpn}
    \end{figure*}
    As shown in Figure \ref{fig:fpn}, a parallelogram-shaped pattern appears with $\sim$180 pixels in width and rapidly diminishes in a low-temperature regime.

\subsection{Photon Transfer Curve \label{sec:PTC}}
    \begin{figure}[ht]
        \centering
        \includegraphics[width=0.98\linewidth]{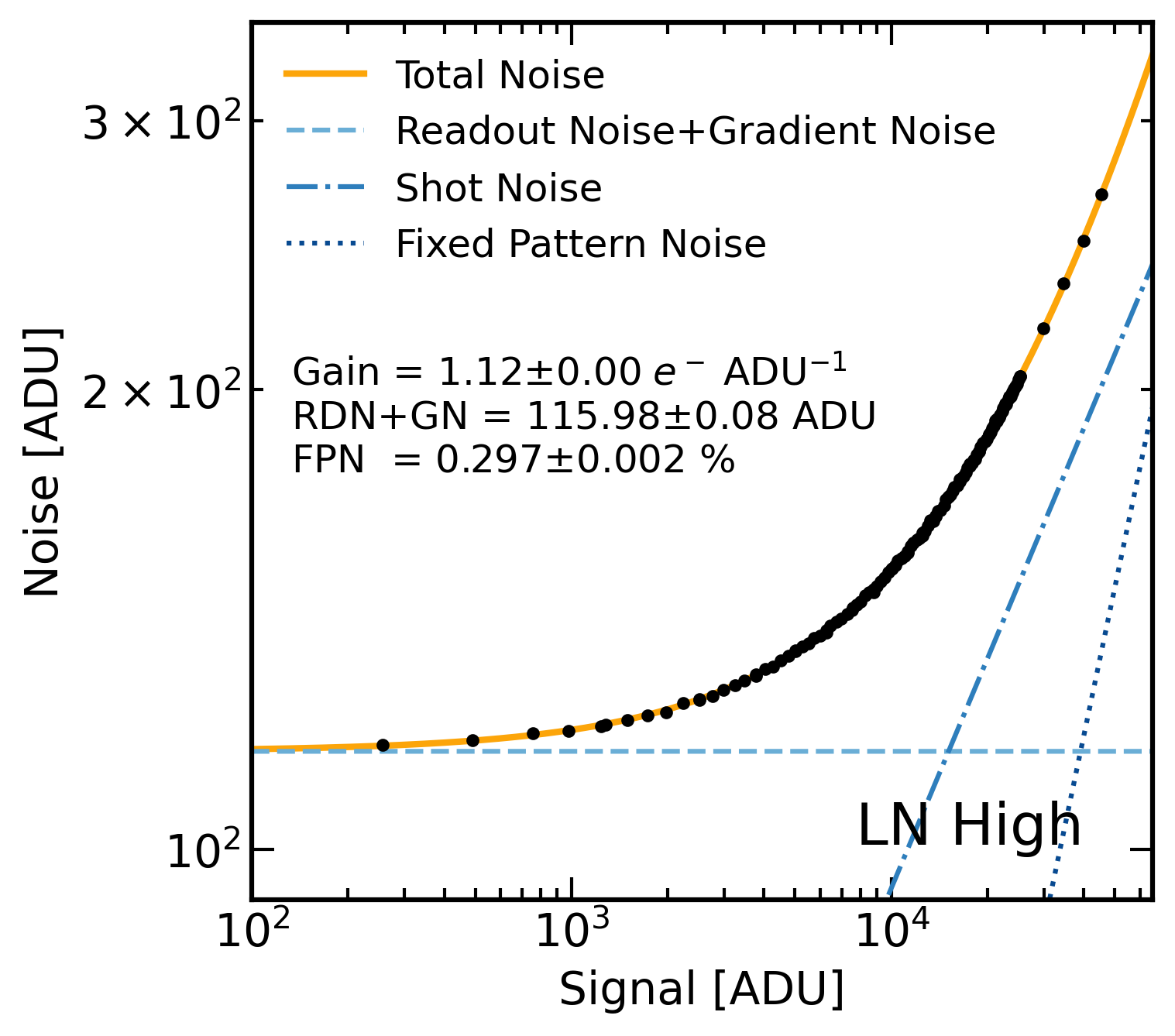}
        \caption{Photon transfer curve at low-noise, high-gain mode are depicted with the fitting line of Equation \eqref{eq:noise}.  Each noise component is plotted with a different color and line style. The parameter values of PTCs are described in caption with $1\sigma$ fitting error.}
        \label{fig:PTC}
    \end{figure}
    A photon transfer curve (PTC), which depicts temporal noise as a function of signal level on a logarithmic scale, provides a powerful tool for characterizing noise behavior across a wide dynamic range. According to \cite{Janesick2007}, the noise of an image contains three main components: readout noise (RDN), shot noise, and fixed pattern noise (FPN). As detailed in previous sections, the excessive exposure during readout induces gradient noise (GN). Modified version of total noise expressed as
    \begin{equation}
        \sigma_{\rm total}=\left(\sigma_{\rm RDN}^2+\sigma_{\rm GN}^2+\frac{S}{G}+(P_{\rm FPN }S)^2\right)^{1/2}
        \label{eq:noise}
    \end{equation}
    where the readout noise is denoted as $\sigma_{\rm RDN}$, the gradient noise is denoted as $\sigma_{\rm GN}$, the FPN factor is denoted as $P_{\rm FPN}$, the signal is denoted as $S$ and the gain of CCD on-chip amplifier is denoted as  $G$. Although gradient noise is fundamentally a form of shot noise, it is treated separately as a constant value here to enable precise analysis of temporal noise behavior as a function of exposure time. A detailed derivation of the gradient noise term is described in Appendix \ref{sec:grad noise}. 

    We collect five independent flat frames each for progressively increasing exposure times---ranging from the minimum exposure (mean signals around $10-100\;\rm ADU$) up to below saturation---and apply bias and gradient corrections. To achieve an accurate measurement of the signal and temporal noise, uniform illumination is crucial. Even with an integrating sphere, notable intensity changes exist between the center and edges of the detector. We locate the true center of the light distribution by fitting a two-dimensional quadratic surface to the gradient-corrected flat frame with exposures adjusted to keep the peak signal below the upper limit of the linear response. Around that centroid, we extract a $256\times256$ pixel cutout region in which the rms non-uniformity is below $1\%$ of the mean level. 
    
    For consistency, we use the identical patch achieved from low-noise, high-gain mode for all illumination levels and acquisition modes. Because the gradient noise highly depends on the spatial location, fixing the patch location across different acquisition settings ensures that our noise comparisons are not biased by spatial non-uniformity. Within the selected patch, we calculate the average signal value (Equation \eqref{eq:mean_signal}) and determine the temporal noise by calculating the standard deviation of the corrected frame deviation from the mean signal value (Equation \eqref{eq:temp_noise}). 
    \begin{equation}
        \label{eq:mean_signal}
        S=\left\langle{1\over N}\sum_{k=1}^N X_k[i,j]\right\rangle_{(i,j)\in \rm patch}
    \end{equation}
    \begin{equation}
        \label{eq:temp_noise}
         \sigma^2={1\over N}\sum_{k=1}^N\langle (X_k[i,j]-S)^2\rangle_{(i,j)\in \rm patch}
    \end{equation}

    We show the PTC of low-noise, high-gain mode in Figure \ref{fig:PTC}. The signal-independent noise mainly contributes to the total noise up to $10,000\;\rm ADU$. Beyond that level and up to roughly $30,000\;\rm ADU$, the shot noise (i.e., which is proportional to signal level) dominates. Above $50,000\;\rm ADU$, the FPN emerges as the largest source of total noise. This observed response, which is dominated by the noise floor due to substantial gradient noise contributions, is unexpected if only considering the low readout noise obtained from the dark bias measurements. Hence, we should take GN into account as a combination of RDN. Our noise model entirely isolates all noise components; the determination of gain and $P_{\rm FPN}$ is not influenced by the existence of GN (see Section \ref{sec:discussion}). We repeated this PTC analysis under various acquisition modes and gain settings (see Appendix \ref{sec:PTC_app}). The best-fit parameter values obtained from these analyses are summarized in Table \ref{tab:ptc}; Our measurements agree closely with the specification of the manufacturer.

\subsection{Linearity \label{sec:linearity}}
\begin{figure*}[ht!]
        \centering
        \includegraphics[width=0.49\linewidth]{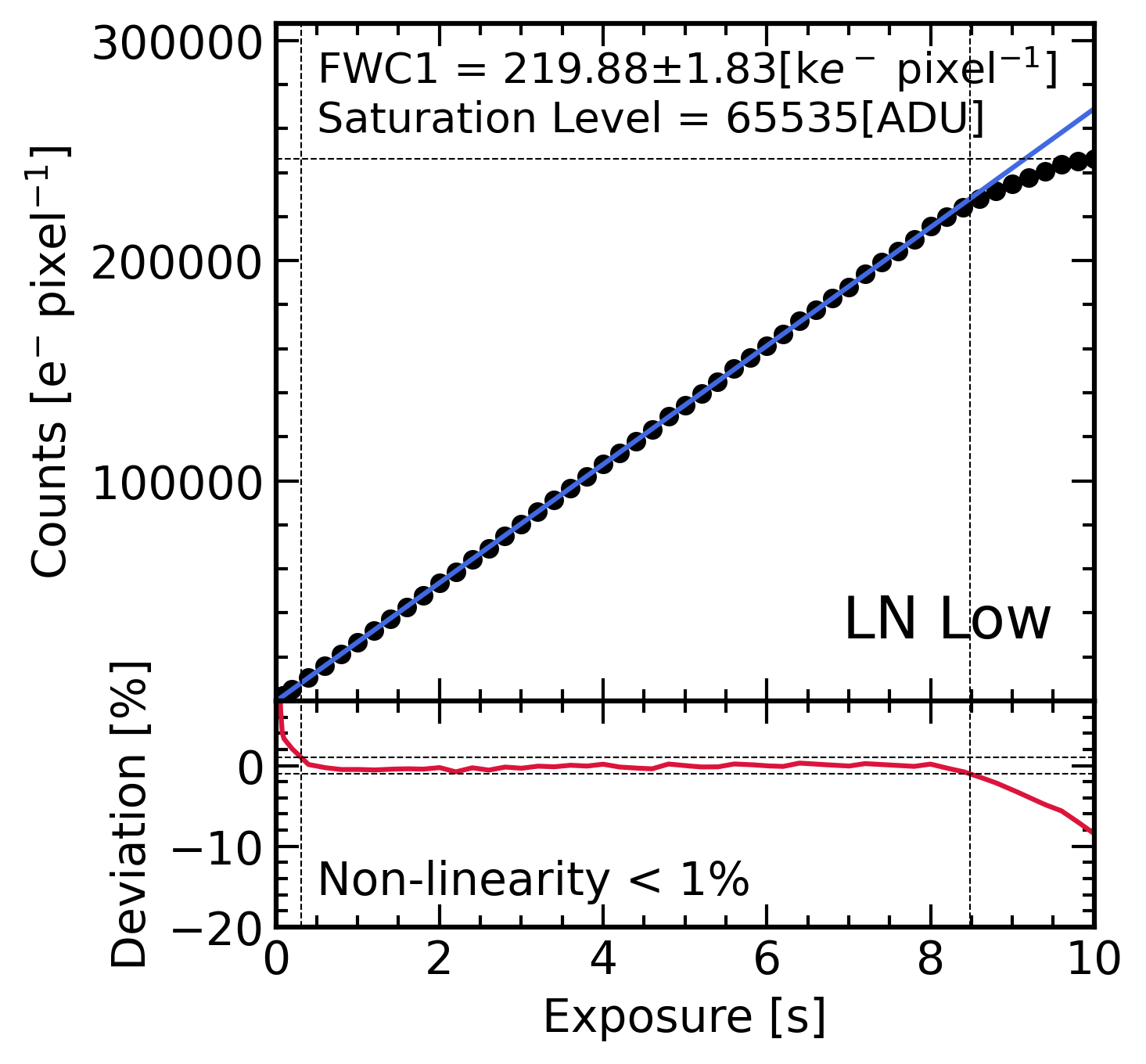}
        \includegraphics[width=0.49\linewidth]{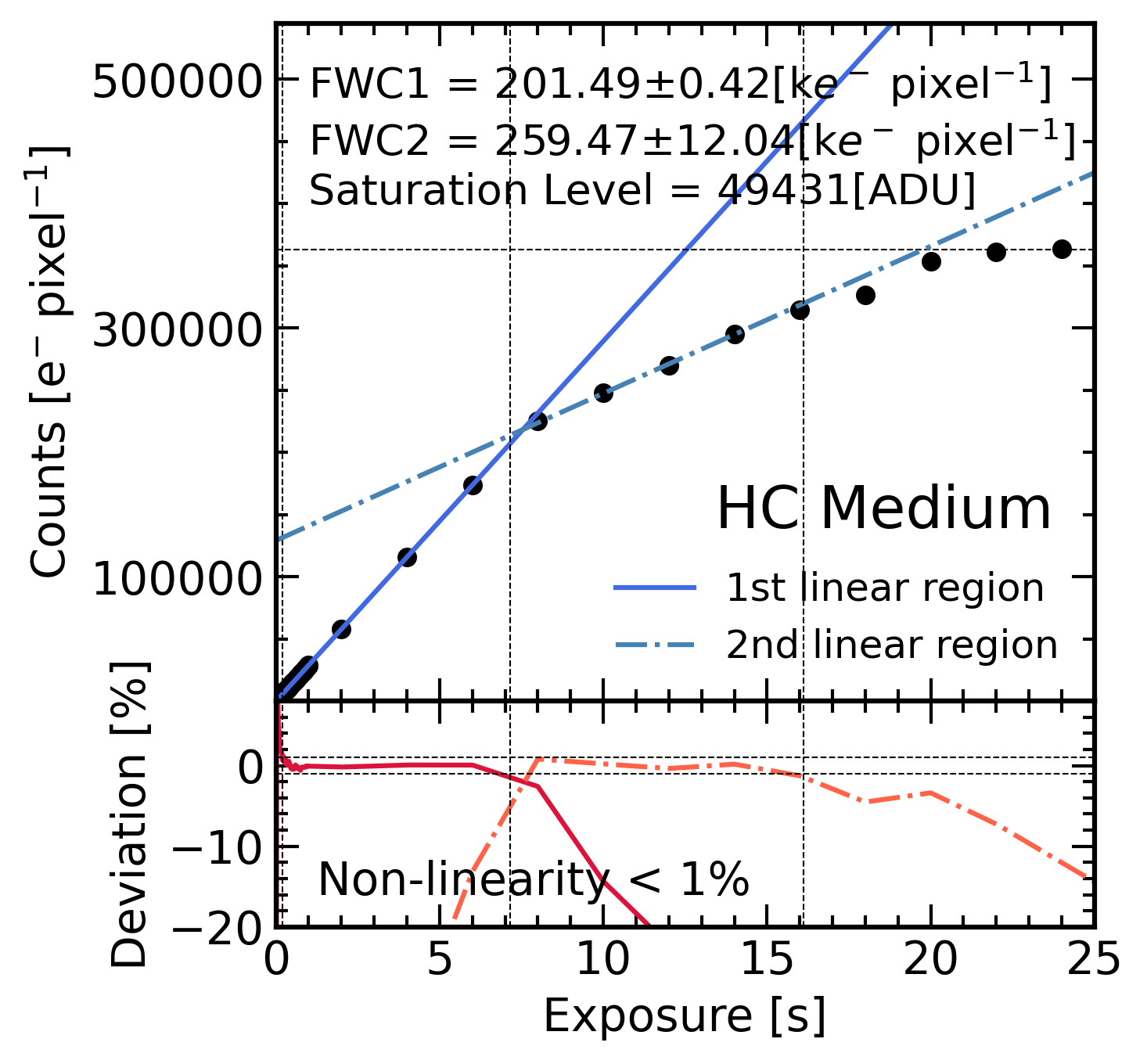}
        \caption{Linearity curve (upper panel) and deviation from linear response (lower panel) at low-noise, low-gain (left) and high-capacity, medium-gain (right) conditions. Vertical dotted lines indicate the boundary of $1\%$ non-linearity regime, and horizontal dotted lines along with the linearity curve represent the saturation level. ADU values are converted to electron units by multiplying the corresponding gain value measured with PTC.}
        \label{fig:FWC}
    \end{figure*}

    Full-well capacity (FWC) refers to the effective size of the potential well that ensures the linear response of the CCD. At low illumination levels, RDN is prevalent, where signal and noise values are balanced. Conversely, at high illumination levels, pixel saturation is crucial. These effects result in a non-linear response to exposure time. We obtained ten frames for each exposure time, which ranged from 0 s to 30 s. We plot the exposure time versus pixel counts in Figure \ref{fig:FWC}. We adjust the maximum exposure time for each exposure setting to a relatively long duration to ensure the signal reaches the saturation level. 

    We define the linear response region as the range over which deviations from a linear fit remain below $1\%$, establishing the boundary for non-linearity. We also determined the saturation level by the average signal value at prolonged exposure time. 
    
    According to Table \ref{tab:technical_info}, the FWC of a single pixel is significantly smaller than the output node. The saturation level depends only on the limitation of a single pixel, and our measurement exhibits good agreement with the specification provided by the manufacturer. Given the 16-bit ADC used in our camera, the maximum digital value achievable is $2^{16}-1=65,535$. Our measured saturation levels nearly approach this limit of the ADC, confirming that the camera response effectively utilizes the full dynamic range available through the A/D conversion process.

    It should be noted that in the case of high-capacity mode, the low/medium gain mode exhibits multiple deflection points (i.e., multiple linear regions). In contrast, other exposure modes (i.e., all gain modes at low-noise mode and high-capacity, high-gain mode) have single linear responses with flat regions at long exposure. The slopes of linear regions decrease at higher exposure levels, indicating a breakdown of the linear response. We divide the linearity curve into two linear regions and the saturation region, and investigate non-linearity separately. 
    This piecewise linear response of image sensors has been employed for high dynamic range, dynamically adjusting the pixel voltage level once a programmed threshold level is reached \citep{Meng2014}.

     \begin{table*}[ht!]
        \caption{Summary of the PIXIS 1300BX characteristics: the readout noise (RDN), the gain ($G$), the FPN factor ($P_{\rm FPN}$), the full well capacity (FWC). The readout noise is derived from the difference between bias frames, and the parameter values are obtained from the photon transfer curve and the linearity curve across different acquisition modes.\label{tab:ptc}}
        \centering
        \begin{tabular}{cccccccc}
        \toprule
            Image quality & Gain & \multicolumn{2}{c}{RDN [$e^-$ rms]} & \multicolumn{2}{c}{G [$e^{-}\;\rm ADU^{-1}$]} & $P_{\rm FPN} \;[\%]$ & FWC [$\rm{k}e^{-}$]\\
            \cmidrule(lr){3-4} \cmidrule(lr){5-6} \cmidrule(lr){7-7} \cmidrule(lr){8-8}
            & & Spec. & Measured$\rm ^a$ & Spec. & Measured & Measured & Measured$\rm ^b$ \\
        \midrule
             & Low & & 16.03$\pm$0.08 & 4.40 & 4.09$\pm$0.02 & 0.281$\pm$0.001 & 221.72$\pm$1.35\\
             Low Noise & Medium & 15.80 & 15.88$\pm$0.08 & 2.24 & 2.10$\pm$0.01 & 0.282$\pm$0.001 & 111.83$\pm$1.71 \\
             & High & &  15.42$\pm$0.14 & 1.17 & 1.12$\pm$0.01 & 0.295$\pm$0.002 & 50.94$\pm$0.49\\
             \hline
             & Low & & 41.34$\pm$0.16 & 14.88 & 15.37$\pm$0.06 & 0.304$\pm$0.001 & 226.76$\pm$2.50\\
             High Capacity & Medium & 34.46 & 37.57$\pm$0.08 & 7.36 & 7.73$\pm$0.01 & 0.293$\pm$0.001 & 197.67$\pm$5.69 \\
             & High & & 35.08$\pm$0.09 & 3.59 & 3.74$\pm$0.01 & 0.291$\pm$0.001 & 209.40$\pm$1.75 \\
        \bottomrule
        \end{tabular}
        \tabnote{$\rm ^a$ The RDNs measured from the standard deviation of the difference between bias pairs are converted to $e^- \rm rms$ units by multiplying the gain from PTC analysis. \\ $\rm ^b$ The FWC values in low-noise mode indicate the full range of the linear response regime. In high-capacity mode, the FWC values represent the first linear region, starting from 0 seconds of exposure up to the first knee point on the linearity curve.}
    \end{table*}
    
\subsection{Quantum Efficiency \label{sec:QE}}
    Quantum efficiency (QE) indicates the reactivity of a detector to incident light as a function of wavelength. Experimental settings are the same as depicted in Figure \ref{fig:Optical_Bench}. We obtained five frames each for different exposure times and selected data between $1,000\;\rm ADU$ and $50,000\;\rm ADU$, ensuring a linear response with non-linearity below $1\%$. The monochromator provides monochromatic light with $3.5\;\rm nm$ bandwidth; the central wavelength varies in the range from $300-950\;\rm nm$ with an interval of $50\;\rm nm$.

    The definition of QE is given as
    \begin{equation}
        \textrm{QE}(\lambda) = \frac{S(\lambda,t_{\rm exp})\;G}{\frac{dN_{\rm ph}}{dt}(\lambda)\;t_{\rm exp}}\times 100
        \label{eq:QE}
    \end{equation}
    where $S$ is the mean signal from the identical $256\times256$ patch used in previous sections, $G$ is the gain, converting signal value to photoelectron units, and $\frac{dN_{\rm ph}}{dt}$ is the photon number-flux at a single pixel (i.e., in $\rm counts\;pixel^{-1}\;s^{-1}$). To calculate the photon number-flux, we need to consider the pixel width, the size of the wand detector, and the distance from the exit face of the integrating sphere to the image sensor.
    \begin{equation}
        \frac{dN_{ph}}{dt} = \frac{P_\lambda/T_\lambda}{hc/\lambda}\left(\frac{w_{\rm pixel}}{w_{\rm sensor}}\right)^2\left(1+\frac{d}{R_0}\right)^{-2}
    \end{equation}
    where $P_\lambda$ is the light power measured by PMM, the pixel width $w_{\rm pixel}=20\;\mu \rm m$, the wand detector width $w_{\rm sensor}=10\;\rm mm$, integrating sphere radius $R_0=83.82\;\rm mm$, distance from exit of integrating sphere and image sensor $d=20\;\rm mm$. We assumed the average transmission of the $\rm MgF_2$ vacuum window $T_\lambda$ as a constant value with $95\%$ for incident flux correction. Its true transmittance slowly varies by $\pm2\%$ across $300-1000\;\rm nm$ \citep{PIXIS1300_specsheet}.

\begin{figure}[ht!]
        \centering
        \includegraphics[width=0.98\linewidth]{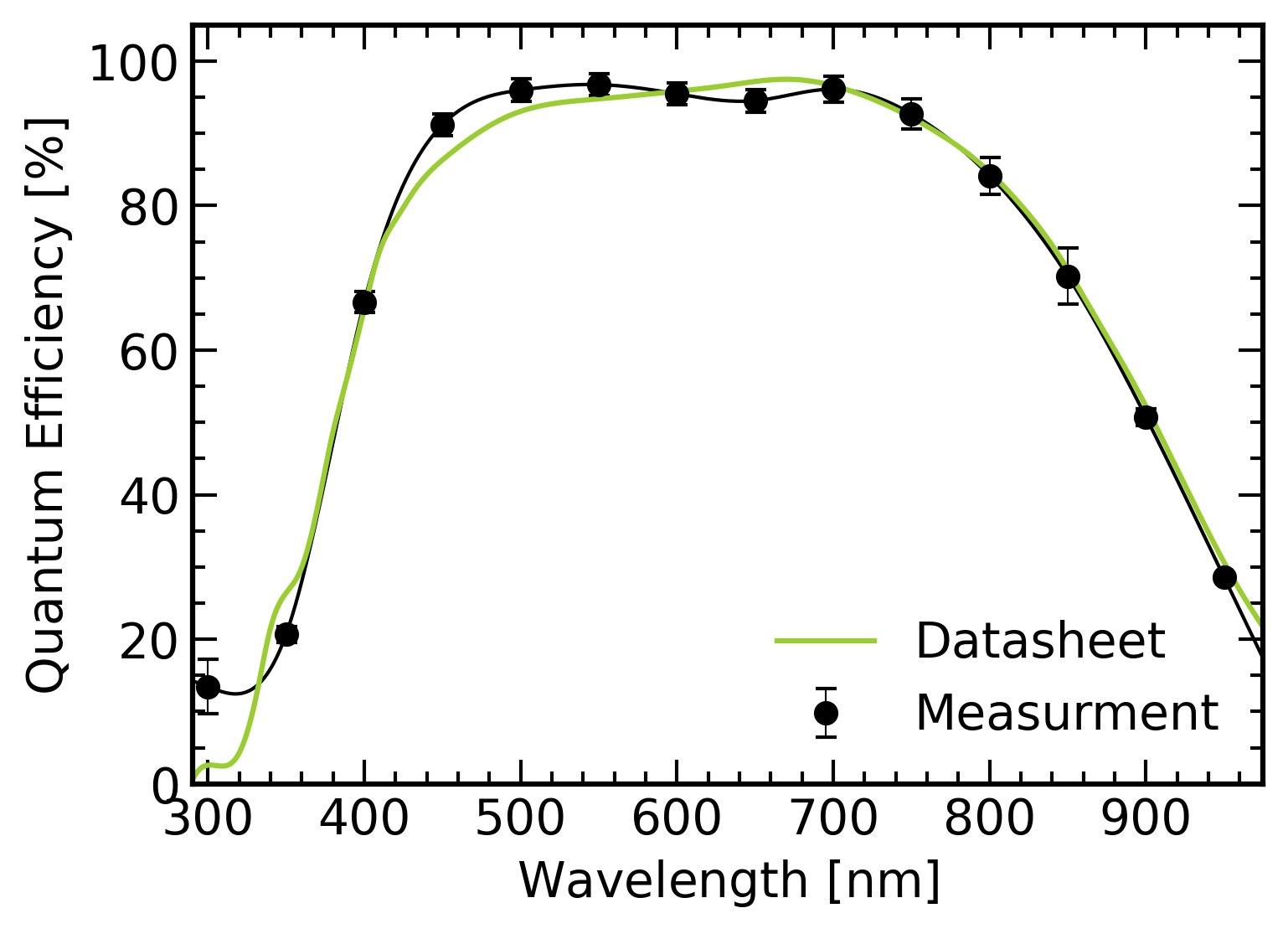}
        \caption{The quantum efficiency curve of the PIXIS 1300BX with cubic spline interpolation curve. Green line represents QE curve from datasheet.}
        \label{fig:QE}
    \end{figure}

    In Figure \ref{fig:QE}, a wide plateau is observed between $400\;\rm nm$ and $800\;\rm nm$ with $\rm QE\geq80\%$ and a gradient of both blue and red wings, indicating relatively slower degradation at longer wavelengths. Our measurement presents a similar overall shape to the one provided by the manufacturer, with only a $2$-$5\%$ offset. This remaining offset can be attributed to a combination of these factors: (1) gain measurement error, (2) pixel-level photon flux calibration error caused by wand detector calibration uncertainty and lamp stability, (3) monochromator bandwidth. Notably, the overestimation of QE value at $300\rm\;nm$ likely arises from the considerably low intensity level of approximately $0.5\rm\;nW$, which reduces reliability at that near-UV wavelength.
        \begin{figure}[ht!]
        \centering
        \includegraphics[width=0.98\linewidth]{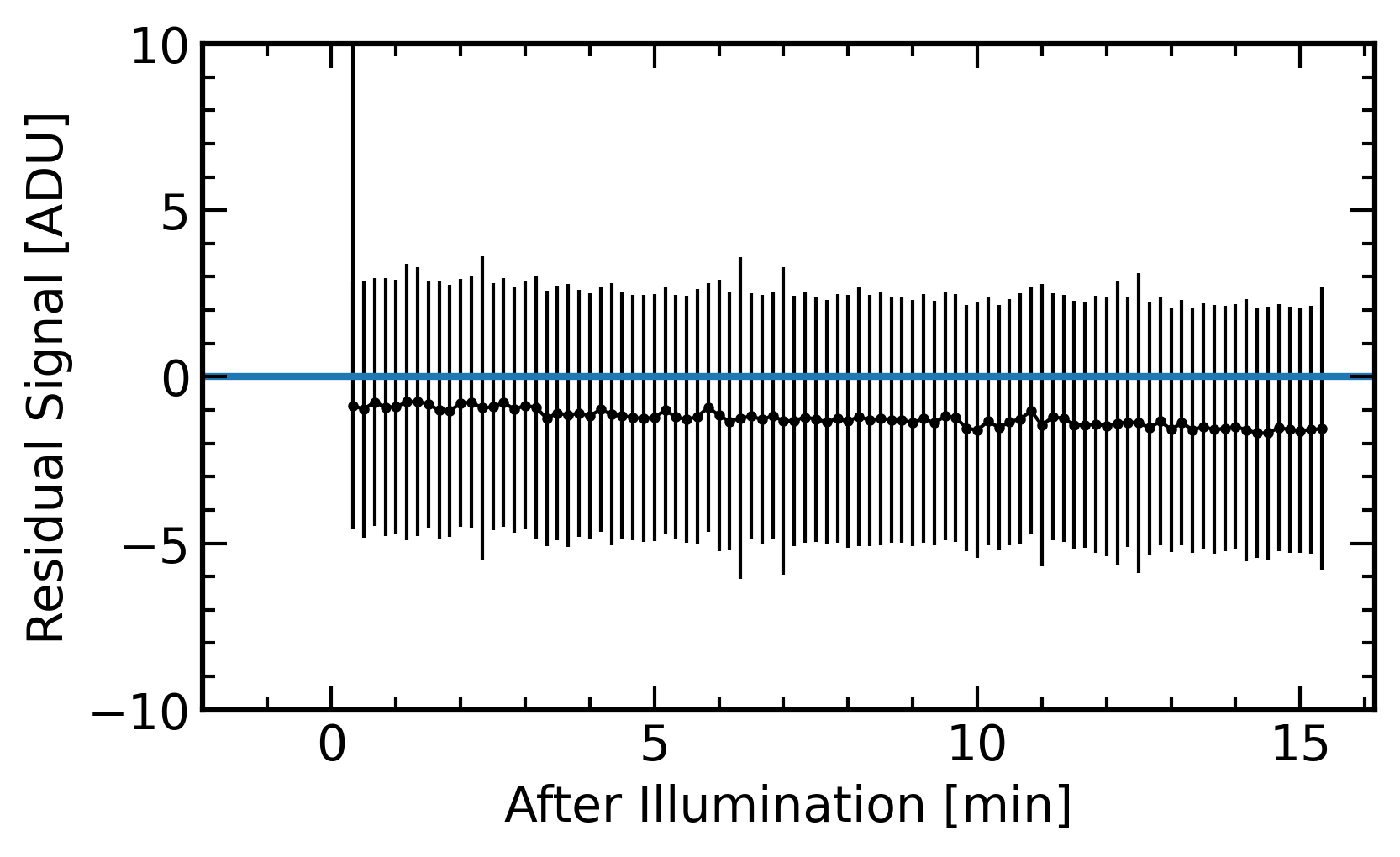}
        \caption{Time variation of mean residual level after illumination with standard deviation between pixels within frame.}
        \label{fig:persistence}
    \end{figure}
    
\subsection{Charge Persistence \label{sec:charge persistance}}
    After illumination by the bright light source, some charge may remain at the exposed location momentarily after the readout process due to imperfect charge transfer. \citep[e.g.,][]{Chromey2010, Karpov2020}
    It can affect subsequent exposure, especially critical contamination when transitioning from immediately after very bright sources to low signal-to-noise sources.
    
    We investigated the charge persistence of the camera after strong illumination close to saturation by taking a sequence of post-illumination frames with 10 s exposure time over a total duration of 15 minutes. Figure \ref{fig:persistence} shows the mean level of the dark-subtracted frames\footnote{The dark frame was obtained in a sufficiently dark environment to avoid any additional illumination.}. The mean signal level drops immediately below to the readout noise level. The negative mean-level drift (few ADUs) is comparable to the mean shift of the bias level shift due to subsequent exposure, and is not an indication of long-lived charge traps after a bright illumination. Because our integration time is the 10 s exposure, we can only place an upper limit on the charge persistence timescale of $\lesssim 10\rm\;s$. We therefore conclude that the PIXIS 1300BX has negligible charge persistence shorter than 10 seconds.

\section{Characterization of the IsoPlane 320A Spectrograph \label{sec:spectrograph test}}
    Characterizing a spectrograph is an important prerequisite for precise spectroscopy, directly influencing the accuracy of wavelength calibration, spectral resolution, and flux calibration of astronomical observations. In particular, fiber-fed spectrographs require careful evaluation of the impact of instrumental effects, such as spectral dispersion, instrumental broadening, and fiber illumination uniformity \citep[e.g.,][]{Barden1988,Barden1995}. This laboratory assessment provides a robust foundation for interpreting subsequent on-sky astronomical observations.
    
    In this section, we outline the characterization of the IsoPlane 320A spectrograph coupled to optical fibers under laboratory conditions.
    
\subsection{Spatial Separability of Multi Spectra \label{sec:multi}}
    Testing in laboratory conditions, we illuminate the inlet of the fiber bundle with a homogeneous light source in wide wavelength coverage using reflected lights from a white LED lamp on a screen with a Lambertian surface. 
    \begin{figure}[ht!]
        \centering
        \includegraphics[width=0.95\linewidth]{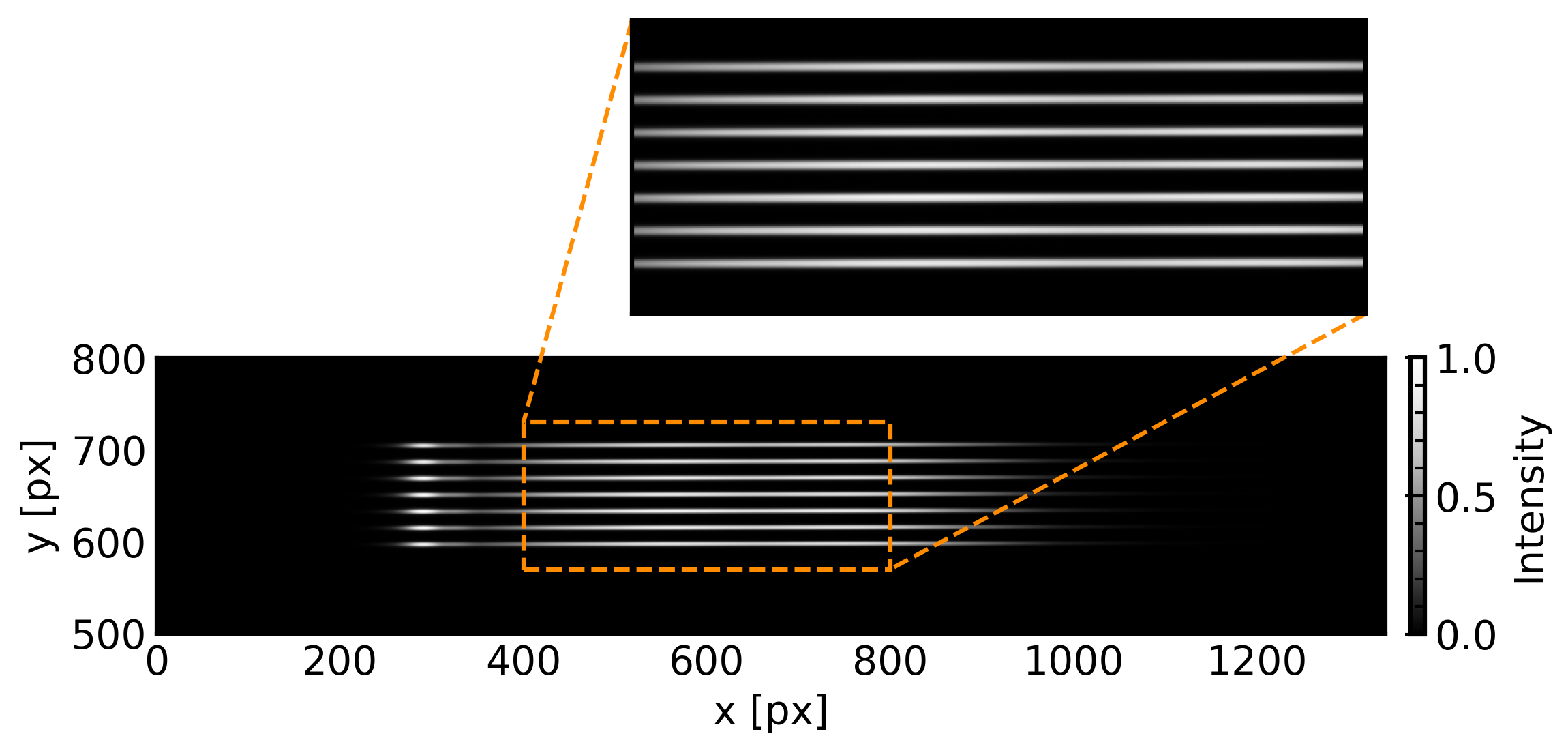}
        \caption{Cut-out image of the multi-fiber spectrum of the homogeneous light source at 150 $\rm gr\;mm^{-1}$ grating. Intensity was normalized with the maximum value.}
        \label{fig:multi-fiber image}
    \end{figure}
    As illustrated in Figure \ref{fig:multi-fiber image}, we distinctly separate the spectrum from each fiber along spatial directions. Despite careful efforts to ensure uniform illumination of the fibers, subtle misalignments lead to inevitable fluctuations in illumination levels across individual spectra. In our setting, all spectra originate from the same light source and share common spectral features. By comparing these features, we can identify some instrumental effects in the spectral profile.
    
    Extracting a spectrum from each fiber along the horizontal direction without considering curvature or tilt could be misleading due to atmospheric refraction and the tilting of the imaging sensor. To address this issue, we trace individual spectra by detecting local maxima along the vertical axis, employing multimodal Gaussian fitting along the vertical directions to pinpoint the precise spatial center along the horizontal axis (i.e., exact dispersion axis). We conduct iterative peak detection along the horizontal axis, applying a 10-pixel median binning to enhance efficiency. We then fit the locations of the peaks using a polynomial fitting with iterative 3-$\sigma$ clipping to precisely determine the spatial center as a function of horizontal pixels. For this purpose, we select a 6th-order Chebyshev polynomial. Next, we digitize the spatial (i.e., cross-dispersion) profile to extract pixel values from the data surrounding a small vertical aperture, which is set to $10$ pixels (i.e., corresponding to roughly two times the FWHM of spatial dispersion). Finally, we take the average of the signal within the aperture and normalize it by exposure time.

\begin{figure}[ht!]
        \centering
        \includegraphics[width=1.0\linewidth]{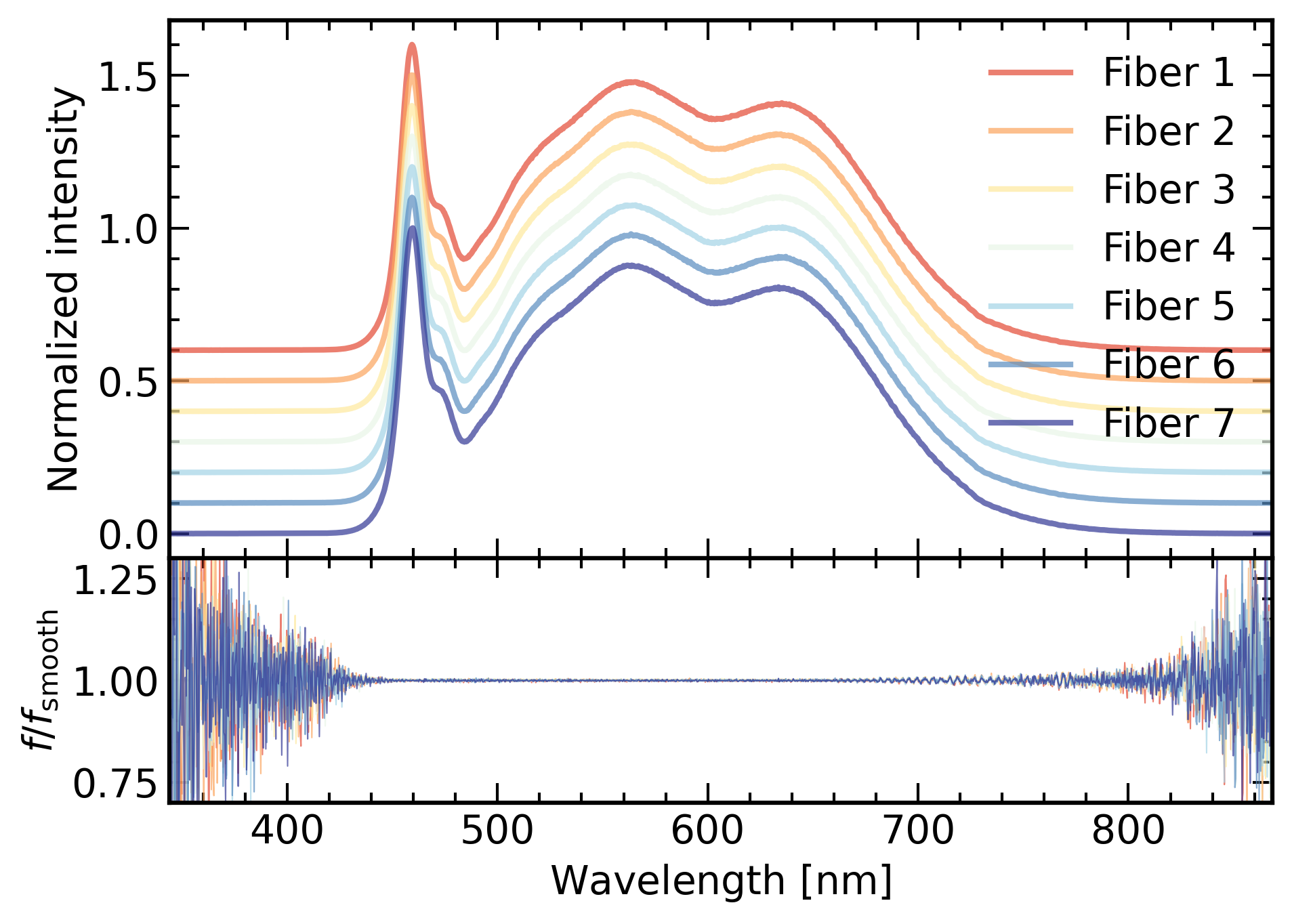}
        \caption{Comparison of spectra among different fibers. The upper panel shows the normalized spectra, while the lower panel displays the ratio between the fiber spectrum and the corresponding smoothed curve. To enhance the visibility of the plot, we vertically shifted each fiber spectrum in the upper panel.}
        \label{fig:multi_comparision}
    \end{figure}

    In Figure \ref{fig:multi_comparision}, we overlay the extracted individual spectra, with every spectrum normalized to its respective peak fluxes, and show their ratios to a fitted B-spline smoothed curve. The pixel-to-wavelength conversion is done by line identification of the well-known atomic lamp spectrum; it will be discussed in more detail in Section \ref{sec:spec_cal}.   
    
    To avoid introducing artificial ripples, we fit each spectrum with a uniform‑knot B‑spline with $2\rm\;nm$ knot spacing. This process compensates for the residual wavelength solution deviation induced by spectrum tracing, enabling fair comparison across fibers. The differences among the spectra are minimal in the central region, where the LED spectrum itself exhibits intense flux; Notable scatter is observed at both ends, yet the mean levels are consistent. However, the signal levels in these regimes are significantly low and lack well-defined profiles, resulting in increased scatter.
    
    Additionally, we observe fluctuations above $660\;\rm nm$, which are well-known fringe effects in back-illuminated CCDs at NIR wavelengths \citep{Malumuth2003}. 
 
\subsection{Spectral Resolution \label{sec:spectral resolution}}
    Spectral resolution is one of the key performance metrics of a spectrograph, determining the limit of separability for adjacent lines. The spectral resolution is defined as the ratio of the wavelength to the instrumental broadening in terms of the line widths \citep{Chromey2010}.
    \begin{equation}
        R={\lambda \over \Delta\lambda}
        \label{eq:spectral_resolution}
    \end{equation}

    In this section, we measure the instrumental broadening using two complementary approaches: the Fourier-based approach and the combination of empirical point-spread function (PSF) with specsheet-based dispersion. The Fourier-based line-stacking method operates entirely in the wavelength space after wavelength calibration. This approach is largely independent of manufacturer specifications and insensitive to unknown geometrical parameters (e.g., grating tilt). The combined image-domain PSF and datasheet dispersion is anchored to the designed performance. It also inherits any inaccuracies due to the discrepancy between the real spectrograph and the ideal design of the manufacturer. 
    
    The intrinsic line widths of the Th-Ar lamp, which we employed for wavelength calibration, are generally much smaller than the instrumental broadening width. Consequently, measured line widths can be taken as direct tracers of the instrumental dispersion. To isolate each emission line, we extract a small ($\pm20$‐pixel) window around well‐separated peaks, ensuring no other spectral line features within $\pm10$ pixels. Despite careful selection, numerous weak lines from the calibration lamp can still contaminate the wings of the primary lines, making a direct line profile fit more challenging. One can stack the isolated lines; the pixelization effect makes the exact center of the line uncertain.
    
    Instead, we use a Fourier‐transform approach. According to the convolution theorem, the observed spectrum $s(\lambda)$ can be understood as the convolution of the individual line profile $f(\lambda)$ and the spectral line distribution function $w(\lambda)$. For a single line centered at $\lambda_0$, $w(\lambda)$ is expressed with a delta function $\delta(\lambda-\lambda_0)$; for multiple lines, $w(\lambda)$ can be expressed with a linear combination of individual lines with different central wavelengths. In Fourier space, convolution becomes a product:
    \begin{equation}
        s(\lambda) = (f*w)(\lambda) \stackrel{\mathcal{F}}{\Leftrightarrow} S(k)=F(k)W(k)
    \end{equation}
    where $k=2\pi/\lambda$ represents an angular wavenumber. $W(k)$ induces oscillation of $|S(k)|$, while the overall envelope is governed by $F(k)$. Based on this framework, we can easily isolate the line profile function, enabling stacking of multiple line profiles to increase the signal-to-noise ratio. Empirically, we can approximate the line profile $f$ with a Gaussian, $f(x)=A\;\text{exp}(-(x-\mu)^2/2\sigma_x^2)$. Because the Fourier transform of a Gaussian is also Gaussian, the dispersion in real space $\sigma_{\lambda}$ relates simply to the dispersion in Fourier space $\sigma_k$ by
    \begin{equation}
        \sigma_\lambda={1\over\sigma_k}
    \end{equation}

\begin{figure*}[ht!]
        \centering
        \includegraphics[width=0.33\linewidth]{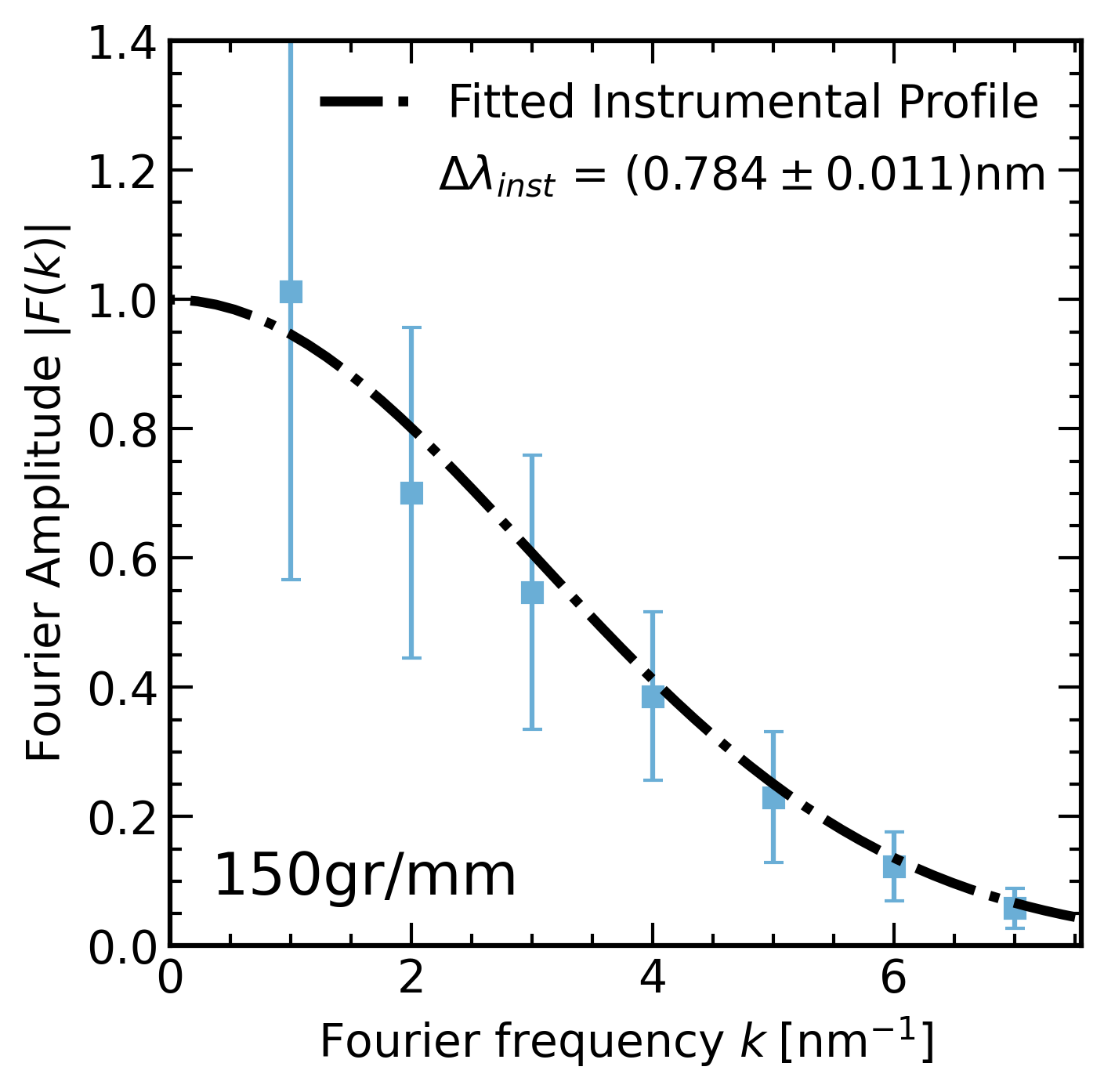}
        \includegraphics[width=0.33\linewidth]{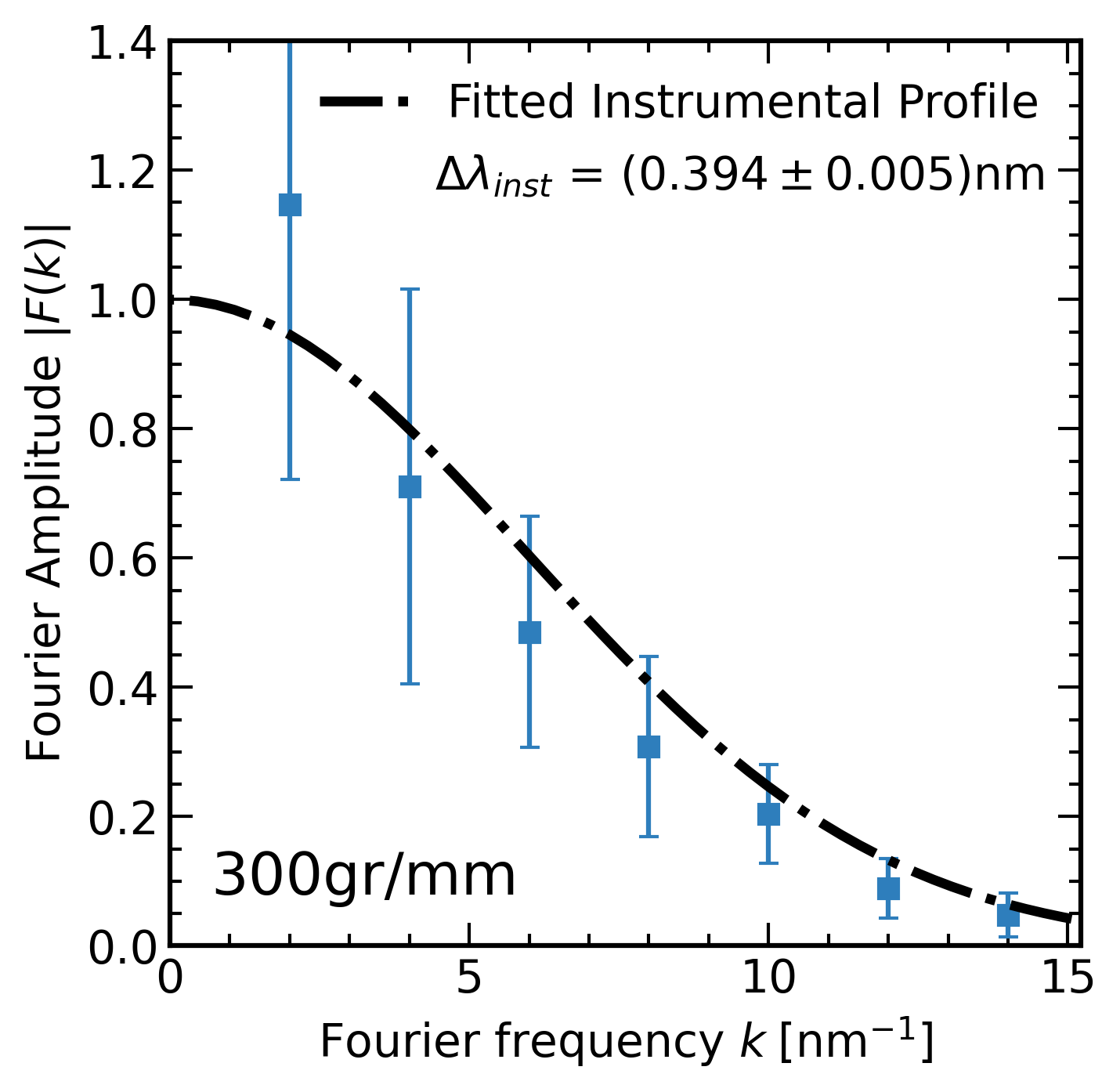}
        \includegraphics[width=0.33\linewidth]{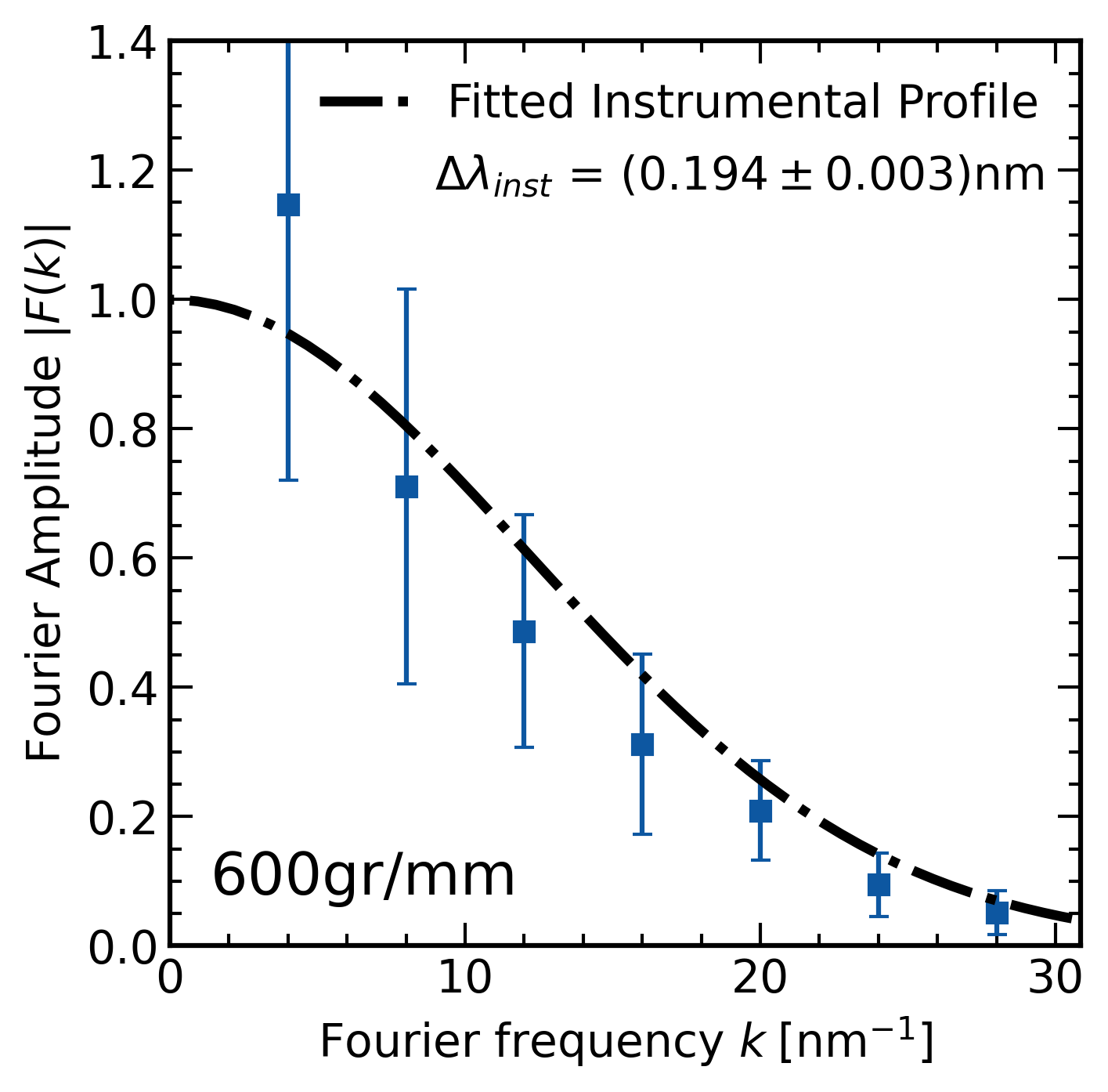}
        \caption{Fourier spectrum of Th-Ar emission lines at 150 $\rm gr\;mm^{-1}$ (left), 300 $\rm gr\;mm^{-1}$ (center), 600 $\rm gr\;mm^{-1}$ grating (right). Blue dots with error bars represent the median Fourier spectrum of instrumental broadening obtained through $k$-space binning of individual line profiles. The dash-dotted line represents a fitted curve with a Gaussian function, presented with the 1$\sigma$ fitting error.}
        \label{fig:fourier}
    \end{figure*}

    In Figure \ref{fig:fourier}, we show the stacked instrumental dispersion profile for all gratings. For each isolated ThAr emission line, we compute its Fourier spectrum as a function of  $k$. We then stack the Fourier spectrum amplitudes by binning in $k$; within each bin, we apply 3-$\sigma$ outlier clipping and take the median as the representative value, while the scatter provides the uncertainty. Although individual lines show considerable scatter because of the limited window size and weak neighbouring features, their ensemble average reveals a smooth, nearly Gaussian profile. We fit this stacked Fourier amplitude with a Gaussian function (dash-dotted curve) and convert the fitted dispersion $\sigma$ into the line width in terms of full-width half maximum (FWHM) equivalent to $2\sqrt{\rm 2ln2}\;\sigma$. This result provides an empirical estimate of the instrumental broadening that is largely insensitive to the exact peak-selection threshold and to pixel-sampling effects in the image domain. It can be compared with the theoretical value, which utilizes the prior knowledge of dispersion from the specification, as discussed next.
    
    Theoretically, one can compute the spectral resolution of a spectrograph exactly from the well-known design parameters \citep[e.g., grating angle, grating groove density, and spectral order;][]{Chromey2010}. However, these details are not always fully accessible or remain uncertain (e.g., commercial ones), particularly when the central wavelength is changed or the setup is adjusted for different wavelength ranges. Moreover, the focal ratio degradation (FRD) inherent to fiber optics \citep{Barden1995} broadens the output beam relative to the input beam, complicating calculations based on simple geometry.
    \begin{figure*}[ht!]
        \centering
        \includegraphics[width=0.33\linewidth]{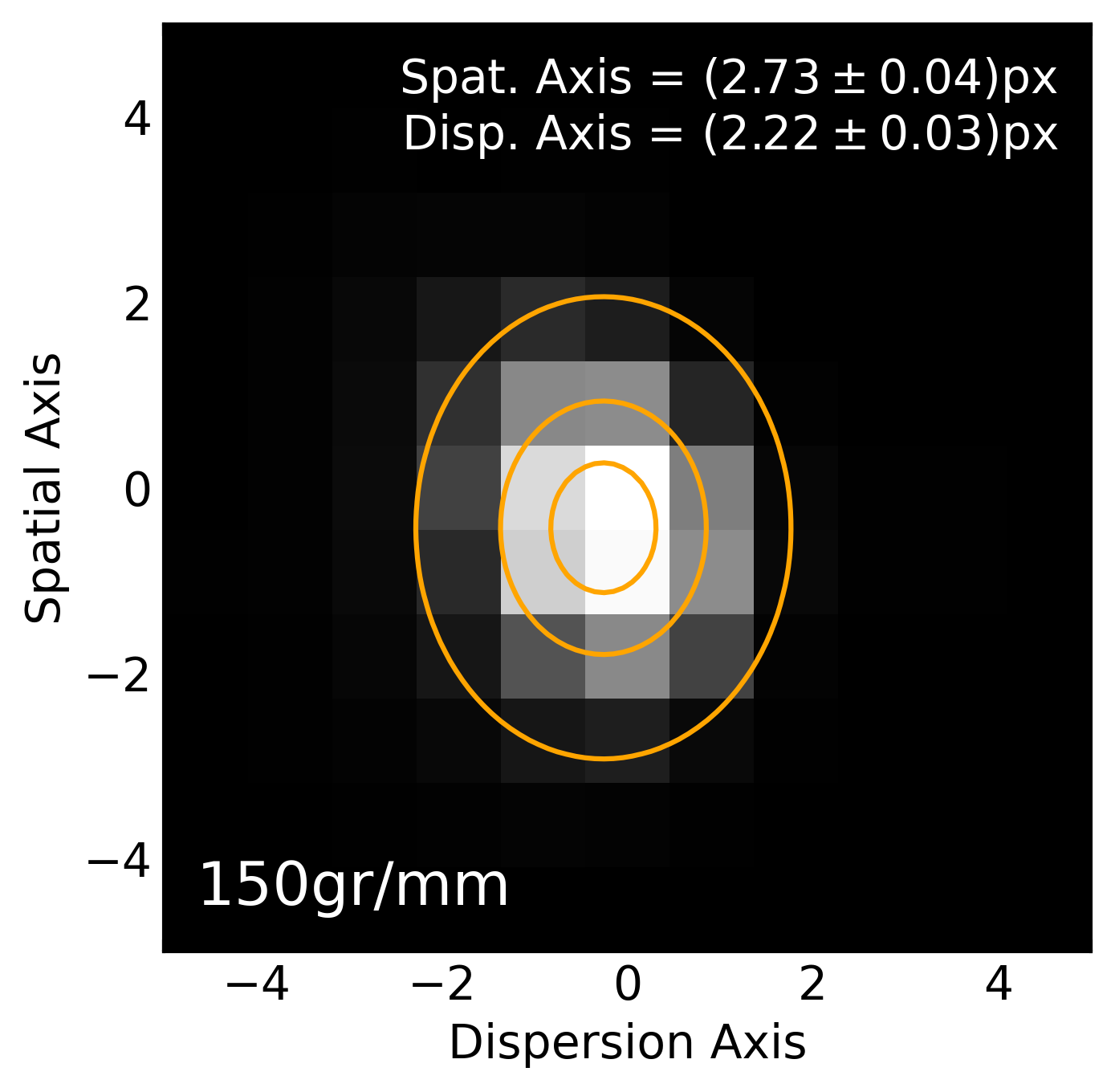}
        \includegraphics[width=0.33\linewidth]{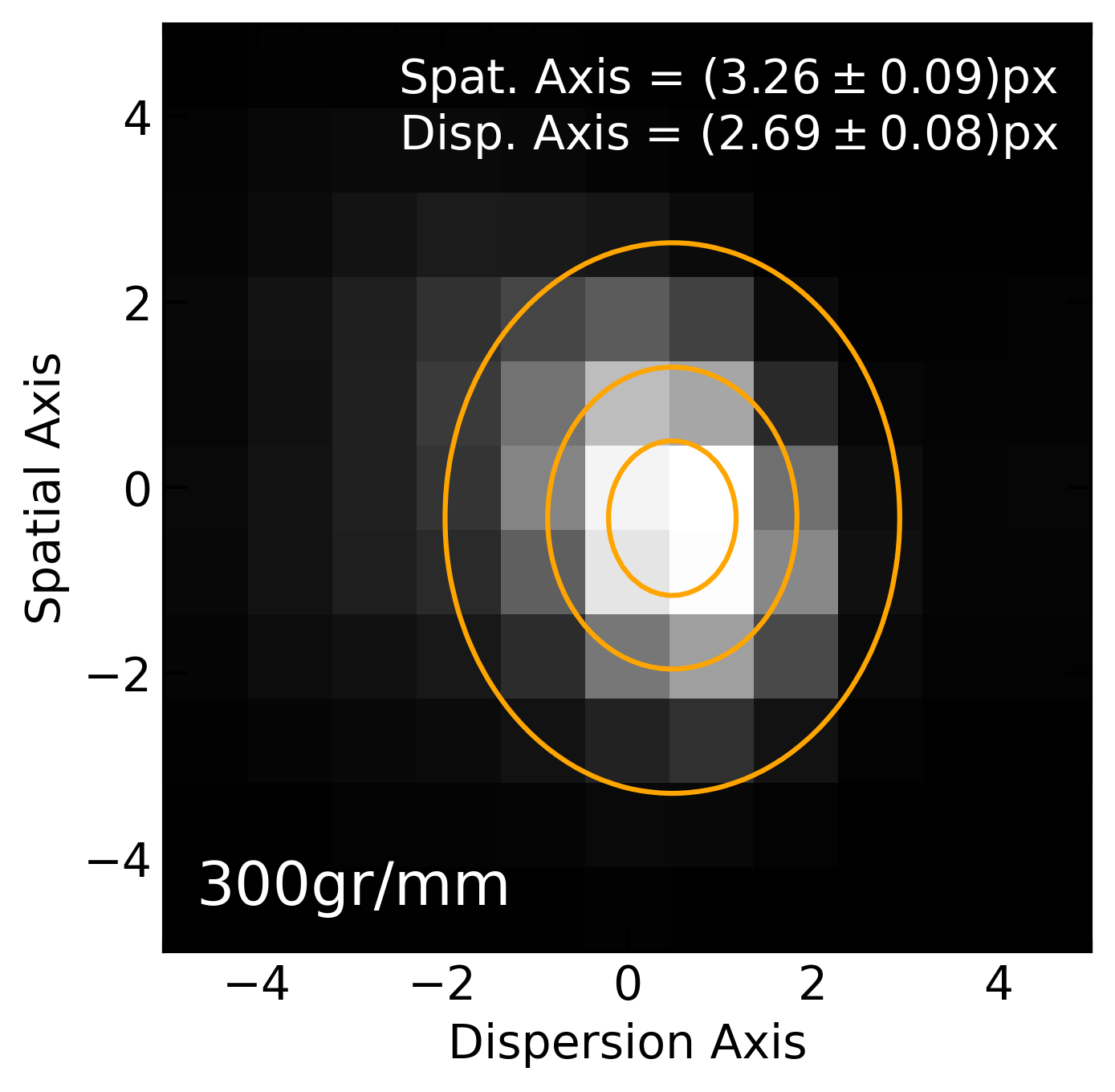}
        \includegraphics[width=0.33\linewidth]{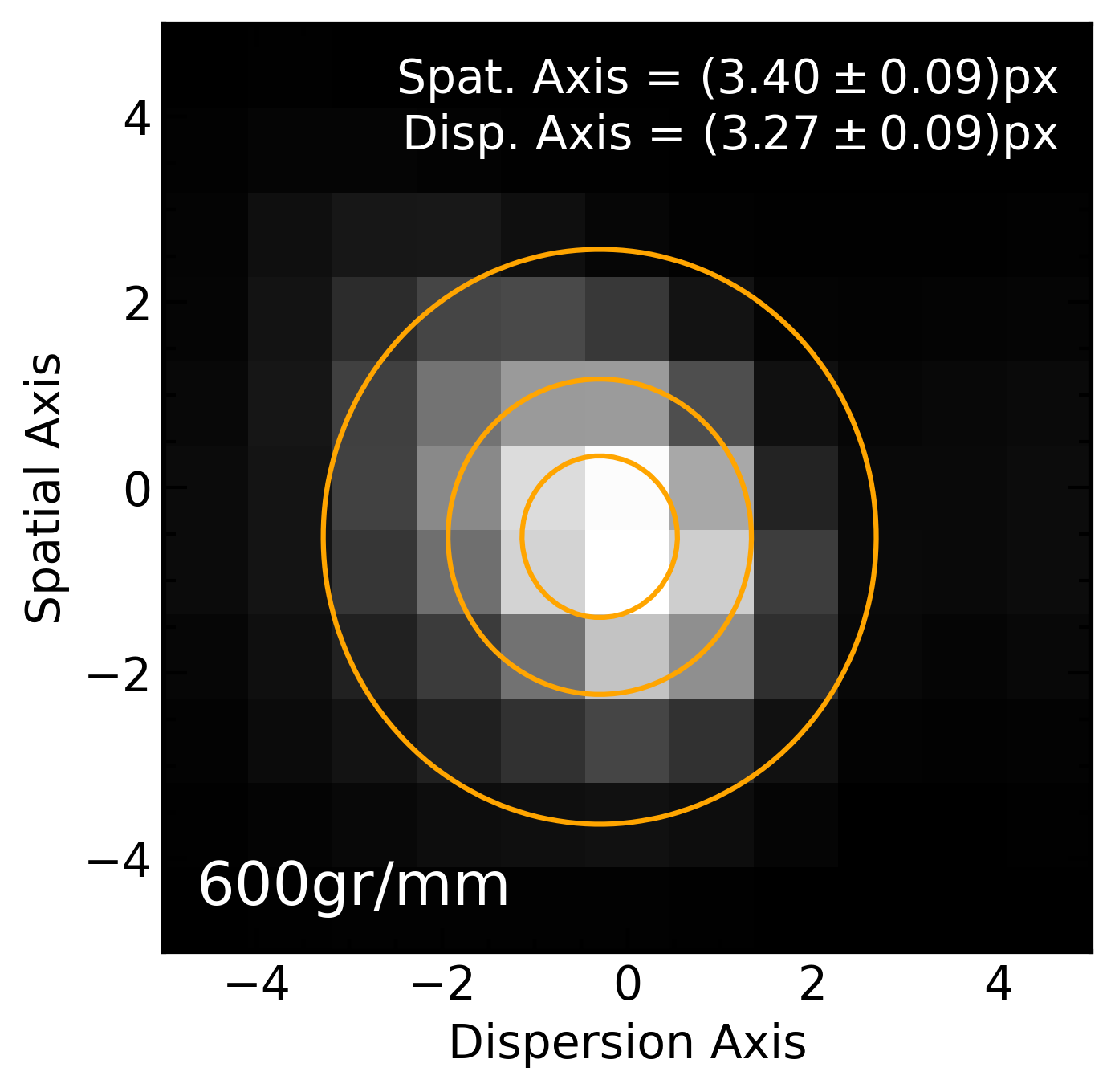}
        \caption{The example of the fiber beam image of Th-Ar spectrum at 150 $\rm gr\;mm^{-1}$ grating (left) and 300 $\rm gr\;mm^{-1}$ grating (right). The orange circles represent the result of 2D-Gaussian fitting (1$\sigma$, FWHM, and 2$\times$FWHM from the smallest one). The numbers in the top right indicate FWHMs of the Gaussian function in both directions with 1$\sigma$ fitting error.}
        \label{fig:psf}
    \end{figure*}
    
    As an alternative, we measure the beam image size directly on the focal plane of the spectrograph and derive the reciprocal linear dispersion from the datasheet provided by the manufacturer. This approach provides a theoretical resolution estimate suitable for comparison with empirical measurements. The reciprocal linear dispersion of a diffraction grating is given by
    \begin{equation}
        {d\lambda \over d\ell}= {s\text{cos}\theta \over m f_{\rm spec}}
    \end{equation}
    where $m$ is the spectral order, $s$ is the groove spacing (i.e., the inverse of groove density, $\rm gr\;mm^{-1}$), $\theta$ is the incident angle, and $f_{\rm spec}$ is the effective focal length of the spectrograph \citep{Schroder1987}. We assume that $\text{cos}\theta/m$ does not vary significantly across different gratings at the same central wavelength. We then derive approximate reciprocal linear dispersions, which are presented in Table \ref{tab:isoplane320a}. Next, we determine the size of the fiber beam image along the dispersion axis by fitting a 2D-Gaussian function to each line as illustrated in Figure \ref{fig:psf}; If we multiply this size by the reciprocal linear dispersion, we can obtain the theoretical line width. 

    In Table \ref{tab:resolution}, we present a comparison between the instrumental widths derived from our Fourier-stacking method against those obtained from direct 2D-Gaussian fits to the fiber-beam image multiplied by the nominal dispersion. At $500\;\rm nm$, the derived resolutions from the Fourier-stacking method are $R\simeq640$, $1270$ and $2580$ for the $150$, $300$, and $600\;\rm gr\;mm{-1}$ gratings, respectively, while the PSF+datasheet approach gives $R\simeq612$, $1010$ and $1660$. These values correspond to rms velocity resolution of about $200$, $100$ and $50\;\rm km\;s^{-1}$ for Fourier-stacking method; $210$, $120$ and $80\;\rm km\;s^{-1}$ for PSF+datasheet method.

    Our measurements reveal the systematic difference between individual methods. While both results agree for the lowest-dispersion grating, the discrepancy increases towards higher dispersion, where the PSF-based estimate becomes more vulnerable to pixel sampling (i.e., uncertainties in the centroid position), and the lower S/N of individual lines, all of which tend to broaden the fitted image profile and thus underestimate $R$.

    In contrast, the Fourier-stacking technique averages over many lines in Fourier space and significantly suppresses random noise and contamination ripples, providing a pure, intrinsic response from our spectrograph. We therefore prioritize Fourier results, which are insensitive to contamination from neighboring lines and to sampling effects, but also present the PSF+datasheet results as an independent cross-check. Also, the discrepancy between the two methods could fall within the expected range, given the finite window size (20 pixels), the pixel sampling limit, and unmodeled effects (e.g., FRD, blaze shifts).
    
    Practically, the $600\;\rm gr\;mm^{-1}$ grating delivers the highest resolution, while the $150\;\rm gr\;mm^{-1}$ setting trades resolution for broader wavelength coverage, beneficial for capturing full-spectral shapes and fainter targets. 
    \begin{table*}[ht]
        \caption{Comparison of spectral resolution of IsoPlane 320A across various diffraction gratings. The instrumental FWHMs are determined by (a) the Fourier line‐profile method and (b) direct 2D‐Gaussian fitting of the beam image on the focal plane. \label{tab:resolution}}
        \centering
        \begin{tabular}{l cccccc}
        \toprule
            \multirow{2}{*}{Grating} &\multicolumn{2}{c}{150 $\rm gr\;mm^{-1}$} & \multicolumn{2}{c}{300 $\rm gr\;mm^{-1}$} & \multicolumn{2}{c}{600 $\rm gr\;mm^{-1}$}\\
            \cmidrule(lr){2-3} \cmidrule(lr){4-5} \cmidrule(lr){6-7}
             & $\Delta \lambda_{\rm inst}\;\rm[nm]$ & $R^{\rm a}$ & $\Delta \lambda_{\rm inst}\;\rm[nm]$ & $R^{\rm a}$ & $\Delta \lambda_{\rm inst}\;\rm[nm]$ & $R^{\rm a}$ \\
        \midrule
            Fourier-based & 0.784$\pm$0.011 & 638$\pm$9 & 0.394$\pm$0.005 & 1269$\pm$16 &  0.194$\pm$0.003 & 2577$\pm$40 \\
            PSF+Datasheet$^{\rm b}$& 0.817$\pm$0.011 & 612$\pm$8 & 0.495$\pm$0.014 & 1010$\pm$29 & 0.301$\pm$0.031 & 1660$\pm$170\\
        \bottomrule
        \end{tabular}
        \tabnote{$^{\rm a}$ Expressed in value at $500\;\rm nm$. \\
                 $^{\rm b}$ Derived value by combining PSF size measurement and the reciprocal linear dispersion in the datasheet.}
    \end{table*}
    
\section{ON-SKY Tests \label{sec:on-sky}}
    To validate the throughput and assess potential systematics of the spectrograph, we carried out test observations and calibrations using a spectrophotometric standard star. In this section, we describe the complete spectroscopic calibrations from raw data to calibrated spectra, including both wavelength and flux calibration. For the wavelength solution, we first fit the Balmer absorption lines in the standard-star spectrum to obtain an initial dispersion relation, and then refine it with Th–Ar arc spectra acquired in the laboratory to achieve higher accuracy. For the flux calibration, we compare the extracted counts of the standard star with its reference spectrum to derive the instrument response (sensitivity function), which we then apply to the target spectra to place them on an absolute flux scale.
    
\subsection{Observations with SAO 1 m telescope \label{sec:observation info}}
    We conducted spectroscopic observations of standard stars with the SAO 1-m telescope using the IsoPlane 320A. We adopted a single-core optical fiber with $50\;\mu\rm m$ diameter, which subtends 1.7 arcseconds on the sky. We chose the well-known bright star Vega ($\alpha$ Lyr) as our spectrophotometric standard for the on-sky test and used its spectrum from CALSPEC \citep{Bohlin2014, Bohlin2020}, which provides stellar spectra for flux standards on the HST system. Observations were carried out on August 24, 2024, on a clear and dark night with an average seeing of 2.1 arcseconds.  
    \begin{table}[ht!]
        \caption{The information of astronomical target for On-Sky test of spectrograph \label{tab:targets}}
        \centering
        \begin{tabular}{lcccc}
        \toprule
            Target & RA & Dec & V [mag] & Desc.\\
        \midrule
            $\alpha$ Lyr & $\rm 18^h 36^m 56^s$ & $\rm +38^\circ 47^\prime 01^{\prime\prime}$ & 0.031 & A0V\\
        \bottomrule
        \end{tabular}
    \end{table}
    
\subsection{Spectroscopic Calibration \label{sec:spec_cal}}
\begin{figure}[ht!]
        \centering
        \includegraphics[width=0.98\linewidth]{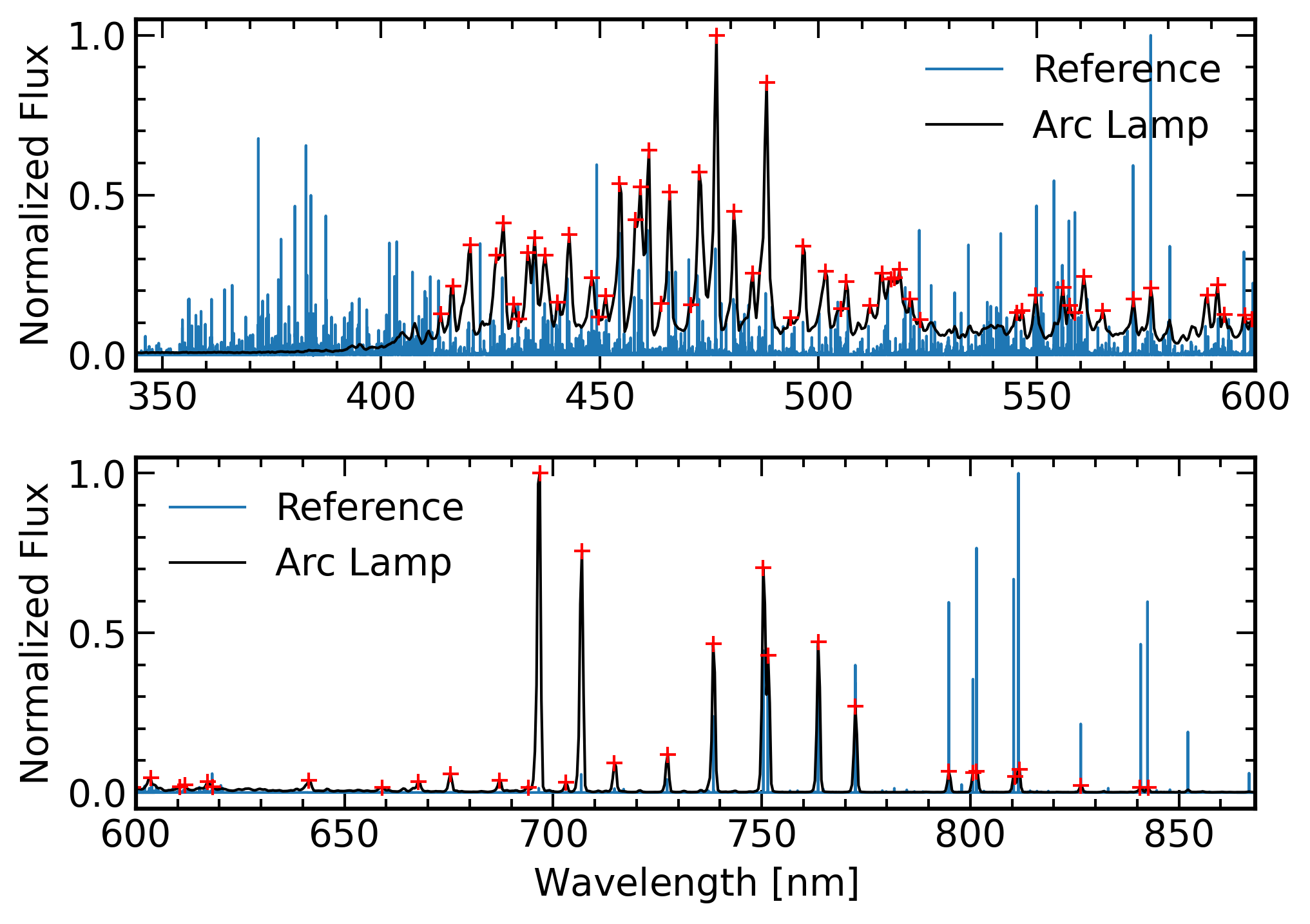}
        \caption{Wavelength calibration result of ThAr lamp spectrum at 150 $\rm gr\;mm^{-1}$ grating. Plots are split at $650\;\rm nm$ to better visualize the dense line regions. Red plus markers represent detected peaks used for wavelength calibration from the observed lamp spectrum.}
        \label{fig:wavelength cal}
    \end{figure}

    We median-combined 5 frames of the target, extracted in 1D spectra as described (Section \ref{sec:multi}), and performed a two-step wavelength calibration: (1) fit of stellar absorption lines with a 2nd-order Chebyshev polynomial, and (2) refit of the Th-Ar lamp spectrum with a 3rd-order polynomial aligned to the NOIRLab spectral atlas\footnote{Line identifications trace back to the Th of \cite{Palmer1983} and the Ar of \cite{Norlén1973}. NOIRLab Th-Ar Spectral Atlas: \url{https://noirlab.edu/science/sites/default/files/media/archives/documents/scidoc2212.fits}}, which yields a 1$\sigma$ calibration error of $0.001\;\rm nm$. Figure \ref{fig:wavelength cal} illustrates that the wavelength calibration was successfully done without any significant offsets along a wide spectral range. The line intensity may differ from the references because the instrumental sensitivity does not account for the property of our lamp being slightly different from the reference. 
    
\begin{figure}[ht!]
        \centering
        \includegraphics[width=1.0\linewidth]{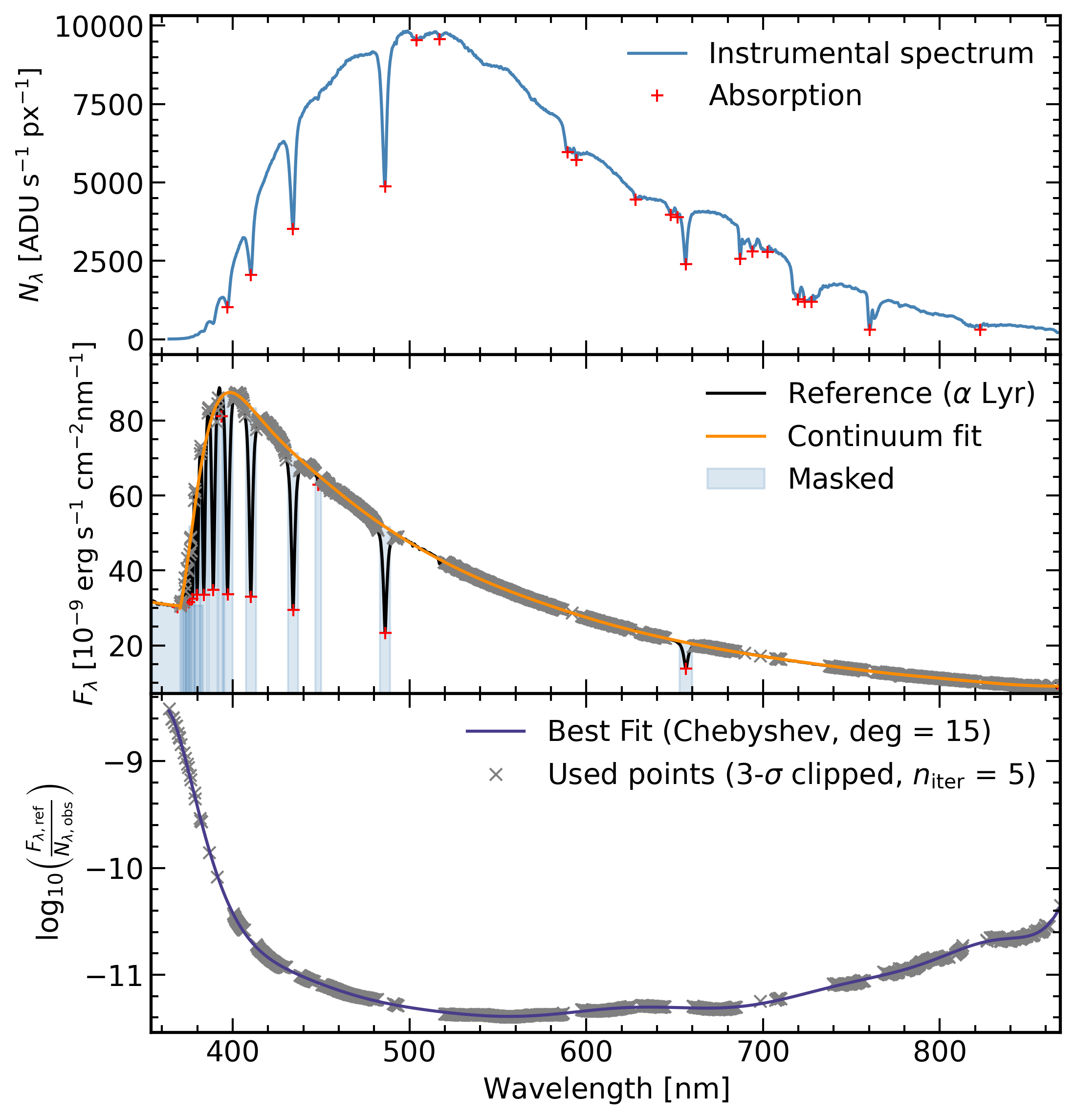}
        \caption{Flux calibration result of Vega with the 150 $\rm gr\;mm^{-1}$ grating. \textbf{Top:} Extracted instrumental spectrum $N_{\lambda,\rm obs}$; red plus markers indicate prominent absorption features. \textbf{Middle:} CALSPEC reference spectrum $F_{\lambda,\rm ref}$ (black) and its smoothed continuum fit (orange) derived from selected continuum points (grey crosses); shaded blue bands indicate masked regions around absorption features. \textbf{Bottom:} Logarithmic response function; grey crosses are the points used after iterative 3-$\sigma$ clipping (five iterations), and the indigo curve is the best-fit Chebyshev polynomial. The fitted curve is adopted as the sensitivity function for flux calibration.}
        \label{fig:flux_cal}
    \end{figure}

    We derive the instrumental response function by comparing the known flux of the reference star to the counts recorded by the spectrograph.
    \begin{equation}
         \mathcal{R}(\lambda)={F_{\lambda,\rm ref}(\lambda) \over N_{\lambda, \rm obs}(\lambda)}
    \end{equation}
    where $F_{\lambda, \rm ref}$ is the reference spectrum and $N_{\lambda,\rm obs}$ is the instrumental spectrum of standard star. We display the comparison between the instrumental and reference spectra in Figure \ref{fig:flux_cal}.
    
    Practically, the instrumental and reference spectra do not share the same resolution and differ due to atmospheric absorption, which can distort the response function. The spectral line fluxes are influenced by instrumental dispersion. The convolution of the instrumental line profile causes the mixing of fluxes from adjacent pixels, resulting in absorption peaks appearing shallower than those in the reference spectrum.  
    
    Additionally, several broad absorption bands at longer wavelengths are attributed to the absorption of molecules in the Earth atmosphere, known as telluric absorption, including molecular oxygen ($\rm O_2$), ozone ($\rm O_3$), and water vapor ($\rm H_2O$). Molecular oxygen absorbs light in specific bands: approximately $628 < \lambda (\rm nm) < 634$ (the $\gamma$ band), $686 < \lambda (\rm nm) < 695$, and $759 < \lambda(\rm nm) < 772$ (the A band) \citep{Smette2015}. Water vapor exhibits absorptions in several bands at wavelengths longer than $620\;\rm nm$ with weaker absorptions \citep{Smette2015}. Ozone absorbs in two main categories: the Huggins bands \citep{Huggins1890} below $400\;\rm nm$ and the Chappuis bands \citep{Chappuis1880} in the range of $500 < \lambda (\rm nm) < 700$. Ozone absorption varies slowly compared to molecular oxygen and water vapor \citep[see Figure 2 of][]{Noll2012}; therefore, ozone absorption survives from outlier masking and is included for obtaining a sensitivity curve. These telluric absorption features can be corrected by published correction pipelines \citep[e.g., \texttt{Molecfit};][]{Smette2015}. A thorough correction of telluric features is beyond the scope of this work; we will focus solely on the overall shape of the spectral response of this spectrograph. 
    
    To isolate the smooth continuum component of the reference spectrum, we model it using a piecewise function made up of two polynomials. For wavelengths below $370\;\rm nm$, a low-order polynomial is used to consider the Balmer break in the stellar spectrum. For the longer wavelength range, a smooth curve is fitted to mask the stellar absorption lines and atmospheric absorption features present in the observed spectrum. We divide the reference spectrum (in physical units) by the observed counts per second to produce a raw response curve and then fit this ratio with a smooth polynomial. At each step, we perform an iterative 3-$\sigma$ outlier clipping on the residuals to discard points contaminated by narrow features. The smooth curve in the bottom panel of Figure \ref{fig:flux_cal} is our final response function, which we then apply to the spectra of scientific targets to convert instrumental counts into absolute flux units.
    
    We note that significant response degradation at the blue end ($<400\;\rm nm$) determines that the lower limit of the reliable spectral range is almost $390\;\rm nm$. This degradation originated from the combination of QE degradation and grating efficiency degradation at wavelengths shorter than $\sim450\;\rm nm$. Meanwhile, the spectral range extends to approximately $900\;\rm nm$, which benefits the accuracy of full-spectral fitting of galaxy spectra.

    We search for 2nd-order contamination in the $>750\rm\;nm$ by exploiting the prominent Balmer break of A0 star at $364.6\rm\;nm$. Any overlap would imprint a spurious feature near $\sim730\rm\;nm$ in the response curve. We see no abrupt change at this wavelength or at other twice the Balmer line positions, so any 2nd-order contribution is below our fit residuals and not significant for this calibration. If it requires a higher precision in the future, it would be good to have a dedicated test with an order-sorting filter isolating the spectrum of the desired order \citep{Stanishev2007}.
    
\section{Discussion \label{sec:discussion}}
    We have conducted a comprehensive evaluation of the PIXIS 1300BX under laboratory conditions. The camera exhibits negligible dark current over typical 15‐minute exposures, maintaining at least an order of magnitude below the readout noise at a sensor temperature of $-55\;^{\circ}\mathrm{C}$. Under these cooling conditions, the spatial thermal noise pattern discussed in Section \ref{sec:pattern noise} also becomes insignificant.

    We extend the noise-separation techniques from Section \ref{sec:PTC} to recover the true photon-transfer behavior of our camera. Although we do not know the exact value of the readout noise and gradient noise separately from the PTC analysis, one can isolate and remove GN by combining with a dark bias analysis. In practice, we take the RDN value derived from the dark bias pair differences (Section \ref{sec:RDN}), subtract it in quadrature from the total noise, and refit the PTC. The resulting gradient-noise-subtracted PTC (Figure \ref{fig:PTC_sep}) reveals that:
    \begin{figure}[ht!]
        \centering
        \includegraphics[width=1\linewidth]{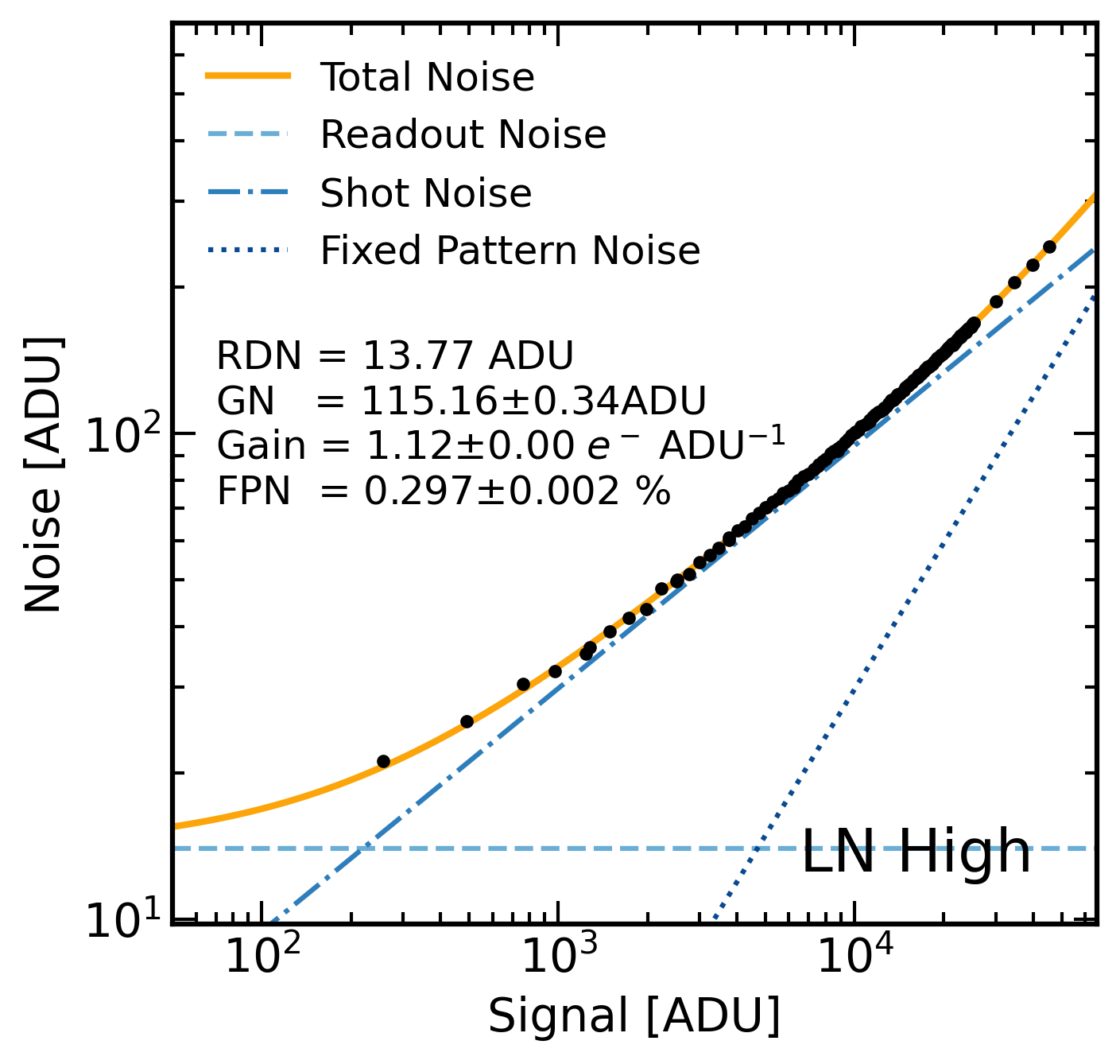}
        \caption{Gradient-noise-subtracted photon transfer curve at low-noise, high-gain mode. The estimated gradient noise is derived by adopting the readout noise from the difference between dark bias frames.}
        \label{fig:PTC_sep}
    \end{figure}
    
    The shot noise dominates the signal level above $\sim100\;\rm ADU$, and fixed pattern noise overtakes at $\sim30,000\;\rm ADU$. Regardless of the gradient noise term, the gain and FPN factor remain identical to the uncorrected fit, confirming our gradient removal only affects the low-signal noise floor. 

    Our camera exhibits excellent linearity, exceeding $99\%$ across the full dynamic range up to near‐saturation, thereby ensuring accurate flux measurements for astronomical sources. A notable feature in high-capacity mode is the piecewise linear response of image sensors for high dynamic range. While beneficial for capturing very bright sources, such functionality introduces complexity due to the multi-modality of the linearity response, complicating data reduction procedures. For observations of faint objects, we recommend employing the low-noise, high-gain mode, which optimizes sensitivity and delivers high signal-to-noise performance for low-level signals. 
    
    Another benefit of our camera is high QE, close to $100\%$ across a broad wavelength range. Compared with commercial scientific CMOS sensors \citep[e.g.,][]{Alarcon2023,Khandelwal2024,Layden2025}, PIXIS 1300BX responses are highly sensitive beyond $500\;\rm nm$; many CMOS devices exhibit significant QE drops at longer wavelengths. Relative to specialized astronomical CCDs used in spectroscopic surveys such as Dark Energy Spectroscopic Instrument \citep[DESI;][]{Bebek2017}, the PIXIS 1300BX achieves comparably flat QE response below approximately $800\;\rm nm$. It begins a gentle decline at roughly $800\;\rm nm$, but DESI detectors typically deteriorate more steeply beyond $900\;\rm nm$. According to \cite{Bebek2017}, the difference in dopant of silicon substrates or AR-coating details leads to a steep near-IR drop. This high, constant QE across wide coverage benefits spectroscopy for extragalactic sources remarkably. High QE allows for obtaining a reliable spectrum of galaxies even close to the Near-IR range ($\sim900\;\rm nm$), which is essential for robust determination of stellar populations and galaxy stellar mass. 
    
    However, several systematic uncertainties are inherent in QE measurements. Obtaining precise QE measurements requires numerous frames across various wavelengths and exposure levels, during which minor fluctuations in the arc‐lamp output ($< 1\%$) occur. Furthermore, our integration sphere setup introduced additional issues: the sphere was quite small relative to the distance to the sensor plane, combined with uncertainties in the exact CCD focal plane positioning behind the vacuum window, both of which introduce small ($< 5\%$) systematic errors in the absolute QE determination. Nonetheless, these issues only alter the QE scale by a few percent and do not significantly change its overall shape.
    
    We have also evaluated the performance of IsoPlane 320A as an astronomical fiber-fed spectrograph, specifically focusing on the spectral resolution. In deriving theoretical spectral resolution, we adopt a simple assumption that the ratio $\text{cos}\theta/m$ does not vary significantly across different central wavelength settings. While such an assumption facilitates a convenient comparison with the designed resolution, the angular responses might deviate from a simple model, due to practical challenges inherent in real spectrograph designs. 
    
    To verify the validity of this assumption, we empirically determined the ratio $\text{cos}\theta/m$, exploiting the slope derived from the pixel-to-wavelength calibration. The ratio exhibits only minor variations across all tested configurations, with deviations limited to approximately 0.01. Such subtle variation confirms that our initial assumption introduces negligible systematic error relative to the overall accuracy of our previous spectral resolution measurements.

    Because the spectrograph is fed by an optical fiber, the achievable spectral resolution is primarily determined by the projected fiber PSF and the internal dispersive element, rather than directly by the aperture of the telescope. The telescope aperture mainly affects the photon-collecting power, while the image quality at the fiber input is set by atmospheric seeing and the telescope optics. The resolving power measured on the SAO 1 m telescope, therefore, provides a good proxy for performance on other telescopes, provided that the seeing conditions and fiber illumination are similar.

    For the A-SPEC survey, the spectrograph will be coupled to the KMTNet-SSO telescope, which typically experiences average seeing conditions of 1.0-2.0 arcsec \citep{Kwon2025}; It is slightly better than the conditions tested at SAO. Using the same fiber diameter and grating configuration, we expect the on-sky resolving power to remain unchanged. Thus, the spectral resolution demonstrated in this commissioning study is representative of, and adequate for, the planned survey observations.

\section{Conclusion}
In this study, we have comprehensively characterized the performance of the PIXIS 1300BX camera in conjunction with the IsoPlane 320A spectrograph, providing key insights relevant for their application in astronomical spectroscopy. Our principal findings are summarized as follows:

\begin{itemize}
    \item When tested standalone, the camera requires the gradient-effect correction. Due to the absence of an internal mechanical shutter, the PIXIS 1300BX exhibits a significant vertical gradient resulting from row-to-row sequential readout. This gradient effect is especially pronounced during illuminated tests and must be carefully corrected for accurate measurements of critical parameters such as linearity, full-well capacity (FWC), photon transfer curves (PTC), and quantum efficiency (QE).

    \item At the typical operating temperature of $-55\;^{\circ}\mathrm{C}$, the camera demonstrates extremely low dark current, at least an order of magnitude below the readout noise over typical astronomical exposure durations (e.g., 15 minutes).

    \item The camera shows excellent linearity, maintaining greater than $99\%$ accuracy up to near-saturation levels, ensuring reliable flux calibration. While the high-capacity, low-to-medium-gain modes exhibit piecewise linear behavior aimed at high-dynamic-range imaging, the low-noise, high-gain mode is optimal for faint-object spectroscopy due to its superior linearity and sensitivity.

    \item The measured QE curve of the PIXIS 1300BX is both high and broad, with $\rm QE\geq80\%$ maintained consistently between $400$ and $800\;\rm nm$, closely matching the specifications provided by the manufacturer. This stable, high QE significantly benefits extragalactic spectroscopy, where sensitivity in the red and near-infrared is critical for robust stellar-population studies.

    \item We found that image persistence effects in the camera are negligible within a short time interval ($\leq10$ s). This indicates minimal impact on sequential observational data.
    \item We performed a comprehensive evaluation of the IsoPlane 320A combined with an optical fiber. The spectrograph showed clear spatial separability of multiple fiber spectra with negligible cross-fiber mixing.
    \item Using the Fourier-domain stacking method of ThAr lamp lines, we measure spectral resolution for different gratings ($150$, $300$, and $600\;\rm gr\;mm^{-1}$). It shows broad agreement with independent PSF-based estimates while revealing a systematic tendency of the latter to underestimate $R$ at the highest-dispersion grating. In practice, $600\;\rm gr\;mm^{-1}$ grating delivers the highest spectral resolution, whereas the $150\;\rm gr\;mm^{-1}$ grating trades resolution for wider wavelength coverage.
    
    \item We calibrated flux with spectrophotometric standard stars using the SAO 1-m telescope. Our on-sky flux calibration agrees well with established reference spectra, successfully reproducing continuum shapes and major absorption features.
\end{itemize}
Overall, the combination of the IsoPlane 320A Spectrograph with the PIXIS 1300BX Camera—when operated with optimal gain and fiber configurations—provides a robust, high-performance instrumental platform, particularly well-suited for astronomical spectroscopic programs with moderate spectral resolution ($R\approx600-2600$).

\section{Code Availability}
The Python codes used in this work are publicly available on GitHub.
The camera-control and data-acquisition software is provided by the \texttt{PICam} package\footnote{\url{https://github.com/jiwon-astro/PICAM}}. The data reduction routines are implemented in \texttt{eval\_CCD}, which provides CCD characterization tools,\footnote{\url{https://github.com/jiwon-astro/eval_CCD}}, and \texttt{eval\_Spec}, which provides routines for spectrograph evaluation.\footnote{\url{https://github.com/jiwon-astro/eval_Spec}}. The data products underlying this article can be shared with the corresponding author upon request.

\acknowledgments
We thank the referee for constructive comments. 
We are grateful to Jinguk Seo, the SAO operator, for helping with the instrumental setup and practical help during the on-sky tests. 
HSH acknowledges the support of the National Research Foundation of Korea (NRF) grant funded by the Korea government (MSIT), NRF-2021R1A2C1094577, 
Samsung Electronic Co., Ltd. (Project Number IO220811-01945-01), and Hyunsong Educational \& Cultural Foundation.
The work of HB was supported by Basic Science Research Program through the National Research Foundation of Korea (NRF) funded by the Ministry of Education (RS-2025-25403440).
DK acknowledges the support of the Global-LAMP Program of the National Research Foundation of Korea (NRF) grant funded by the Ministry of Education (No. RS-2023-00301976).
This work was partially supported by the Korea Astronomy and Space Science Institute under the R\&D program (Project No. 2025-1-831-00), supervised by the Korea Aerospace Administration.
A-SPEC is an all-sky spectroscopic survey of nearby galaxies conducted with the K-SPEC instrument. A-SPEC is managed by the Korean Spectroscopic Survey Consortium for the Participating Institutions including Korea Astronomy and Space Science Institute (KASI), Korea Institute for Advanced Study (KIAS), and Seoul National University (SNU).

\appendix
\counterwithin{figure}{section}
\section{Gradient Noise \label{sec:grad noise}}
A gradient noise is inevitable when characterizing shutter-less imaging sensors in laboratory conditions. As introduced in Section \ref{sec:PTC}, this effect has been incorporated into the total noise model presented in Equation \eqref{eq:noise}. 
 
Let $F[i,j]$ denote the incident flux at pixel $(i,j)$, where $i$ and $j$ indicate the row and column indices, respectively. In addition to the nominal exposure time $t_{\rm exp}$, each pixel is exposed for an additional time, $t_{\rm excess}(i)$ that depends on the row index. Consequently, the detected signal $S[i,j]$ at pixel $(i,j)$ is given by:
\begin{equation}
    \begin{aligned}
        S[i,j]&=F[i,j]\left(t_{\rm exp}+t_{\rm excess}(i)\right) \\
        &=F[i,j](t_{\rm exp}+t_{\rm offset})+\sum_{k<i} F[k,j]t_{\rm row}
    \end{aligned}
 \label{eq:signal}
\end{equation}
where $t_{\rm offset}$ is an additional constant exposure time, $t_{\rm row}$ is the time interval between photon transfer to the next row, corresponding to $N_{\rm col}/f_{\rm ADC}$. This equation models the accumulation of the photoelectrons during charge transfer along the serial register of the CCD, which results in a signal dependent on both pixel location and readout speed.  

The shot noise $\sigma_{\rm shot}$ is proportional to the square root of the signal. For clarity, we separate the total signal into two terms, $S_0$ and $S_{\rm excess}$, defined by the product of $F[i,j]$ with $t_{\rm exp}$ and $t_{\rm excess}$, respectively. The shot noise at pixel $(i,j)$ can be written as
\begin{align}
\sigma_{\rm shot}& =\sqrt{{S \over G}} \nonumber\\
&= \sqrt{{1\over G}(S_0+S_{\rm excess})} \nonumber\\
& = \sqrt{{S_0\over G}+\sigma_{\rm GN}^2} 
\label{eq:shot_noise}
\end{align}
where $G$ represents the gain of imaging sensor and $\sigma_{\rm GN}$ denotes the gradient noise. Within this framework, the additional shot noise contribution from $S_{\rm excess}$ constitutes the gradient noise term introduced in Section \ref{sec:gradient}. This noise behaves similarly to readout noise, making it difficult to distinguish between the two due to their degeneracy.

\section{Photon Transfer Curve\label{sec:PTC_app}}
\begin{figure}[ht!]
    \centering
    \includegraphics[width=0.94\linewidth]{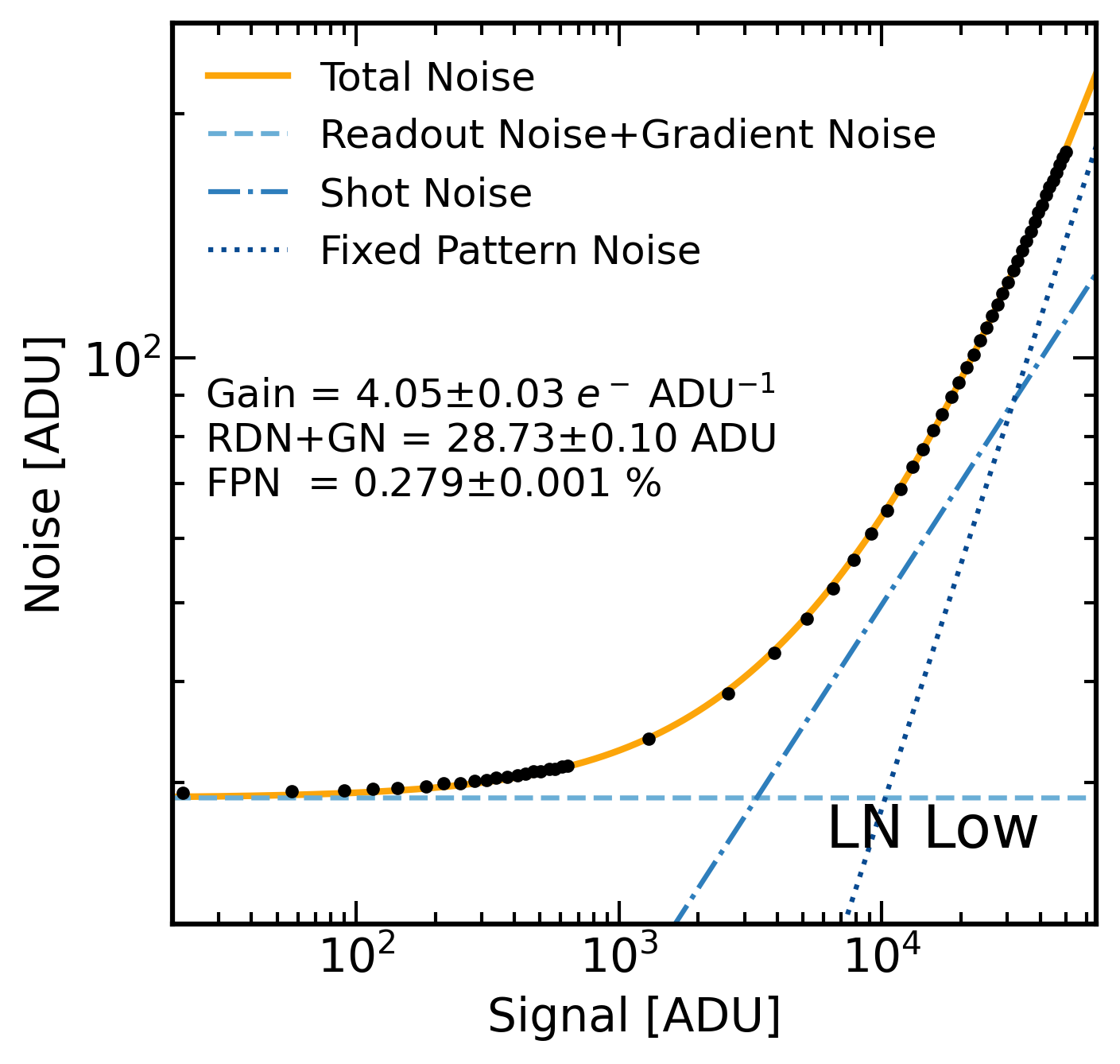}
    \includegraphics[width=1.0\linewidth]{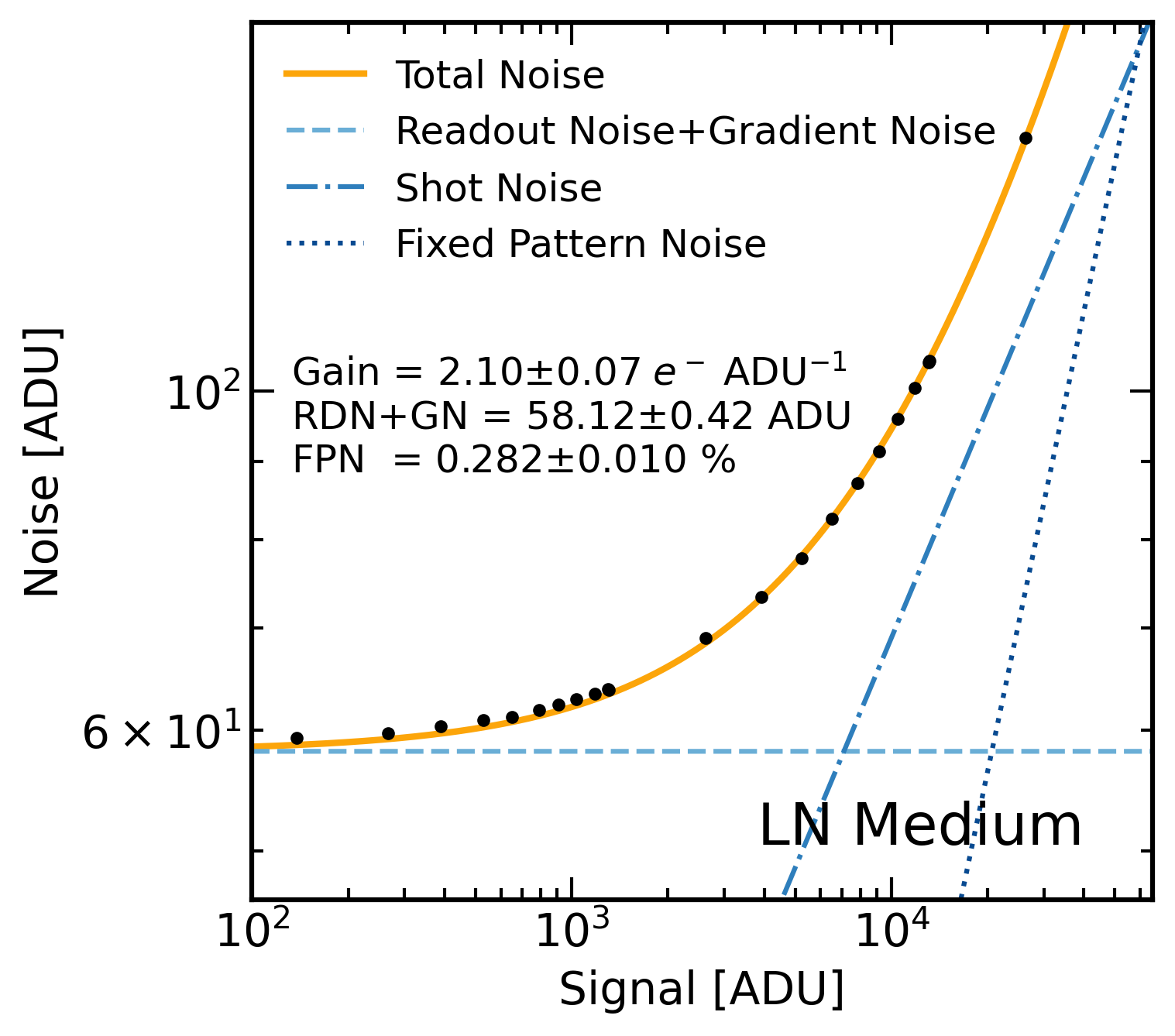}
    \caption{Photon transfer curve at low-noise acquisition mode with low (upper) and medium gain (lower) setup.}
    \label{fig:PTC_app1}
\end{figure}

\begin{figure}[ht!]
    \centering
    \includegraphics[width=0.98\linewidth]{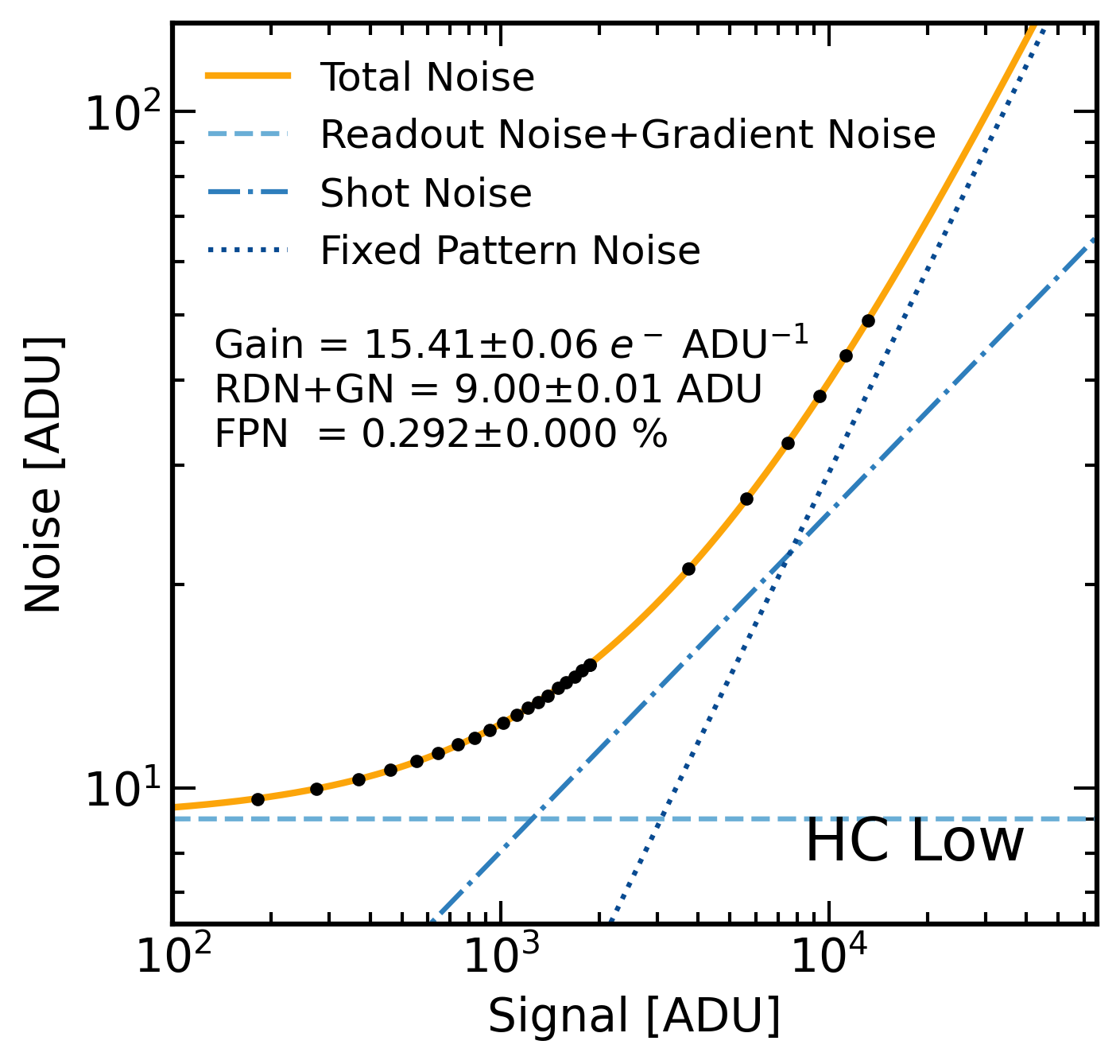}
    \includegraphics[width=0.98\linewidth]{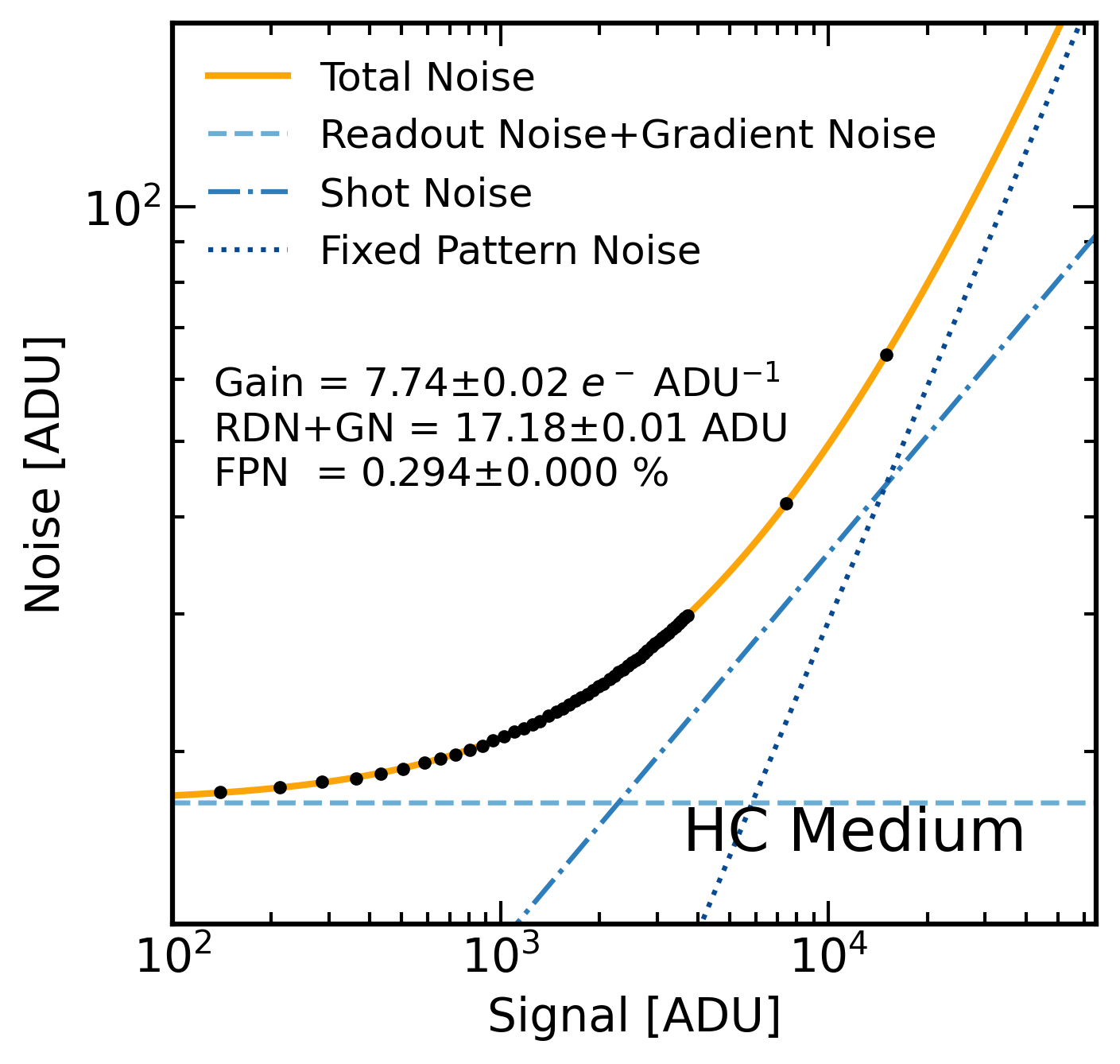}
    \includegraphics[width=0.98\linewidth]{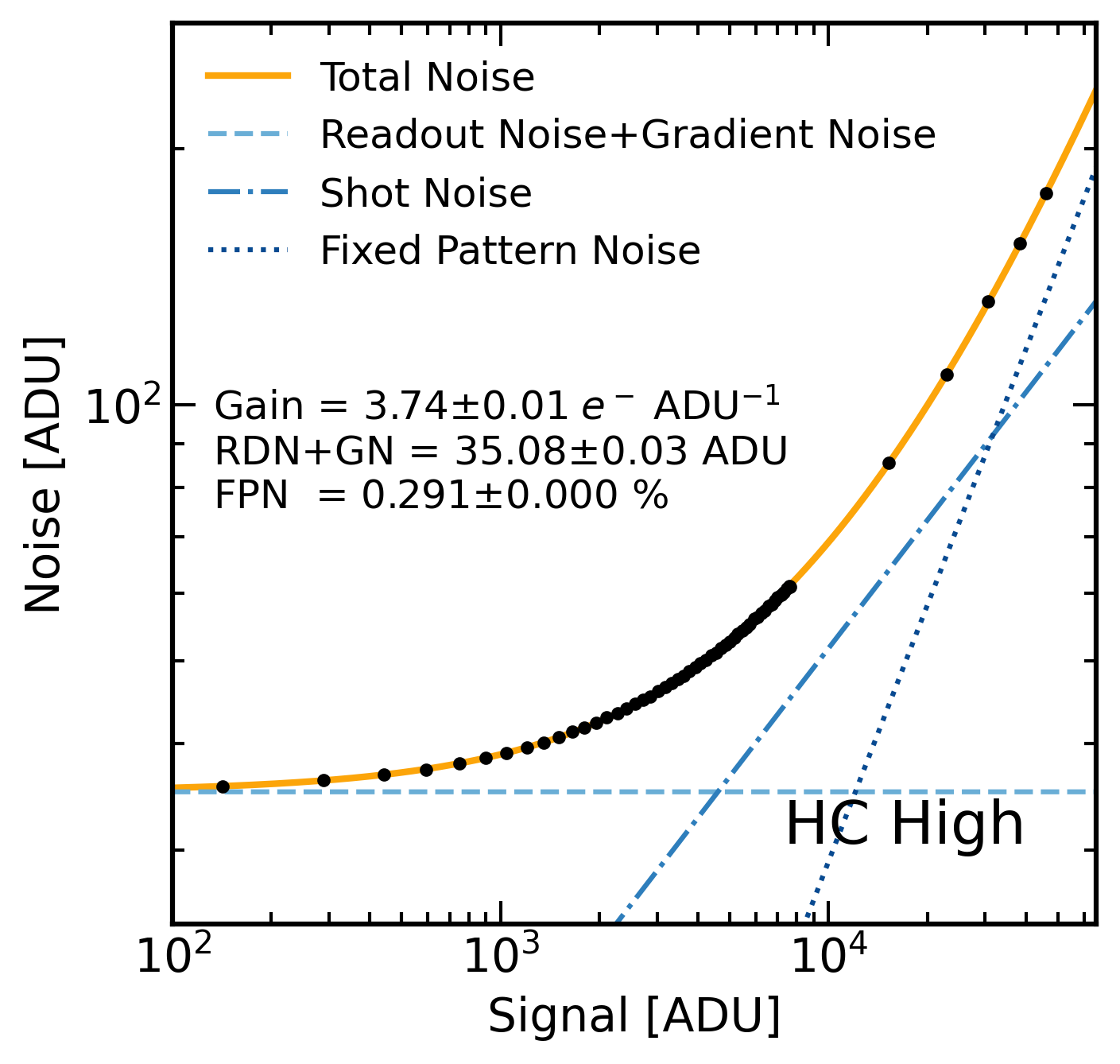}
    \caption{Photon transfer curve at high-capacity acquisition mode with low (upper), medium (center), and high-gain (lower) setup.}
    \label{fig:PTC_app2}
\end{figure}

Figures \ref{fig:PTC_app1} and \ref{fig:PTC_app2} plot the photon transfer curve for the remaining acquisition modes and gain selections. In each case, we performed the same procedure as in the main text. The best-fit parameters are summarized in Table \ref{tab:ptc}. 

\section{Linearity Curve\label{sec:FWC_app}}
\begin{figure}[ht]
    \centering
    \includegraphics[width=0.98\linewidth]{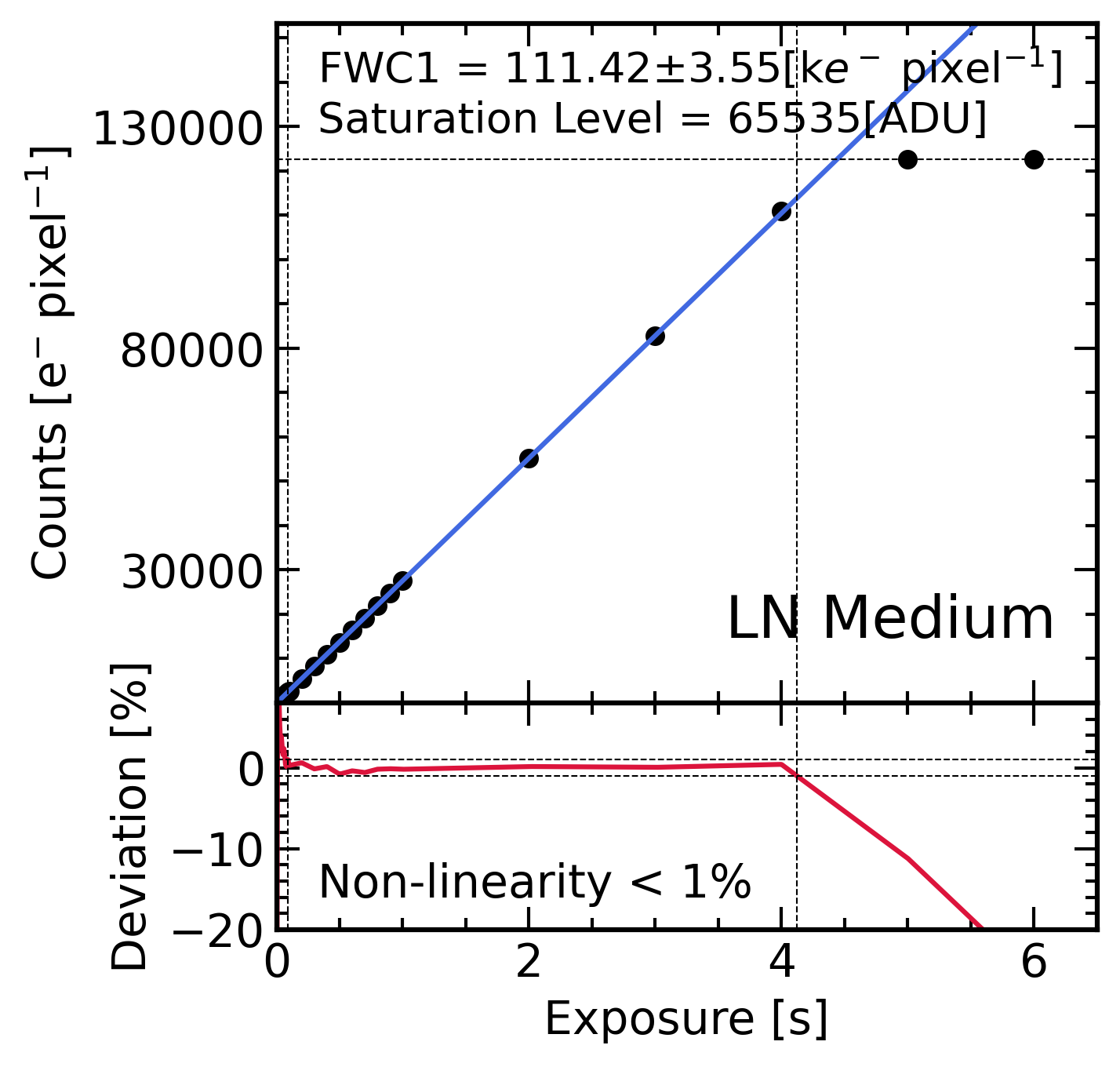}
    \includegraphics[width=0.99\linewidth]{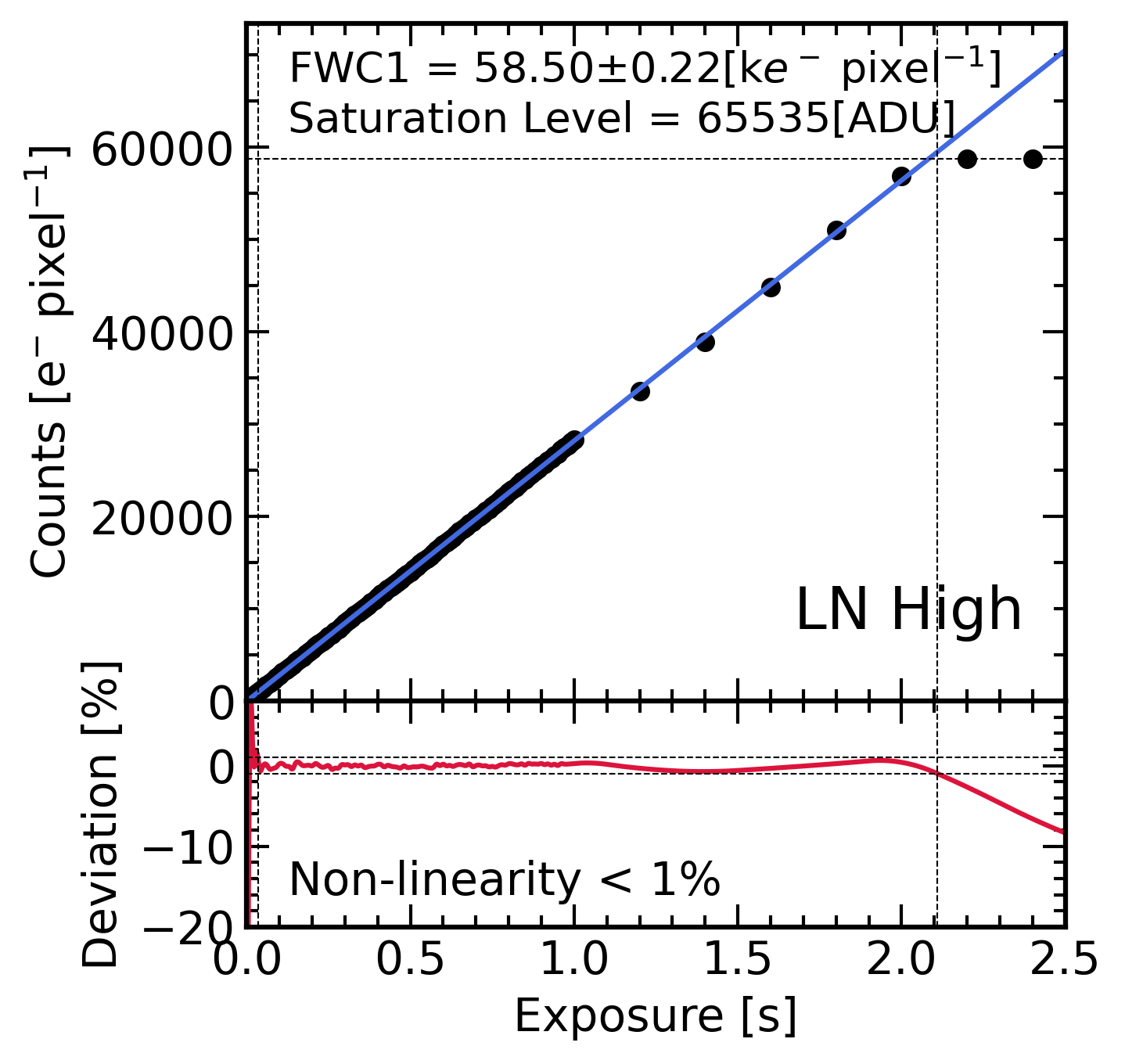}
    \caption{Linearity curve at low-noise acquisition mode with medium (upper) and high-gain (lower) setup.}
    \label{fig:FWC_app1}
\end{figure}

\begin{figure}[ht]
    \centering
    \includegraphics[width=0.99\linewidth]{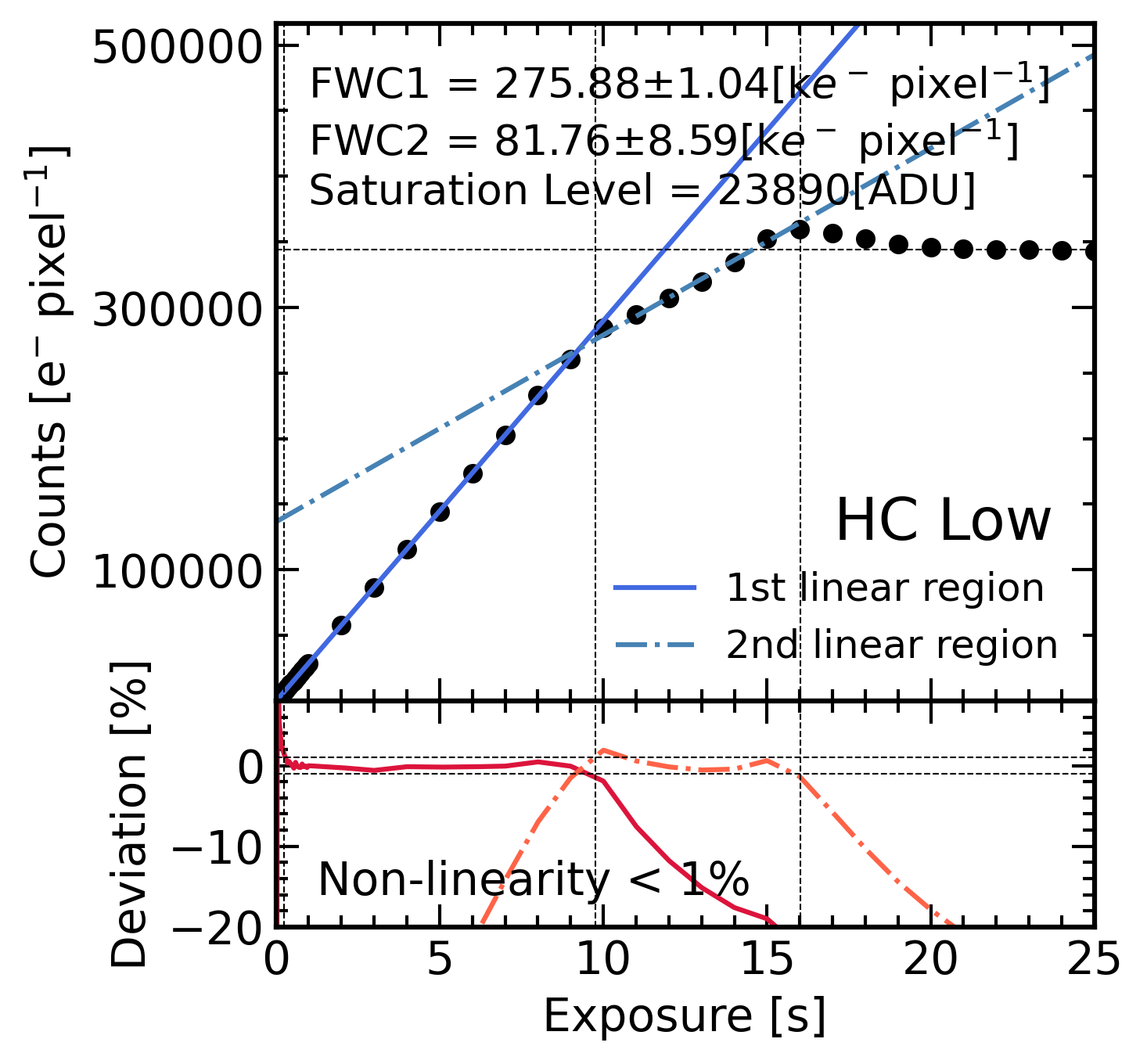}
    \includegraphics[width=0.98\linewidth]{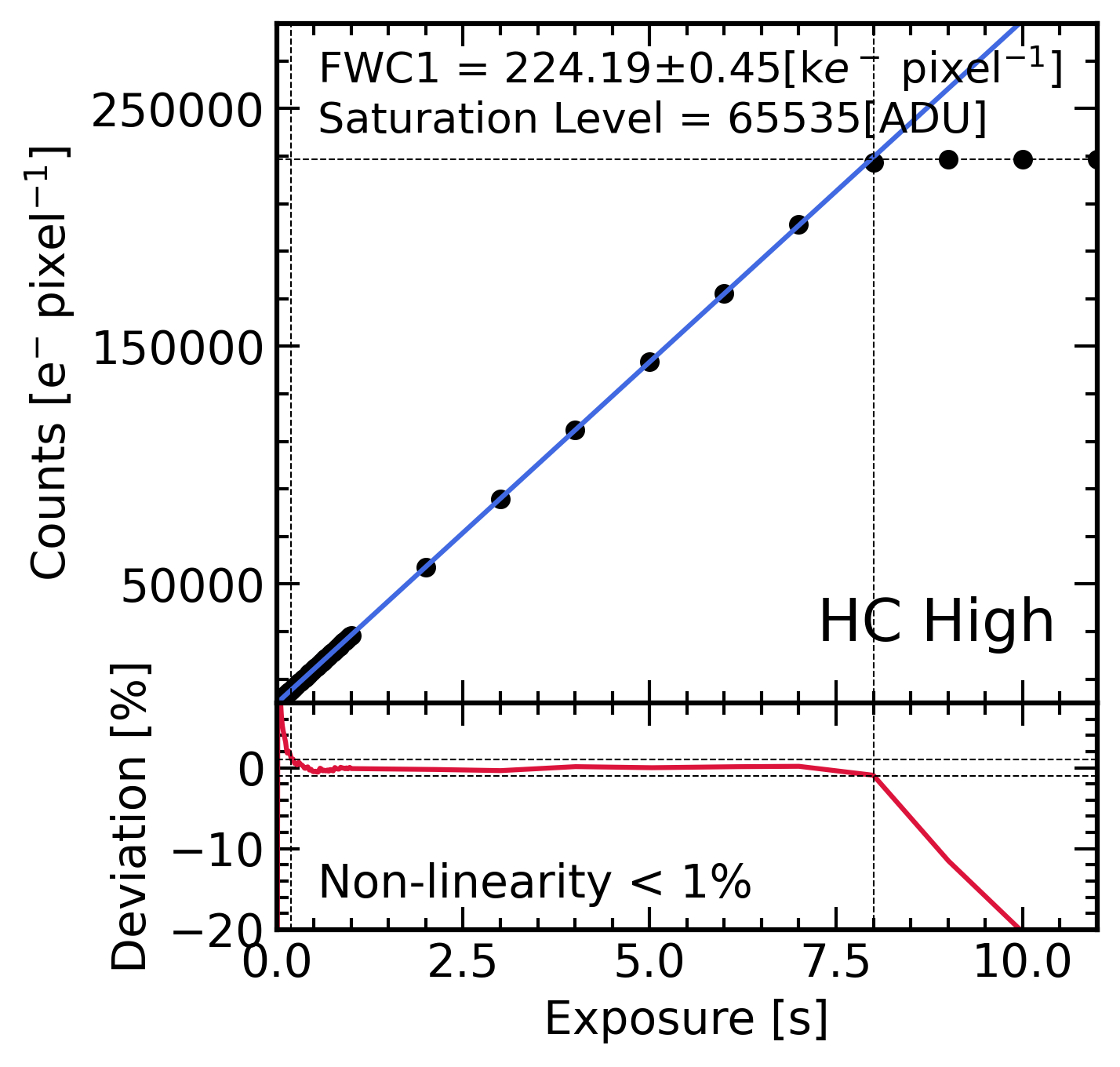}
    \caption{Linearity curve at high-capacity acquisition mode with low (upper) and high-gain (lower) setup.}
    \label{fig:FWC_app2}
\end{figure}

Figures \ref{fig:FWC_app1} and \ref{fig:FWC_app2} display the linearity curves for each acquisition setting. The bottom panel illustrates the deviation from the best linear fit across the full dynamic range. The full-well capacity, defined by a non-linearity of $<1\%$, is summarized in Table \ref{tab:ptc}.



\bibstyle{aasmacros+jkas}
\bibliography{ref}

\end{document}